\documentstyle[12pt]{article}

\def\hybrid{\topmargin -20pt    \oddsidemargin 0pt
        \headheight 0pt \headsep 0pt
        \textwidth 6.25in       
        \textheight 9.5in       
        \marginparwidth .875in
        \parskip 5pt plus 1pt   \jot = 1.5ex}

\hybrid

\def\baselinestretch{1.2}

\catcode`\@=11

\def\marginnote#1{}
\newcount\hour
\newcount\minute
\newtoks\amorpm
\hour=\time\divide\hour by60
\minute=\time{\multiply\hour by60 \global\advance\minute by-\hour}
\edef\standardtime{{\ifnum\hour<12 \global\amorpm={am}%
        \else\global\amorpm={pm}\advance\hour by-12 \fi
        \ifnum\hour=0 \hour=12 \fi
        \number\hour:\ifnum\minute<10 0\fi\number\minute\the\amorpm}}
\edef\militarytime{\number\hour:\ifnum\minute<10 0\fi\number\minute}

\def\draftlabel#1{{\@bsphack\if@filesw {\let\thepage\relax
   \xdef\@gtempa{\write\@auxout{\string
      \newlabel{#1}{{\@currentlabel}{\thepage}}}}}\@gtempa
   \if@nobreak \ifvmode\nobreak\fi\fi\fi\@esphack}
        \gdef\@eqnlabel{#1}}
\def\@eqnlabel{}
\def\@vacuum{}
\def\draftmarginnote#1{\marginpar{\raggedright\scriptsize\tt#1}}

\def\draft{\oddsidemargin -.5truein
        \def\@oddfoot{\sl preliminary draft \hfil
        \rm\thepage\hfil\sl\today\quad\militarytime}
        \let\@evenfoot\@oddfoot \overfullrule 3pt
        \let\label=\draftlabel
        \let\marginnote=\draftmarginnote
   \def\@eqnnum{(\theequation)\rlap{\kern\marginparsep\tt\@eqnlabel}%
\global\let\@eqnlabel\@vacuum}  }

\def\preprint{\twocolumn\sloppy\flushbottom\parindent 2em
        \leftmargini 2em\leftmarginv .5em\leftmarginvi .5em
        \oddsidemargin -.5in    \evensidemargin -.5in
        \columnsep .4in \footheight 0pt
        \textwidth 10.in        \topmargin  -.4in
        \headheight 12pt \topskip .4in
        \textheight 6.9in \footskip 0pt
        \def\@oddhead{\thepage\hfil\addtocounter{page}{1}\thepage}
        \let\@evenhead\@oddhead \def\@oddfoot{} \def\@evenfoot{} }

\def\numberbysection{\@addtoreset{equation}{section}
        \def\theequation{\thesection.\arabic{equation}}}

\def\underline#1{\relax\ifmmode\@@underline#1\else
        $\@@underline{\hbox{#1}}$\relax\fi}

\def\titlepage{\@restonecolfalse\if@twocolumn\@restonecoltrue\onecolumn
     \else \newpage \fi \thispagestyle{empty}\c@page\z@
        \def\thefootnote{\fnsymbol{footnote}} }

\def\endtitlepage{\if@restonecol\twocolumn \else \newpage \fi
        \def\thefootnote{\arabic{footnote}}
        \setcounter{footnote}{0}}  

\catcode`@=12
\relax

\def\figcap{\section*{Figure Captions\markboth
        {FIGURECAPTIONS}{FIGURECAPTIONS}}\list
        {Figure \arabic{enumi}:\hfill}{\settowidth\labelwidth{Figure
999:}
        \leftmargin\labelwidth
        \advance\leftmargin\labelsep\usecounter{enumi}}}
 \relax
\def\tablecap{\section*{Table Captions\markboth
        {TABLECAPTIONS}{TABLECAPTIONS}}\list
        {Table \arabic{enumi}:\hfill}{\settowidth\labelwidth{Table
999:}
        \leftmargin\labelwidth
        \advance\leftmargin\labelsep\usecounter{enumi}}}
 \relax
\def\reflist{\section*{References\markboth
        {REFLIST}{REFLIST}}\list
        {[\arabic{enumi}]\hfill}{\settowidth\labelwidth{[999]}
        \leftmargin\labelwidth
        \advance\leftmargin\labelsep\usecounter{enumi}}}
 \relax
\makeatletter
\newcounter{pubctr}
\def\publist{\@ifnextchar[{\@publist}{\@@publist}}
\def\@publist[#1]{\list
        {[\arabic{pubctr}]\hfill}{\settowidth\labelwidth{[999]}
        \leftmargin\labelwidth
        \advance\leftmargin\labelsep
        \@nmbrlisttrue\def\@listctr{pubctr}
        \setcounter{pubctr}{#1}\addtocounter{pubctr}{-1}}}
\def\@@publist{\list
        {[\arabic{pubctr}]\hfill}{\settowidth\labelwidth{[999]}
        \leftmargin\labelwidth
        \advance\leftmargin\labelsep
        \@nmbrlisttrue\def\@listctr{pubctr}}}
 \relax
\makeatother
\newskip\humongous \humongous=0pt plus 1000pt minus 1000pt

\newif\ifdtup

\relax

\def\be{\begin{equation}}
\def\ee{\end{equation}}
\def\ba{\begin{eqnarray}}
\def\ea{\end{eqnarray}}

\def\G{\Gamma}

\def\e{\epsilon}

\def\no{\noindent}

\def\IR{\relax{\rm I\kern-.18em R}}

\def\pa{\partial}

\def\IR{\relax{\rm I\kern-.18em R}}
\def\inv{^{\raise.15ex\hbox{${\scriptscriptstyle -}$}\kern-.05em 1}}

\def\sp{\partial\!\!\!\!/}

\def\sv{\varphi\!\!\!\!/}

\def\scF{{\cal F}\!\!\!\!/}

\begin{document}

\renewcommand{\theequation}{\thesection.\arabic{equation}}

\newcommand{\beq}{\begin{equation}}
\newcommand{\eeq}[1]{\label{#1}\end{equation}}
\newcommand{\ber}{\begin{eqnarray}}
\newcommand{\eer}[1]{\label{#1}\end{eqnarray}}
\newcommand{\eqn}[1]{(\ref{#1})}
\begin{titlepage}
\begin{center}

\hfill hep--th/0211257\\
\vskip -.1 cm
\hfill CERN-TH/2002-210\\
\vskip -.1 cm
\hfill November 2002\\

\vskip .4in

{\large \bf On integrable models from pp-wave string backgrounds}

\vskip 0.3in

{\bf Ioannis Bakas}

\vskip 0.1in
{\em Department of Physics, University of Patras \\
GR-26500 Patras, Greece\\
\footnotesize{\tt bakas@ajax.physics.upatras.gr}}\\

\vskip0.3in

{\bf Jacob Sonnenschein}
\vskip 0.1in
{\em  School of
Physics and Astronomy\\
Beverly and Raymond Sackler Faculty of Exact
Sciences\\
Tel Aviv University, Ramat Aviv, 69978, Israel\\
\footnotesize{\tt cobi@post.tau.ac.il}}\\
\vskip 0.1in
{\em Theory Division, CERN \\
CH-1211 Geneva 23, Switzerland\\
\footnotesize{\tt jacob.sonnenschein@cern.ch}}\\\end{center}

\vskip .3in

\centerline{\bf Abstract}

\no
We construct solutions of type IIB supergravity with non-trivial
Ramond-Ramond 5-form in ten dimensions by replacing the transverse flat space
of pp-wave backgrounds with exact $N=(4,4)$ $c=4$ superconformal field theory
blocks. These solutions, which also include a dilaton and (in some cases)
an anti-symmetric tensor field, lead to integrable models on the
world-sheet in the light-cone gauge of string theory. In one instance
we demonstrate explicitly the emergence of the complex sine-Gordon model,
which coincides with integrable perturbations of the corresponding
superconformal building blocks in the transverse space. In other cases we
arrive at the supersymmetric Liouville theory or at the complex sine-Liouville model.
For axionic instantons in the transverse space, as for the (semi)-wormhole
geometry, we obtain an entire class of supersymmetric pp-wave backgrounds
by solving the Killing spinor equations as in flat space, supplemented by the
appropriate chiral projections; as such, they generalize the usual Neveu-Schwarz
five-brane solution of type IIB supergravity in the presence of a
Ramond-Ramond 5-form.
We also present some further examples of interacting light-cone models and we
briefly discuss the role of dualities in the resulting string theory backgrounds.

\vfill
\end{titlepage}
\eject

\def\baselinestretch{1.2}
\baselineskip 16 pt
\noindent

\section{Introduction}
\setcounter{equation}{0}

String theory on plane-fronted wave backgrounds, such as pp-waves, has been
studied extensively in the past for two main reasons. First, under certain
circumstances, they provide solutions to string theory that are exact to all
orders in $\alpha^{\prime}$, and hence their properties are fully captured
by the lowest order effective supergravity \cite{ama}, \cite{hor}.
Second, in many cases, they provide
exactly solvable string models, where the gravitational background is
non-trivial and yet tractable, as in the simpler case of strings propagating in
flat Minkowski space \cite{nw}, \cite{kkl}, \cite{tsey1}.
The main advantage of these backgrounds is given by the
existence of a null isometry, which is adapted to the light-cone gauge,
where string theory can be quantized in relatively simple terms. Plane-fronted
wave geometries can be written using Brinkman (instead of Rosen)
coordinates in the form
\be
ds^2 = -4dx^+ dx^- - F(y; x^+) \left(dx^+\right)^2 + dy_i dy^i ~,
\ee
where the front factor $F$ may depend in general on both $x^+$ and the transverse
coordinates $y^i$. Thus, in the light-cone gauge,
$x^+ = \tau$, one obtains a two
dimensional action that involves interactions among the world-sheet
bosons $y^i$, which are summarized by $F$; if the front factor depends
explicitly on $x^+$, the corresponding potential will also be world-sheet time
dependent. As a result, and also depending on which other fields are turned
on in the given background, the light-cone quantization can be carried
out explicitly for those solutions that admit a rather simple form for
the front factor, typically quadratic in the transverse space coordinates,
or even in some more complicated situations that will be encountered later.

Recently, there has been revived interest in understanding string theory on
pp-wave backgrounds following a recent proposal that the Penrose limit of
superstring theory on $AdS_5 \times S^5$, which is described by a maximally
supersymmetric pp-wave in ten dimensions \cite{blau},
selects on the gauge theory side
of the $SU(N)$ super-Yang-Mills theory (with
coupling constant $g_{\rm YM}$) certain operators with large $R$-charge $J$
with $g_{\rm YM}^2 N/J^2$ kept fixed \cite{bmn}. This is a special limit
of the more general AdS/CFT correspondence that was proposed earlier
\cite{malda1} and reduces gauge theory computations to a supergravity
framework via the holographic principle.
Subsequently, there has been a lot
of activity in understanding the physical implications of the Penrose
limit in more general terms, and in favor of exploring further the
connection between supergravity and
gauge theories (see, for instance, \cite{bmno1}, \cite{bmno2}),
as any gravitational solution admits a plane-wave limit
by zooming along a suitably selected null geodesic of the background
geometry \cite{pen}; see also \cite{other} that cover some of the more
recent extensions.
This general result makes it even more attractive to study
string theory on pp-waves, as these solutions provide a consistent limit of
all gravitational backgrounds of string theory that are otherwise impossible
to treat in exact terms within
our current understanding of the whole subject.
The Penrose limit is a universal limit that
may also give hints on how to proceed with the quantum theory of strings on
more general backgrounds. Thus, it is not surprising that this area of research
has received considerable share of attention in the recent developments in string
theory and gauge theories.

One important new aspect of the latest
developments in string theory is the appearance of pp-wave geometries that
also carry a non-trivial Ramond-Ramond (R-R) 5-form in the case of type IIB
critical superstring models. Since the presence of an R-R form is
necessary to establish the supersymmetric properties of the
corresponding pp-wave solutions, as for the maximally supersymmetric
plane wave solution, and they also play a prominent role in
the string theory/gauge theory correspondence, it has been proposed to
consider exactly solvable models with R-R fields in the light-cone gauge,
and try to quantize them. In the maximally supersymmetric solution, the
front factor is independent of $x^+$ and provides positive mass-squared
terms for all bosons on the world-sheet, thus leading to a rather
simple, yet quite interesting, quantization problem \cite{tsey2} (but see
also \cite{green} in the presence of D-branes). In more complicated
cases, where the front factor describes non-linear interactions of the
string model, the situation is in general very difficult to handle in
exact terms unless the corresponding two dimensional field theory
becomes integrable. Then, modulo some technicalities that have to do with
the definition of the theory on a two dimensional cylinder rather than
a plane, the quantization can be carried out rather explicitly by
relying on known facts about the quantization of integrable field
theories.

A complete list of gravitational backgrounds that lead to supersymmetric
interacting field theories on the world-sheet has been recently worked
out by Maldacena and Maoz by solving the Killing spinor equations
directly on pp-wave backgrounds in the presence of a non-trivial R-R
5-form \cite{malda2}. Their solutions involve only the metric, which can be flat or
more generally Ricci flat in the transverse space, and a front factor
that is independent of $x^+$, i.e., $F(y)$.
It was further shown in general terms that all such supersymmetric
solutions provide exact backgrounds of string theory \cite{malda3}. In a related
recent development, a generalization was also considered by turning
on a null Neveu-Schwarz--Neveu-Schwarz (NS-NS) and/or an R-R 3-form in the
target space, in the sense that these fields have
non-vanishing components along the
$x^+$ direction, as for the R-R 5-form \cite{tsey3}; these supersymmetric
backgrounds were also shown to be exact to all orders in $\alpha^{\prime}$
in close analogy with ordinary pp-wave solutions with null torsion.
Certainly, a systematic generalization of these results becomes
important when more background fields are turned on, and in many different
ways, in the presence of a non-trivial self-dual R-R 5-form.

In the present work we study systematically new solutions that can
be obtained by replacing (at least) four of the eight transverse
directions of a pp-wave by exact conformal field theory blocks with
$N=(4,4)$ world-sheet supersymmetry. These models have the advantage of
being exact to all orders in $\alpha^{\prime}$, thus rendering the
full ten dimension solution exact in string theory. Also, they include
a number of exact conformal field theory backgrounds that can be
formulated as Wess-Zumino-Witten models, and as such they can be
studied systematically from the world-sheet point of view, which is
important for the light-cone formulation of the problem; the front
factor that is induced by the presence of a non-trivial R-R 5-form
can then be understood as deformation of the corresponding
superconformal field theories, which in many interesting cases can
be integrable. Thus, to conform with the general program of the
light-cone quantization of string theory, one may employ
results from the integrable deformations of two dimensional
conformal field theories in order to provide exactly solvable
models of interacting quantum theories on the string world-sheet.

Our focus is put on models with a large amount of world-sheet supersymmetry,
which in some cases can also be manifestly space-time supersymmetric
in the presence of non-trivial dilaton and NS-NS anti-symmetric tensor
fields. The implications of various dualities within the class of
the selected models can also be studied in order to generate some
new solutions. There is a special type of such models that can be
constructed by embedding axionic instantons in the transverse space,
which are also space-time supersymmetric. For this particular class,
the transverse space metric is conformally flat and the solution
of the Killing spinor equations reduces to the Maldacena-Maoz solution
as in flat transverse space, after imposing the appropriate chiral
projections.
Then, one expects in general that more arbitrary
$N=4$ superconformal theories with torsion, which have their target space
metric conformally equivalent to a hyper-Kahler metric, will
also admit systematic solutions by reducing the problem
to the Killing spinor equations of a purely gravitational
hyper-Kahler background
by conformal transformation; this generalization, however, will not be spelled out
in any detail in the present work. One way to think of the solutions
that we construct explicitly here is as being
generalizations of the usual five-brane solutions
\cite{cal} in the presence of
an R-R 5-form, in effect, this induces a $(dx^+)^2$ perturbation of the metric
driven by the
corresponding front factor that transforms the remaining six dimensional
Minkowski space of the usual five-brane solution into a (special) pp-wave form.
In fact, we will be able to obtain some interesting integrable systems on the
string world-sheet by putting the (semi)-wormhole geometry in the transverse
directions or, for that matter,
any other exact $N=(4,4)$ background related to it by various
dualities.

It is important to realize
that there can be interesting examples of exactly solvable models
in our general class of superconformal theories, which
will not have manifest space-time supersymmetry if their $N=(4,4)$
world-sheet supersymmetry is non-locally realized in terms of
parafermion fields \cite{koun}, \cite{bs1}.
In those cases, we will only be able to obtain
some sporadic solutions by solving directly the second order equations,
as it does not make much sense to look for solutions of the Killing
spinor equations in their background geometries. Since all
models with non-locally realized world-sheet supersymmetry can be
mapped into axionic instantons, using T-duality transformations with respect to
non-triholomorphic Killing vector fields, one may further inquire whether
the classification of all space-time supersymmetric solutions induced by
axionic instanton backgrounds in the transverse space can also provide a
systematic list of solutions for their T-dual faces.
Clearly, in some cases, it might be
possible to generate certain new solutions by T-duality
within the type IIB theory,
but in general the dual background will
not permit the consistent choice of an R-R 5-form for any prescribed front
factor; we will encounter some
examples of when this can happen or cannot happen later.
In any case, it is known that T-duality is a
symmetry that relates type IIB with type IIA theory \cite{dine}, and so in general one
may only use it systematically to construct new solutions with a
consistent type IIA interpretation (see also \cite{berg}).

The remaining paper is organized as follows. In section 2, we review the
basic elements of interacting light-cone models by considering the
classification of all space-time supersymmetric solutions in the
simplest case of purely gravitational backgrounds carrying a self-dual
R-R 5-form. We will expose some aspects
of the general construction, which are important for the more
general class of solutions that will be presented later. In section 3,
we study generalized gravitational R-R backgrounds for pp-wave geometries in the
presence of non-trivial dilaton and NS-NS anti-symmetric tensor fields in the
transverse space, but without having any components or dependence
of these fields on
the light-cone coordinate $x^+$. We also
formulate our proposal to use $N=(4,4)$ superconformal building blocks as
internal theories in the transverse space,
and outline a short proof for the exact nature of such
gravitational backgrounds in string theory. In section 4, we present a list
of superconformal building blocks that admit an exact description as
Wess-Zumino-Witten models, and which can be put in the transverse space
to construct exact pp-wave solutions with interesting geometrical properties.
Among these exact solutions, the (semi)-wormhole geometries are also
included to provide generalized five-brane solutions in the presence of
R-R fields. In section 5, we construct some explicit solutions of the
second order field equations by making appropriate (yet consistent) choices
for the front factor $F$ and the R-R field. Here, we also interpret the
resulting interacting light-cone models as integrable perturbations of the
underlying exact Wess-Zumino-Witten models, and consider some special cases where
T-duality can be employed as a solution generating transformation within
type IIB theory. The integrable systems that we encounter include
the complex sine-Gordon model, the supersymmetric Liouville theory and the complex
sine-Liouville model in two dimensions. In section 6, we study systematically
the case where axionic instantons are placed in the transverse space and
show that the classification of space-time supersymmetric solutions reduces
to the simpler problem of supersymmetric pp-wave backgrounds with flat
transverse space by a conformal rotation and appropriate chiral projections.
Thus, new examples of
interacting light-cone models can be constructed with support on axionic
instanton spaces. In section 7, we briefly discuss the possibility to place
double axionic instanton solutions in the full eight dimensional transverse
space and present some simple solutions. We also briefly discuss some general
aspects of the type IIA--IIB duality for pp-wave backgrounds in view of future
applications of the current work.
Finally, in section 8, we present our conclusions and indicate some
general future directions of research.

\section{Interacting light-cone string models}
\setcounter{equation}{0}

In this section we review the construction of solutions of type IIB supergravity in
ten dimensions with non-trivial Ramond-Ramond 5-form in spaces with a null Killing
isometry. More precisely, following \cite{malda2},
we consider the ansatz for the metric $g$ and the
self-dual R-R 5-form ${\cal F}$,
\ba
ds^2 & = & -4dx^+ dx^- - F(y) \left(dx^+ \right)^2 + h_{ij}(y) dy^i dy^j ~, \nonumber\\
{\cal F} & = & dx^+ \wedge \varphi(y) ~,
\ea
whereas all other fields of the theory are set equal to zero. Later, we will generalize
the construction to spaces with non-trivial dilaton and anti-symmetric NS-NS tensor
fields. Here, we assume that the front factor $F$ and the metric $h$ of the transverse
space depend only on the transverse coordinates $y^i$ with $i=2, 3, \cdots , 9$.
Also, self-duality of ${\cal F}$ in ten dimensions implies that $\varphi (y)$, which is
also taken to depend on the transverse coordinates, is an anti-self-dual 4-form in the
transverse space.

It is known that the field equations in this purely gravitational target space simply
read as follows,
\be
d\varphi = 0 ~, ~~~~ R_{ij}[h] = 0 ~, ~~~~ \nabla^2 F = {1 \over 2 \cdot 4!}
\varphi_{ijkl}\varphi^{ijkl} ~,
\ee
where $\nabla^2$ is the Laplacian in the transverse space and the normalization of
$\varphi$ is fixed accordingly. Therefore, in the absence
of any other fields, it follows that the transverse space has to be Ricci flat and
the anti-self-dual 4-form $\varphi$ has to be closed, and hence co-closed.
Then, the problem is to
find solutions of the second order equation that relates the norm of $\varphi$ with the
front factor $F$ for this general class of pp-wave type backgrounds\footnote{We use
the terminology of pp-waves
rather loosely, even if the transverse space is not flat, i.e., as for the
usual plane geometry.}. These solutions will in turn give rise to interacting
string models in the light-cone formulation, as the front factor $F$ can be a
highly complicated function of $y$ by making appropriate choices of the R-R form.

\subsection{Anti-self-dual closed 4-forms}

We present first some basic facts about the classification of anti-self-dual closed
4-forms in eight dimensions, which will be of general value in the sequel, as well as
in all remaining sections. For this, it is useful to consider transverse spaces that admit
complex structure and denote the complex coordinates by $u, w, v, z$ plus their complex
conjugates. Then, in complex notation, we have in general the occurrence of
$(4, 0)$ or $(0,4)$ or
$(1, 3)$ or $(3, 1)$ or $(2,2)$ forms. In order to examine which of them can be
anti-self-dual, we introduce the fully anti-symmetric tensor in eight dimensions with
${\epsilon_{uwvz}}^{uwvz}=1$,
which is fully covariantized and its indices can be raised or lowered by the Kahler
metric $h$ of the ambient space.
Since the Poincare dual of the $(4,0)$ form equals to itself, the
condition of anti-self-duality makes it vanish and there can be no anti-self-dual
forms of this kind; likewise, there can be no non-vanishing anti-self-dual $(0,4)$
form. As for the remaining forms, one has to consider appropriate combinations that
select the anti-self-dual components, since any form can always be written as a sum of
self-dual and anti-self-dual terms.

To motivate the presentation, it is useful to compared
this situation with the
anti-self-dual equations for Yang-Mills fields in four dimensions, as it was
originally done by Yang in his R-gauge formulation of the problem in
flat Euclidean space \cite{yang}. Introducing
complex coordinates $u$ and $w$ and their complex conjugates, one finds that the
$(2,0)$ and $(0,2)$ forms are self-dual, i.e., $^{\star}F_{uw} = F_{uw}$,
$^{\star}F_{\bar{u}\bar{w}} = F_{\bar{u}\bar{w}}$; therefore, imposing anti-self-duality
on the field strength makes these components vanish. On the other hand, the
other components of the field strength, which are represented by $(1,1)$ forms
in complex notation, obey $^{\star}F_{u\bar{u}}=F_{w\bar{w}}$ and so
the combination $F_{u\bar{u}} - F_{w\bar{w}}$ is an anti-self-dual form.
Then, in this context, the anti-self-dual Yang-Mills equations read $F_{u\bar{u}}+
F_{w\bar{w}}=0$, i.e., the complementary self-dual $(1,1)$ form vanishes. As for the
remaining $(1,1)$ form $F_{u\bar{w}}$ and its complex conjugate,
$F_{\bar{u}w}$, it can be easily seen
that they are self-dual and hence vanish by imposing anti-self-duality once more.
It turns out in Yang-Mills theory that setting all self-dual forms equal to zero,
which is equivalent to considering anti-instanton configurations,
is also consistent with
the second order classical field equations.

In the present higher dimensional setting, with a curved metric $h$ in general,
the classification is made easy by
introducing the short-hand notation, as in \cite{malda2},
\be
\varphi_{mn} = {1 \over 6} \varphi_{m\bar{l}\bar{p}\bar{q}}
\epsilon^{\bar{l}\bar{p}\bar{q}\bar{n}}h_{n\bar{n}} ~, ~~~~
\varphi_{m\bar{n}} = {1\over 2} \varphi_{m\bar{n}p\bar{p}} h^{p\bar{p}}
\ee
using the corresponding $(1,3)$ and $(2,2)$ forms; the $(2,2)$ forms
$\varphi_{m\bar{n}p\bar{q}}$ with all $m,n,p, q$ different from each other are
all self-dual, and hence they have been omitted from the short-hand notation above.
First, as far as the
anti-self-duality of $(1,3)$ forms is concerned, it simply translates
to having symmetric matrices $\varphi_{mn}$, thus determining the ten
($=16-6$) different linear combinations
of them, whereas the anti-symmetric part of $\varphi_{mn}$, which consists
of six self-dual forms,
is dismissed. Likewise, one may consider ten different anti-self-dual
$(3,1)$ forms by simply taking the complex conjugate of the anti-self-dual
$(1,3)$ forms that have just been discussed. Then, taking the 4-form $\varphi$ as
a sum of the anti-self-dual $(1,3)$ forms plus their complex conjugates, yields
a real expression for the R-R 5-form ${\cal F}$. Next, as far as the $(2,2)$
forms are concerned, it turns out that the
condition of anti-self-duality simply translates to having only those linear combinations
of the $\varphi_{m\bar{n}p\bar{p}}$ that render the matrix $\varphi_{m\bar{n}}$ traceless;
thus, there can be fifteen ($=16-1$) independent components of this kind.
Also, since the R-R 5-form ${\cal F}$ has to be real, the matrix that summarizes
$\varphi_{m\bar{n}}$ has to be Hermitian.

Finally, the closure of the 4-form $\varphi$ introduces additional
dynamical constraints on
the components of the corresponding anti-self-dual forms that have to be taken
into account in the construction of consistent type IIB supergravity backgrounds
with non-trivial R-R 5-form fields. Actually, for 5-forms ${\cal F} = d C^{(4)}$ that
are derived from an R-R 4-form potential $C^{(4)}$, as in all cases that
we considered here, the condition of self-duality is sufficient to insure the
closure and co-closure of ${\cal F}$.

\subsection{Maldacena-Maoz supersymmetric solutions}

Following the recent developments in the subject, we present a brief account of the
classification of all supersymmetric solutions found in the presence of a non-trivial
R-R 5-form, provided that only the metric field is taken into account. In this case,
one has to solve the Killing spinor equations in ten dimensions,
\be
D_\mu \epsilon \equiv (\nabla_\mu + {i \over 16} \scF ~ \Gamma_\mu)
\epsilon = 0 ~,
\ee
where the covariant derivative includes the contribution from the spin connection,
as usual, and $\Gamma_{\mu}$ are the standard Dirac matrices.
We also use the normalization $\scF = \Gamma^{ijklr} {\cal F}_{ijklr} / 5!$, and furthermore
note that the second order equations are inert to the change of sign of the R-R field
${\cal F} \rightarrow - {\cal F}$. The equations above
are required by space-time supersymmetry of the bosonic class of pp-wave
string backgrounds, as they can be used to set the variation of the gravitino fields
consistently equal to zero. Supersymmetric conditions reduce the second order
equation for the front factor $F$ into a simpler set of first order equations and
provide solutions for $F$ and ${\cal F}$ whenever Killing spinors exist.
The number of Killing spinors depends on the amount of space-time supersymmetry, and
as a result the structure of the solutions becomes richer as the amount of
supersymmetry increases. Maldacena and Maoz were able to determine the Killing
spinors explicitly and characterize completely the allowed form of the
functions $F$ and ${\cal F}$ in a systematic way \cite{malda2}. We summarize their results
below, without giving details of their proof, in a way that depends on the amount of
supersymmetry.

(i) \underline{$(2,2)$ supersymmetry}: In this case the solutions are parametrized by
a holomorphic function ${\cal W}$ and a real Killing potential ${\cal U}$, which gives
rise to a Killing vector field with components $V_m = i\nabla_m {\cal U}$
and $V_{\bar{m}} = -i \nabla_{\bar{m}} {\cal U}$, so that $V^m$ is holomorphic and
$V^{\bar{m}}$ is anti-holomorphic. We also have to impose the conditions, which are
additionally required by supersymmetry,
\be
\nabla_m V^m = 0 ~, ~~~~~ \nabla_n (V^m \nabla_m {\cal W}) = 0 ~.
\ee
Then, the supergravity solution is
described in complex coordinates as follows:
\ba
2F & = & h^{m\bar{n}} (\nabla_m {\cal W}) (\nabla_{\bar{n}} \bar{\cal W}) +
h_{m\bar{n}} V^m V^{\bar{n}} ~, \nonumber\\
\varphi_{mn} & = & \nabla_m \nabla_n {\cal W} ~, ~~~~ \varphi_{\bar{m}\bar{n}} =
\nabla_{\bar{m}} \nabla_{\bar{n}} \bar{{\cal W}} ~, ~~~~
\varphi_{m\bar{n}} = \nabla_{m} \nabla_{\bar{n}} {\cal U} ~.
\ea
All solutions in this class have $(2,2)$ supersymmetry or more.

Simpler solutions are obtained when ${\cal U} = 0$, in which case there is no
contribution to the R-R field from the anti-self-dual $(2,2)$ forms.

(ii) \underline{$(1,1)$ supersymmetry}: In this case the solutions are parametrized
by a {\em real} {\em harmonic} function $U$ and the corresponding expressions
assume the form
\ba
2F & = & h^{m\bar{n}} (\nabla_m U) (\nabla_{\bar{n}} U) ~, \nonumber\\
\varphi_{mn} & = & \nabla_m \nabla_n U ~, ~~~~ \varphi_{\bar{m}\bar{n}} =
\nabla_{\bar{m}} \nabla_{\bar{n}} U ~, ~~~~
\varphi_{m\bar{n}} = \nabla_{m} \nabla_{\bar{n}} U ~.
\ea
thus receiving contributions from both $(2,2)$ and $(1,3)$ forms and their complex
conjugates.

Note that when the transverse directions are flat,
and so we may choose $h_{ij} = \delta_{ij}$,
the solutions with $(2,2)$ supersymmetry (or more) become simplified, even when
${\cal U} \neq 0$. In this case, the real Killing potential can be chosen as
\be
{\cal U} = C_{m\bar{n}}z^m z^{\bar{n}} ~,
\ee
where $z^m$ and $z^{\bar{m}}$
denote collectively all four
complex coordinates and their complex conjugates in the flat transverse space,
and $C_{m\bar{n}}$ is a constant
Hermitian matrix with zero trace. This is required in order for the corresponding
holomorphic Killing vector field $V^n= -i {C_m}^n z^m$ to be covariantly
constant, which
here translates to $\partial_mV^m = -i{C_{m}}^m = 0$. Then, according to the general
form of the supergravity solution, the anti-self-dual $(2,2)$ forms
$\varphi_{m\bar{n}}$ are simply
given by the constant elements
\be
C_{m\bar{n}} = \varphi_{m \bar{n}} ~.
\ee
Hence, their contribution to
the front factor $F$ is only quadratic in the complex coordinates $z$, as
$h_{m\bar{n}}V^m V^{\bar{n}} = {C_m}^q C_{q \bar{n}} z^m z^{\bar{n}}$.
Finally, in the flat space case, there should be an additional constraint on
${\cal W}$ provided by the condition
\be
\partial_n ({C_q}^m z^q \partial_m {\cal W}) = 0
\ee
for all $n$. As for the class of solutions with $(1,1)$ supersymmetry, there
is nothing special happening in the flat space limit apart from the obvious fact
that covariant derivatives turn into ordinary ones.

One may also verify directly that all these are solutions of the second order equations
without making any reference to supersymmetry.

The maximally supersymmetric pp-wave background of type IIB supergravity \cite{blau} can be
obtained as a special case of the general scheme that was described above by
choosing
\be
{\cal W} = {1 \over 2}\left(u^2 + w^2 + v^2 + z^2 \right) ; ~~~~ {\cal U} =0
\ee
with flat transverse space. Then, the solution is described in complex coordinates
by the following functions:
\ba
F & = & {1 \over 2} \left(|u|^2 + |w|^2 + |v|^2 + |z|^2\right) ~, \nonumber\\
\varphi & = & du \wedge d\bar{w} \wedge d\bar{v} \wedge d\bar{z} +
dw \wedge d\bar{u} \wedge d\bar{v} \wedge d\bar{z}   \nonumber\\
& + & dv \wedge d\bar{u} \wedge d\bar{w} \wedge d\bar{z} +
dz \wedge d\bar{u} \wedge d\bar{w} \wedge d\bar{v} ~ + ~~ cc ~.
\ea
This simple background gives rise to
massive free bosons on the string world-sheet in the light-cone formulation of string
theory, which can be easily quantized \cite{tsey2}, and one can
also incorporate the presence
of D-branes upon quantization \cite{green}.
More complicated solutions with non-constant R-R fields and arbitrary front factors
arise by considering more general holomorphic functions ${\cal W}$.
We will briefly mention some interesting examples of the more general solutions
shortly. In those cases,
the two dimensional action that describes string propagation in the
light-cone gauge, may contain non-linear interaction terms among the world-sheet
bosons, which can be studied exactly when there is an integrable system at work
on the string world-sheet.

\subsection{World-sheet light-cone action}

Gravitational backgrounds with a null Killing isometry, as in the general class of
the string models we are studying here, admit a light-cone formulation as in \cite{hor}
for ordinary plane-fronted wave backgrounds. The bosonic part of the
two dimensional action of a string
propagating in such backgrounds is
\be
S = { 1 \over 4\pi \alpha^{\prime}}
\int d\sigma d\tau \left( -4 \partial_a X^+ \partial^a X^- -F(y) \partial_a X^+
\partial^a X^+ + h_{ij}(y) \partial_a y^i \partial^a y^j \right) ,
\ee
where $\sigma$ and $\tau$ are the world-sheet coordinates, and the world-sheet metric
is chosen to be flat by imposing the conformal gauge. Note that the front factor $F$
is only determined up to an additive constant\footnote{This amounts to the freedom
of shifting the vacuum energy of the light-cone sigma models by an arbitrary
constant.}, as we can always reparametrize
$X^-$ to $X^- + c X^+$. As usual, extremizing with respect
to $X^-$, yields the Laplace equation $\partial_a \partial^a X^+ = 0$, and so we may fix
the remaining gauge invariance by choosing the time-like parameter of the world-sheet
as $X^+ = 2\pi \alpha^{\prime} p^+ \tau$, where the constant $p^+$ is the $+$ component
of the momentum density. From now on, we normalize the parameters so that the light-cone
gauge simply reads $X^+ = \tau$.

We may check the consistency of the light-cone gauge for this
generalized class of pp-wave backgrounds with non-trivial transverse space metric by
writing the equations of motion for the fields $X^-$ and $y^i$, together with the two
Virasoro constraints,
\ba
T_{++} & \equiv & -4 \partial X^- - F(y) + h_{ij}(y) \partial y^i \partial y^j =
0 ~, \nonumber\\
T_{--} & \equiv & -4 \bar{\partial} X^- - F(y) + h_{ij}(y) \bar{\partial} y^i
\bar{\partial} y^j = 0 ~,
\ea
where we have used derivatives with respect to the light-cone coordinates on the
string world-sheet. It can be easily checked, but we spare the details which are
similar to the case of transverse spaces with flat metric (see for instance
\cite{gsw}, but also \cite{hor}), that the Virasoro constraints
provide two first order equations for the field $X^-$, which are compatible provided
that $y^i$ satisfy their own classical equations of motion; thus, they may be used to
uniquely determine $X^-$ up to an arbitrary zero mode. Summarizing, the general class of
pp-wave geometries with arbitrary metric in their transverse space can all be
studied systematically by imposing the light-cone gauge and then quantizing.

The world-sheet action becomes in the light-cone gauge
\be
S_{\rm lc} = \int_{\Sigma} \left(h_{ij}(y) \partial y^i \bar{\partial} y^j - F(y) \right) ,
\ee
where $\Sigma$ is a two dimensional cylinder due to the periodicity of the $\sigma$
variable.
The full action is obtained by also adding the contribution of the fermionic
fields, which are omitted here.
For generic (but consistent) choices of $F$ we obtain interacting
light-cone models, which generalize the simple case of having massive free bosons
on the world-sheet for the maximally supersymmetric pp-wave string background \cite{tsey2}.
In case that we also have an anti-symmetric tensor field $B_{ij}$ (torsion) in the
classical background, as in subsequent sections, the story repeats itself, but
with $h_{ij}(y)$  being replaced by $h_{ij}(y) + B_{ij}(y)$ in the light-cone action.

An interesting and characteristic example of string backgrounds with R-R fields
that lead to integrable theories on the string world-sheet is provided by the
$(2,2)$ supersymmetric solution in flat transverse space, with
\be
{\cal W} = {\rm cos}u ~, ~~~~~ {\cal U} = 0
\ee
within the Maldacena-Maoz framework. Here, the superpotential is taken to depend only
on one complex coordinate, which can be parametrized as
$u = \theta + i \rho$. In this case, we find that the front factor $F$, which provides
the non-linear interactions in the light-cone gauge is given by
\be
2F= |{\rm sin}u|^2 \equiv {\rm sin}^2 \theta + {\rm sinh}^2 \rho ~.
\ee
The underlying integrable system is the $N=2$ sine-Gordon model \cite{sine}, but
here it is defined on a cylinder rather than a two dimensional plane. A ten
dimensional solution of type IIB supergravity can be constructed by choosing
the anti-self-dual closed 4-form $\varphi$ as the following $(1,3)$ form,
plus its complex conjugate \cite{malda2}:
\be
\varphi = -({\rm cos}u) du \wedge d\bar{w} \wedge d\bar{v} \wedge d\bar{z} ~~ + ~ cc ~.
\ee
Another interesting example is provided by the choice
\be
{\cal W} = u -{1 \over 3} \alpha^2 u^3 ~, ~~~~~ {\cal U} = 0 ~,
\ee
which yields a $(2,2)$ solution with quartic interactions having
\be
2F = |1 - \alpha^2 u^2|^2
\ee
and an anti-self-dual closed 4-form
\be
\varphi = -(2\alpha^2 u) du \wedge d\bar{w} \wedge d\bar{v} \wedge d\bar{z} ~~ + ~ cc ~.
\ee

Light-cone quantization proceeds by employing the solvability of the underlying
integrable systems, thus determining in principle the allowed string spectrum.
It should be realized, nevertheless, that many results about integrable systems
have to be generalized accordingly, in order to take into account the periodicity
of the world-sheet coordinate $\sigma$. Thus, apart from the ordinary periodic
solutions there can be other topological sectors in the model with non-trivial
winding or twisting, depending on discrete symmetries; for example, for the sine-Gordon
model, we have invariance of the equations under $u \rightarrow u + 2\pi$ and
$u \rightarrow -u$, whereas for the quartic potential we only have invariance under
$u \rightarrow -u$. Imposing strict periodicity destabilizes the soliton solutions,
but stability can be regained by winding and/or twisting (see, for instance, \cite{bak6}).
In any case, the quantization
of integrable systems on the cylinder is a less studied problem, due to the
technicalities associated with periodicity, and they have to be shorted out before the
spectrum of the corresponding string models can be computed in closed form. We
do not really have anything more concrete to say about these particular issues in the
present work, apart from a few extra clarifying remarks in the conclusions of the paper.

\section{Generalized gravitational string backgrounds}
\setcounter{equation}{0}

We begin our study of more general gravitational backgrounds of type IIB
supergravity with a null Killing vector field  and a R-R 5-form
by turning on a non-trivial
dilaton as well as a NS-NS anti-symmetric tensor field. As we will see
later, this provides an interesting generalization of the pure
gravitational backgrounds with R-R fields, which is still tractable and yields
a variety of integrable systems in the light-cone gauge of string
theory. We will first describe an ansatz that leads to consistent reduction
of the $\beta$-function equations for the background fields, and then
formulate a proposal for constructing some new exact solutions. The
class of backgrounds that will be used in the transverse space of the
ten dimensional theory will exhibit $N=(4,4)$ world-sheet supersymmetry,
which will in turn imply that the proposed solutions are actually
exact to all orders in $\alpha^{\prime}$; a short proof of this fact
will also be given in this section following some standard arguments
about the exactness of pp-wave geometries. Thus, although our starting
point is the lowest order effective theory of type IIB supergravity,
the results we describe are also  exact in string theory.

\subsection{Type IIB string background equations}

We consider the basic equations of type IIB supergravity in ten dimensions in the
presence of a metric $g$, a dilaton $\Phi$, an anti-symmetric
NS-NS tensor field that
is represented by a 2-form B with field stregth $H=dB$,
and an R-R 5-form ${\cal F}$. In general, type IIB supergravity also has an R-R
scalar $C^{(0)}$ and an R-R 2-form $C^{(2)}$ that accompany the NS-NS fields
$\Phi$ and $B$, in which case the R-R 5-form ${\cal F}$ is given by the
expression
\be
{\cal F} = \partial C^{(4)} + B \partial C^{(2)} - C^{(2)} \partial B ~,
\ee
where $C^{(4)}$ is the corresponding R-R 4-form potential appropriately normalized.
In all cases, ${\cal F}$ is a self-dual 5-form in ten dimensions, i.e.,
$^{\star}{\cal F} = {\cal F}$; self-duality does not admit a natural derivation from a
covariant action principle that can select only the self-dual piece of ${\cal F}$ to
the physical propagating degrees of freedom. Thus, this condition has to be implemented
as an on-shell constraint when an action principle is used to describe all other
equations of motion.

Setting $C^{(0)} = 0 = C^{(2)}$ for the remaining R-R fields, the supergravity equations
can be derived from the effective action
\cite{gsw}, \cite{schw} (but see also \cite{berg}),
\be
S_{{\rm eff}} = \int d^{10}x \sqrt{-g} e^{-2\Phi} \left( R + 4 (\nabla_\mu \Phi)
(\nabla^\mu \Phi) -{1 \over 12} H_{\mu \nu \rho}H^{\mu \nu \rho} -
{e^{2\Phi} \over 4 \cdot 5!}
{\cal F}_{\mu \nu \rho \sigma \tau} {\cal F}^{\mu \nu \rho \sigma \tau} \right) ,
\ee
which is written in the $\sigma$-model frame, and they assume the form
\ba
R_{\mu \nu} & = & -2 \nabla_{\mu} \nabla_{\nu} \Phi +
{1 \over 4} H_{\mu \rho \sigma}{H_\nu}^{\rho \sigma} + {e^{2\Phi} \over 4\cdot 4!}
\left({\cal F}_{\mu \kappa \lambda \rho \sigma} {{\cal F}_{\nu}}^{\kappa \lambda \rho \sigma}
-{1 \over 10} g_{\mu \nu} {\cal F}_{\kappa \lambda \rho \sigma \tau}
{\cal F}^{\kappa \lambda \rho \sigma \tau} \right) , \nonumber\\
0 & = & \nabla_{\mu} \nabla^{\mu} \Phi - 2 (\nabla_{\mu} \Phi) (\nabla^{\mu} \Phi)
+ {1 \over 12} H_{\mu \nu \rho}H^{\mu \nu \rho} ~,\nonumber\\
0 & = & \nabla_{\mu} \left(e^{-2 \Phi} H^{\mu \nu \rho} \right) ~, \nonumber\\
0 & = & \nabla_{\mu} {\cal F}^{\mu \nu \kappa \lambda \rho} ~,
\ea
where the sef-duality of ${\cal F}$ is also imposed as an additional constraint.
Here, Greek indices take the values $0, 1, 2, \cdots , 9$.
The string equations arise as conditions for the
vanishing  of the $\beta$-functions to
lowest order in ${\alpha}^{\prime}$ and provide gravitational backgrounds that are
consistent with the requirement of
conformal invariance on the string world-sheet. We note for notational
purposes that the corresponding equations in the Einstein frame can be obtained by
passing to the ten dimensional
metric $g^{\prime}_{\mu \nu} = {\rm exp}(-\Phi/2) g_{\mu \nu}$. For backgrounds
that exhibit enough supersymmetry, typically backgrounds with $N=(4,4)$ supersymmetry
on the world-sheet, the solutions that one obtains are also exact to all orders in
${\alpha}^{\prime}$ and, hence, the present framework is sufficient for their complete
description. It is also useful to recall in this context that the equation for the
dilaton field, which in the Einstein frame of the ten-dimensional theory
simply reads $ e^{\Phi} \nabla^2 \Phi
+ H^2/12 = 0$, is
derived under the assumption that there is no potential term for $\Phi$, as we are
considering critical string theory in $D=10$ dimensions.

Next, we simplify the string background equations by restricting attention to
metrics with a null Killing isometry,
\be
ds^2 = -4dx^+ dx^- -F(y)(dx^+)^2 + h_{ij}(y)dy^i dy^j ~; ~~~~ i,j= 2, 3, \cdots , 9 ~,
\label{metric}
\ee
where the front factor $F$ is taken to depend only on the coordinates $y$ that
parametrize the transverse space with metric $h_{ij}(y)$, as before.
We also assume that the
dilaton $\Phi$ depends only on $y$, while the 2-form $B=B(y)$ is assumed to live entirely
in the transverse space having zero components in the $x^{\pm}$ directions.
As for the R-R 5-form, we
also assume, as before, that
\be
{\cal F} = dx^+ \wedge \varphi(y) ~,
\ee
where $\varphi(y)$ is an anti-self-dual closed 4-form in the transverse
eight-dimensional space, i.e.,
\be
\varphi(y) = - {}^{\star} \varphi(y) ~, ~~~~~ d \varphi(y) = 0 ~,
\ee
in order to insure the self-duality of ${\cal F}$ in ten
dimensions together with its conservation law. Taking the $(++)$-component of the
equation for the Ricci tensor, we obtain
\be
\nabla_i \left(e^{-2\Phi} \nabla^i F \right) = {1 \over 2 \cdot 4!}
\varphi_{ijkl} \varphi^{ijkl} ~, \label{main}
\ee
where all Latin indices and covariant derivatives are refering to the transverse space.
In order to simplify the expressions in the sequel, we will use the the short-hand
notation for the norm of the 4-form
\be
|\varphi|^2 = {1 \over 4!} \varphi_{ijkl} \varphi^{ijkl} ~.
\ee
The remaining equations we obtain in the transverse space are
\ba
R_{ij}[h] & = & -2 \nabla_i \nabla_j \Phi + {1 \over 4} H_{ikl}{H_j}^{kl} ~,
\nonumber\\
0 & = & \nabla_i \nabla^i \Phi - 2 (\nabla_i \Phi)(\nabla^i \Phi)
+ {1 \over 12} H_{ijk} H^{ijk} ~,\nonumber\\
0 & = & \nabla_i \left(e^{-2 \Phi} H^{ijk} \right) ~,
\ea
and therefore we have non-vanishing curvature in the presence of non-trivial dilaton
and anti-symmetric tensor fields, namely
\be
R[g] = R[h] = -2 \nabla^2 \Phi + {1 \over 4} H^2 ~.
\ee

We observe that if we have a metric $h(y)$, a dilaton field $\Phi(y)$ and a 2-form $B(y)$
that solve the
string background equations in $D-2 = 8$ dimensions with local coordinates $y$, then
a solution of the type IIB supergravity can also be obtained in $D=10$ dimensions with a null
Killing isometry as given by the ansatz \eqn{metric}. The front factor $F(y)$
and the anti-self-dual closed
4-form $\varphi(y)$ that characterize the resulting ten-dimensional background are
simply related to each other by equation \eqn{main}, but they remain arbitrary
otherwise. Here, we assume (at least to first order in ${\alpha}^{\prime}$) that the
central charge deficit of the $(D-2)$-dimensional theory is zero, $\delta c_{D-2} =0$,
so that the total central charge in $D$ dimensions is critical, i.e., $c=2+8=10$.
Moreover, taking into account some special circumstances that are described below,
it can be shown that if the $(D-2)$-background is exact to all orders in ${\alpha}^{\prime}$,
the $D$-dimensional background will also be exact to all orders.

At this point it is useful to recall briefly the history of the subject concerning
the exactness of various classes of pp-wave solutions.
When all other fields but the metric are set equal to zero, we have the field equation
$\nabla^2 F(y) =0$ in the transverse space. For $h_{ij} = \delta_{ij}$ and quadratic
front factor $F(y)= A_{ij} y^i y^j$, we obtain the traceless condition for the symmetric
matrix $A$, ${\rm Tr}A = 0$, which yields pp-waves with constant polarization as
solutions of superstring theory; these solutions are known to be exact string backgrounds,
as all higher order terms in the string equations of motion are automatically zero \cite{ama}
(see also \cite{hor} for a class of generalized pp-wave backgrounds, but without R-R fields).
Relaxing the traceless condition on the matrix $A$ for quadratic front
factors, amounts to turning on a non-trivial R-R 5-form.
Indeed, the simplest such background with zero
dilaton and anti-symmetric tensor fields,
is given by the maximally supersymmetric pp-wave solution \cite{blau},
\ba
ds^2 & = & -4dx^+ dx^- - \left(\sum_{i=2}^{9} {y_i}^2\right) (dx^+)^2 +
\delta_{ij}dy^i dy^j ~, \nonumber\\
 \varphi & = & dy^2 \wedge dy^3 \wedge dy^4 \wedge dy^5 -
dy^6 \wedge dy^7 \wedge dy^8 \wedge dy^9 ~,
\ea
which is written here using real coordinates;
equivalently, we may use here the complex coordinates
\be
u = y^2 + i y^6 ~, ~~~~ w = y^3 + i y^7 ~, ~~~~ v = y^4 + i y^8 ~, ~~~~
z = y^5 + i y^9
\ee
to conform with the complex notation of the general supersymmetric backgrounds given
earlier. As we have already seen, many other backgrounds with flat or curved space metric
$h_{ij}$, as those that have been
proposed by Maldacena and Maoz by relying on a simpler set of first
order equations \cite{malda2} (but see also \cite{tsey3}),
are also supersymmetric solutions, which are exact to all orders in $\alpha^{\prime}$.

In our attempt to find generalized solutions in the presence of non-trivial
dilaton and NS-NS anti-symmetric tensor fields, we will use the observation that
the transverse space fields decouple from the rest and satisfy their own
$\beta$-function equations, at least within the ansatz that has been made.
Space-time supersymmetry will not be a concern at first sight, but as we will
find out later, when the transverse space has manifest space-time supersymmetry,
the full ten dimensional solutions of pp-wave type will also be
space-time supersymmetric. Here, we focus instead on a more general
class of models that exhibit a
large amount of world-sheet supersymmetry, as we think that this is more
fundamental for the world-sheet formulation of superstring theory in the light-cone
gauge; in some cases, where all world-sheet supersymmetries can be locally
realized, we will also get solutions with manifest space-time supersymmetry
as bonus.

We conclude the general discussion of the string background equations by examining
the possible restrictions on the components of our background fields, which are
imposed by the choice of R-R fields $C^{(0)} = 0 = C^{(2)}$. In particular, if
these two R-R fields are set equal to zero, the general field equations will admit a
consistent truncation provided that
\be
{\cal F}_{\mu \nu \rho \kappa \lambda} H^{\rho \kappa \lambda} = 0
\ee
for all space-time indices $\mu$, $\nu$. These additional conditions are obtained by
considering the field equation for the R-R field $C^{(2)}$, and in turn they imply
within our ansatz that
\be
\varphi_{ijkl} H^{jkl} = 0
\label{bingo}
\ee
for all transverse space indices. Clearly, they restrict the structure of the
non-vanishing components of $\varphi$ when there is torsion in the transverse
space of the generalized pp-wave solutions; some implications will be examined
shortly in the next subsection.

\subsection{Classes of exact solutions}

In the following, we consider the possibility to construct new type IIB
gravitational backgrounds in ten dimensions with non-trivial dilaton and anti-symmetric
tensor fields, using a suitable
embedding of exact $N=(4, 4)$ superconformal field theory blocks in the transverse space,
namely we consider models of the form
\be
ds^2 = -4 dx^+ dx^- - F(y)(dx^{+})^2 + [N=(4,4), ~ c=4]_1 + [N=(4,4), ~ c=4]_2 ~,
\label{idea}
\ee
which is schematic way of saying that the flat transverse space is replaced by exact
superconformal field theories with $(4,4)$ supersymmetry on the world-sheet. Such
four dimensional building blocks have central charge deficit equal to zero to all
orders in ${\alpha}^{\prime}$ and therefore, they serve our purpose provided that
we also include the contribution of their dilaton fields (and in some cases their
anti-symmetric tensor fields). Solutions for the front factor $F$ and the 4-form $\varphi$
will be determined later. We only note here that for
the special case $F(y) = 0$, and hence $\varphi = 0$,
these backgrounds were studied extensively in the past as exact and stable string
solutions in both type II and heterotic superstring theories with non-trivial
dilaton and anti-symmetric tensor fields \cite{cal}, \cite{koun}, and in many cases the
full spectrum of string excitations was derived in a modular invariant way \cite{afk}.

Turning on a
non-vanishing R-R 5-form amounts to switching on a front factor $F$ in the
ten dimensional metric, thus generalizing the previous class of solutions. Actually,
since the perturbation of the metric driven by the front factor $F$ depends only on
the transverse coordinates $y$, it is natural to expect that the presence of this term
in the light-cone gauge of
string theory can be interpreted as a perturbation of the $N=(4,4)$
superconformal building blocks that drives them
away from criticality. In the cases that we will
consider in the sequel these perturbations correspond to integrable field theories
on the string world-sheet, and hence one expects
that these will also provide tractable interacting
models with calculable string spectrum (although
this particular aspect of the problem is lying beyond the scope of the present work).

In a way, among other things, our proposal generalizes
the construction of five-brane solutions in
ten dimensions by turning on a non-trivial R-R 5-form, and goes further beyond it.
Recall first that the standard
construction of five-brane solutions relies on the concept of axionic instantons
\cite{rey} in order to have manifest space-time supersymmetry. Axionic instantons are special
backgrounds in four dimensions with a large amount of world-sheet supersymmetry,
which provide consistent bosonic solutions of the first order supersymmetric conditions
\be
\left(\gamma^{\mu} \partial_{\mu} \Phi \mp
{1 \over 12} H_{\mu \nu \rho} \gamma^{\mu \nu \rho} \right) \epsilon = 0 ~,~~~~
\left(\partial_{\mu}  +
{1 \over 4} \Omega_{\mp \mu}^{\alpha \beta} \gamma_{\alpha \beta} \right) \epsilon = 0
\ee
associated to the variations of the dilatino and the gravitino fields
respectively, \cite{cal}. Here, $\Omega_{\mp \mu}^{\alpha \beta}$ is
the usual spin connection that has been improved by subtracting
or adding the torsion term for
axionic instanton or anti-instanton backgrounds. The dilatino variation can be made zero by
choosing the NS-NS 3-form to be Poincare dual (or anti-dual) to the derivative of the
dilaton field. Then, for conformally flat metrics, with the conformal factor being equal
to ${\rm exp}(2\Phi)$, the axionic instantons also provide solutions of the gravitino
equation, where half of the supersymmetries are unbroken; the other half, by standard
reasoning, will be associated with fermionic zero modes bound to the solitonic five-brane.
Thus, axionic instantons provide a particular class of $N=(4,4)$ models that can also be
used in the present context to replace four of the transverse coordinates of a pp-wave
configuration.

In those cases that there is an axionic instanton block sitting in the transverse space of
the generalized pp-wave configurations, the non-vanishing components of $\varphi$ cannot
be arbitrary, as they have to satisfy the special algebraic conditions $\varphi_{ijkl}H^{jkl} = 0$,
according to equation \eqn{bingo}. Then, it is reasonable to assume that
$\varphi_{ijkl} \neq 0$ if two of its transverse space indices take values in the first
building block and the remaining two in the second block. Otherwise, if all indices of
$\varphi$ take values in the axionic instanton block, we will have $\varphi_{ijkl} = 0$
for $H \neq 0$,
in which case the components of the dual form with all four indices taking values
in the other block will
also be zero; since this is also conversely true, we rule out the possibility to have
$\varphi$ with all its components living only in one four-dimensional block. Likewise,
if three indices take values in the axionic instanton block, it will be consistent to set
those components of $\varphi$
equal to zero together with the components of the corresponding dual form having only one
index taking values in the axionic instanton block. The $2+2$ splitting of the
transverse space indices that we are advocating here for $\varphi$ amounts to a
block-diagonal structure for the matrices $\varphi_{mn}$ and $\varphi_{m \bar{n}}$ that
describe the 4-form $\varphi$ in short-hand notation;
we will make further use of this structure in section 6 while describing explicit
pp-wave solutions in axionic instanton backgrounds.

There is a further generalized class of candidate backgrounds with $N=4$ supersymmetry
that include torsion, which have manifest space-time supersymmetry and can also be
placed appropriately in the transverse space. They provide generalizations of the
standard axionic instantons, in the sense that their target space metric is conformally
equivalent to an arbitrary hyper-Kahler manifold \cite{hull},
\be
h = \Omega \tilde{h} ~, ~~~~~ \tilde{\nabla}^2 \Omega = 0 ~,
\ee
where the conformal factor $\Omega$ satisfies the Laplace equation with respect to the
conformally equivalent hyper-Kahler metric $\tilde{h}_{ij}$. They can also
be viewed as generalizations of
the usual hyper-Kahler geometry \cite{alv} in the presence of anti-symmetric tensor fields.
These backgrounds will not be
studied here in detail, but they certainly deserve separate investigation, as they can
provide quite general supersymmetric solutions in our framework, which are closely
related to the Maldacena-Maoz solutions in curved transverse space by conformal
transformation using $\Omega$.

Finally, there is yet another class of $N=(4,4)$ superconformal blocks with an exact four
dimensional interpretation, which arise as gauged Wess-Zumino-Witten models and they are
T-dual to axionic instantons. These models will be studied separately, together with
their T-dual counterparts, which are (semi)-wormholes with thin or fat throats, as they
constitute the only backgrounds of this kind with an exact description as
Wess-Zumino-Witten models. However, unlike axionic instantons, their T-dual faces
only have a dilaton but no torsion fields, which in turn imply that the dilatino
variation cannot be made zero. Thus, space-time supersymmetry cannot be seen in
these case despite the fact that in all these backgrounds
there is an underlying extended
superconformal symmetry, which makes them good candidates for building blocks of
the transverse space. We will say more about this peculiar situation towards the
end of section 4 by including a few extra technical details.
Yet another general class of $N=4$
superconformal theories consists of the T-dual versions of all conformally hyper-Kahler
spaces with torsion; however, we will not delve into
the details of these particular models in the present work, as they will be left
open for future
investigation, together with their T-dual faces.

\subsection{On higher order $\alpha^{\prime}$ corrections}

We now turn to the exactness of the pp-wave backgrounds in string theory.
The classical equations of motion for the metric in string theory, which
follow from the vanishing condition of the $\beta$-function $\beta(g_{\mu \nu})$
by imposing conformal invariance of the two dimensional theory, can be
expressed in terms of sigma model perturbation theory as
\be
0=R_{\mu \nu} + {1 \over 2} \alpha^{\prime} R_{\mu \rho \sigma \lambda}
{R_{\nu}}^{\rho \sigma \lambda} + \cdots ~,
\ee
where the dots denote derivatives and higher powers of the curvature that can
in principle occur to all orders in $\alpha^{\prime}$. In the presence of
other background fields, as the dilaton, torsion and so on, there are also
contributions coming from them to $\beta(g_{\mu \nu})$ to all orders
in $\alpha^{\prime}$, and similarly
there can be higher order corrections to their own $\beta$-functions,
$\beta(\Phi)$, $\beta(B_{\mu \nu})$, etc, involving derivatives and higher
powers. The details of the general structure that appears to higher loops
are well known (see, for instance, \cite{exact1}, \cite{exact2}),
and they will not be presented
here. The remarkable thing about pp-wave geometries is that without even having
precise knowledge of the higher order corrections, they can be shown to vanish
under some general conditions.

Simple plane wave geometries with $h_{ij} = \delta_{ij}$ and no R-R fields,
are known to be exact solutions to all orders in $\alpha^{\prime}$, as it was
first shown in \cite{ama} for purely metric backgrounds and in \cite{hor} in the
presence of dilaton and NS-NS anti-symmetric tensor fields. The proof of the absence
of higher order corrections relies
entirely on the presence of a null isometry generated by
a covariantly constant Killing vector field $l^{\mu}$ that is orthogonal to the
Riemann tensor. More recently it has been extended to prove the exactness of the
supersymmetric pp-wave backgrounds found by Maldacena and Maoz, as well as in some
generalizations, by employing various methods \cite{malda3}, \cite{tsey3}.
In our case, the proof goes in two steps: the first
also relies on the presence of a
null isometry, as for the simpler class of backgrounds that have been considered
in the past, and the second uses the exactness of the $N=(4,4)$
superconformal theory blocks that have been proposed to build their transverse space.
It is well known for them, like for all Wess-Zumino-Witten models, that one can
choose a scheme (or else a local covariant $\alpha^{\prime}$-dependent field
redefinition) that relates the exact string background to its leading order form
\cite{exact1}.

Taking the first step, it follows by
explicit calculation that in the general class of metrics \eqn{metric} we have,
\ba
R_{+ \mu + \nu} [g] & = & {1 \over 2} \delta_{\nu, i} \delta_{\mu, j}
\nabla_i \nabla_j F ~, \nonumber\\
R_{+ \mu \nu i} [g] & = & {1 \over 2} \delta_{\nu, +} \delta_{\mu, j}
\nabla_i \nabla_j F ~, \nonumber\\
R_{- \mu \nu \lambda} [g] & = & 0 = R_{\mu \nu \lambda -} [g] ~,
\ea
where the covariant derivatives are taken with respect to the transverse space
metric $h$. We also have
\be
R_{ijkl} [g] = R_{ijkl}[h] ~,
\ee
whereas all other components follow by the symmetry properties of the Riemann curvature
tensor under the interchange of its indices.

Using these expressions, as well as the fact that $g^{++} = 0$ for this class of metrics,
we arrive immediately at the following result,
\be
R_{\mu \rho \sigma \lambda} {R_{\nu}}^{\rho \sigma \lambda} =
R_{\mu ijk} {R_{\nu}}^{ijk} ~,
\ee
which in turns implies that there can be no $R^2$ correction terms of this type to the
components of the $\beta$-functions $\beta(g_{++})$, $\beta(g_{+-})$, $\beta(g_{--})$
and $\beta(g_{i \pm})$, but only to $\beta(h_{ij})$. Thus, the corrections, if any,
which they assume the general form $R_{iklm}{R_j}^{klm}$,
are only restricted to the conformal field theories that live in the transverse space.
Likewise, any possible higher order corrections with more Riemann tensors reduce to their
transverse space part, whereas all other components vanish.
Possible higher order curvature terms of the form
$R_{\mu \nu} R_{\lambda \rho \sigma \tau}R^{\lambda \rho \sigma \tau}$ also vanish
in all cases, as before, due to the contractions of the Riemann curvature tensors,
modulo terms that can be removed by appropriate field redefinitions. Finally, expressions
that may involve derivatives of the curvature, like
$\nabla^{\rho} \nabla^{\sigma} R_{\mu \rho \nu \sigma}$, or higher, which seem to
contribute not only
to the transverse space equations but to the $\beta(g_{++})$ component, can be easily
removed by field redefinitions as described in \cite{exact1}, \cite{tsey3}.
Summarizing the results, we see that {\it all higher order curvature terms of the
string equations of motion $\beta(g_{\mu \nu})$ can be non-vanishing only for the
transverse space components $\beta(h_{ij})$, as all metrics of pp-wave type admit a covariantly
constant null Killing vector field.}

As for the possible correction terms coming from the dilaton and the NS-NS anti-symmetric
tensor fields, they can be easily seen to contribute to higher orders, if at all, only in the
transverse space components of the $\beta$-function equations,
because they do not have any components or dependence on
the remaining space-time coordinates. The R-R 5-form
${\cal F} = dx^+ \wedge \varphi(y)$, which also depends on the coordinates of the
transverse space, can also be seen to contribute
no higher order correction terms in the string equations
of motion. Actually, the situation here is a bit subtle and requires careful investigation.
Following \cite{tsey3}, it can be seen that higher order correction terms that involve
multiple covariant derivatives of ${\cal F}$ can be put equal to zero by working in a natural
scheme where the 2- and 3-point terms in the effective action are not modified from their
supergravity values\footnote{Unlike the bosonic string theory, the on-shell superstring
amplitudes for massless modes do not contain $\alpha^{\prime}$ corrections, i.e., the
supergravity amplitudes are exact, \cite{gsw}.}. Likewise, there can not be any terms that
involve derivatives of ${\cal F}$ contracted with curvature terms. Finally, higher order
terms in ${\cal F}$ consisting of multiple products are also zero due to the ``null"
properties of the pp-wave geometries. All these arguments can be made more elaborative and
exhaustive as in \cite{tsey3}.

In any case, having arranged for all possible higher order correction
terms in $\alpha^{\prime}$
to appear only in the $\beta$-functions of the transverse space theory, which, thus, are
decoupled from the rest to all orders in $\alpha^{\prime}$, one may
further take a second step
and show that for all $N=(4, 4)$ superconformal building blocks of
the transverse space, these
higher order  corrections are actually absent. In fact, one may note independently that
for these lower dimensional models there exists a renormalization group scheme that
makes them exact. Fortunately, this procedure, which is well known but a bit technical to
carry out in detail, has already been
used for the proposed class of four dimensional string backgrounds with dilaton and
torsion in \cite{exact1},
where more details can be found. Hence, the pp-wave geometries we are considering
here are not only solutions
of the lowest order effective theory, but they also provide a large class of
exact string backgrounds.

In conclusion, the generalizations we are considering here
can be very rich in producing a wide range of
non-trivial gravitational pp-wave backgrounds with R-R
fields, but all of them share the same essential stringy properties with the
simplest plane wave
solution, which is exact to all orders in $\alpha^{\prime}$, as in \cite{ama}.

\section{Exact $N=(4,4)$ superconformal building blocks}
\setcounter{equation}{0}

Having presented the main idea behind our construction,
we may now proceed to explicit calculations
by listing some simple examples of $N=(4, 4)$ superconformal field theories,
which all have central
charge $c=4$ to all orders in ${\alpha}^{\prime}$. We first recall that the Wess-Zumino-Witten
models
\be
SU(2)_k ~, ~~~~~ SU(2)_k/U(1) ~, ~~~~~ SL(2)_k/U(1)
\ee
are superconformal theories with $(2,2)$ supersymmetry on the world-sheet and central charge
$c = 3k/(k+2)$, $2(k-1)/(k+2)$  and $2(k+1)/(k-2)$ respectively; we may also append
the two-dimensional flat space $F_2$ in the above list of models, with central charge $c=2$.
Then, the following four models, which are obtained by suitable tensor products,
\ba
F_4 & = & F_2 \times F_2 ~, ~~~~~ \Delta_k = (SU(2)_k/U(1))\times (SL(2)_{k+4}/U(1)) ~,
\nonumber\\
C_k & = & (SU(2)_k/U(1)) \times U(1) \times U(1)_Q ~, ~~~~~ W_k = SU(2)_k \times U(1)_Q
\ea
all have enhanced supersymmetry to $N= (4, 4)$ and central charge $c=4$ independent of
$k$,\footnote{Recall that ${\alpha}^{\prime} \sim 1/k$ and hence the large level $k$
limit is the semi-classical limit that
corresponds to supergravity backgrounds to lowest order in ${\alpha}^{\prime}$.
Higher curvature terms typically arise to higher orders in ${\alpha}^{\prime}$, but for
the $N=(4,4)$ superconformal models these terms are vanishing by supersymmetry.}
provided that the background charge of the $U(1)_Q$ factor is taken to be
$Q=\sqrt{2/(k+2)}$ (see, for instance, \cite{koun} for a detailed exposition from where
we also borrow the notation).
The flat space $F_4$ will not be discussed, as it is a trivial example of a
superconformal field theory; if the transverse space of the proposed ten dimensional
backgrounds \eqn{idea} is built from two $F_4$ superconformal blocks, we will obtain
the class of the usual pp-wave backgrounds that have already been discussed in the
literature \cite{malda2}.

Hence, we will restrict attention to models where at least one of the $N=(4,4)$ superconformal
building blocks is any one from the list of non-trivial gravitational backgrounds
appearing above, namely $\Delta_k$, $C_k$
or $W_k$. We will also extend our presentation to include yet another exact model, the so
called fat throat model, from which all other spaces (including $F_4$)
can be obtained by dualities and/or
appropriate contractions in its parameter space.
In the following, we provide a summary of their target space description using complex coordinates
or equivalently using real coordinates.

(i) \underline{$\Delta_k$}: This model describes semi-classically the geometry of a
two-dimensional bell times a two dimensional cigar (the Euclidean two-dimensional black-hole)
and has non-trivial metric and dilaton fields in four dimensions. In this case, the {\em negative}
curvature of the compact coset cancels the {\em positive} curvature of the non-compact coset so that
the resulting central charge deficit is $\delta c = 0$ to all orders in $1/k$. In particular,
we have
\ba
{1 \over k}ds^2 &=& {du d\bar{u} \over 1 - u \bar{u}} + {dw d\bar{w} \over w \bar{w} + 1} ~,
\nonumber\\
-2\Phi & = & {\rm log} (1- u\bar{u}) + {\rm log} (w \bar{w} + 1)
\ea
provided that $|u|^2 \leq 1$. In this model there is no anti-symmetric tensor field.
Parametrizing the complex coordinates as
\be
u = {\rm sin}\theta e^{i \phi} ~, ~~~~~ w = {\rm sinh}\rho e^{i \chi} ~,
\ee
we obtain the following description of the solution in real coordinates,
\ba
{1 \over k}ds^2 & = & \left(d\theta^2 + {\rm tan}^2 \theta d\phi^2\right) +
\left(d\rho^2 + {\rm tanh}^2 \rho d\chi^2\right) , \nonumber\\
-2\Phi & = & {\rm log}({\rm cos}^2 \theta) + {\rm log}({\rm cosh}^2 \rho) ~.
\ea
The semi-classical geometry of the $SL(2)/U(1)$ coset model
was first studied in the literature in order to interpret
solutions of two dimensional string theory as black holes \cite{black}; the Lagrangian
formulation of the $SU(2)/U(1)$ coset was also studied along similar lines
(see, for instance, \cite{kir}).

There is a subtle point that is worth emphasizing for later use, namely that the cigar geometry
of the $SL(2)/U(1)$ factor corresponds to the axial gauging of $U(1)$. The vector gauging
is simply related to it by a $T$-duality transformation with respect to the Killing vector
$\partial_{\chi}$ and turns ${\rm tanh} \rho$ to ${\rm coth} \rho$ in the metric and
${\rm cosh} \rho$ to ${\rm sinh} \rho$ in the dilaton \cite{kir}.
Thus, we obtain a second version of the
$\Delta_k$ model, which is given by
\ba
{1 \over k}{\tilde{ds}}^2 &=&
{du d\bar{u} \over 1 - u \bar{u}} + {dw d\bar{w} \over w \bar{w} - 1} ~,\nonumber\\
-2\tilde{\Phi} & = & {\rm log} (1- u\bar{u}) + {\rm log} (w \bar{w} - 1)
\ea
and describes the geometry of a two dimensional bell times a two dimensional trumpet with infinite
curvature at the boundaries of the allowed range of values $|u|^2 \leq 1$, $|w|^2 \geq 1$. Note in
this case that the appropriate parametrization of $w$ changes to
$w = {\rm cosh}\rho {\rm exp}(i\chi)$. On the other hand, $T$-duality has no effect on the
topology of the $SU(2)/U(1)$ coset, as the bell interpretation remains unchanged under
$\theta \rightarrow \pi/2 - \theta$ that summarizes the action of the corresponding $T$-duality
on the $SU(2)/U(1)$ factor with respect to the isometry generated by the Killing vector
field $\partial_{\phi}$.

(ii) \underline{$C_k$}: This model describes semi-classically the geometry of a
two-dimensional bell times a two-dimensional cylinder with metric and dilaton fields
given by
\ba
{1 \over k}ds^2 & = & {du d\bar{u} \over 1-u\bar{u}} + dw d\bar{w} ~, \nonumber\\
-2 \Phi & = & {\rm log}(1 - u\bar{u}) + w + \bar{w} ~,
\ea
provided again that $|u|^2 \leq 1$. In this model there is also no anti-symmetric tensor
field. Parametrizing the complex coordinates as
\be
u = {\rm cos} \theta e^{i \phi} ~, ~~~~~ w = \rho + i \chi ~,
\ee
we obtain the following description of the solution in real coordinates,
\ba
{1 \over k}ds^2 & = & d\rho^2 + d\chi^2 + d\theta^2 + {\rm cot}^2 \theta \phi^2 ~,
\nonumber\\
-2 \Phi & = & 2\rho + {\rm log}({\rm sin}^2 \theta) ~.
\ea
We observe a linear dependence of the dilaton field on $\rho$, which is characteristic of
the Liouville-like boson $U(1)_Q$ with background charge;
the other $U(1)$ boson of the model parametrized
by $\chi$ is compactified on a circle of radius $\sqrt{k}$.

(iii) \underline{$W_k$}: This model describes semi-classically the well-known (semi)-wormhole
background, which is an axionic instanton with manifest space-time supersymmetry. We have
a conformally flat metric, a dilaton, as well as an anti-symmetric tensor field given by
\ba
& & {1 \over k}ds^2 = {du d\bar{u} + dw d\bar{w} \over u\bar{u} + w \bar{w}} ~, ~~~~
-2 \Phi = {\rm log}(u\bar{u} + w \bar{w}) ~, \nonumber\\
& & H = {1 \over 2(u\bar{u} + w\bar{w})^2} \left((\bar{u} du - u d\bar{u})
\wedge dw \wedge d\bar{w} + (\bar{w} dw - w d\bar{w}) \wedge du \wedge d \bar{u} \right) .
\ea
Introducing real coordinates
\be
u = e^{\rho + i \phi} {\rm cos}\theta ~, ~~~~~ w = e^{\rho + i \psi} {\rm sin}\theta ~,
\ee
we obtain the equivalent expressions
\ba
{1 \over k}ds^2 & = & d\rho^2 + d\theta^2 + {\rm sin}^2 \theta d\psi^2 + {\rm cos}^2
\theta d\phi^2 , \nonumber\\
-2 \Phi & = & 2 \rho ~, \nonumber\\
B_{\phi \psi} & = & {\rm cos}^2 \theta ~, ~~~~ {\rm with} ~~
H_{\theta \psi \phi} = {\rm sin}2\theta ~.
\ea
Unlike the previous two backgrounds, $W_k$ has a non-vanishing torsion
that originates from the Wess-Zumino-Witten term of the $SU(2)_k$ model.
In this case, for large $k$, the three coordinates of $SU(2)_k$ define a three-dimensional
space with topology $S^3$, while the fourth coordinate $\rho$, which is non-compact
with background charge,
parameterizes the scale factor of the $S^3$ sphere, thus forming a four-dimensional
(semi)-wormhole geometry
as $\rho$ varies along its throat; the throat becomes infinitely thin as
$\rho \rightarrow -\infty$, where the dilaton field blows up.
Also note that the dilaton is linear depending only
on $\rho$ and satisfies the axionic instanton condition
\be
{1 \over 2}H_{ijk} - {{\epsilon}_{ijk}}^l \partial_l \Phi = 0 ~,
\ee
where ${\epsilon_{ijk}}^{l}$ is the covariantized fully anti-symmetric tensor in
four dimensions; it supplies
the vanishing condition for the supersymmetric variation
of the dilatino field $\lambda$. Actually, $W_k$, which describes only the throat of the
wormhole, can be promoted to the complete wormhole solution by performing a simple S-duality
transformation that shifts $e^{2\Phi}$ by a constant without affecting the axionic
instanton solution.

Finally, we note for completeness that $C_k$ can be obtained from the (semi)-wormhole
background by a $T$-duality transformation with respect to the Killing vector
field $\partial_\phi$ of $W_k$, supplemented by a simple change of coordinates,
\be
\phi = \tilde{\phi} - {\tilde{\chi} \over 2} ~, ~~~~~ \chi = \tilde{\phi} +
{\tilde{\chi} \over 2} ~. \label{coc}
\ee

(iv) \underline{Fat throat model}: This model provides a generalization of the
standard (semi)-wormhole solution with a throat that never becomes infinitely thin.
As such, it may be viewed as a regularized version of $W_k$ by switching on another
moduli, say $r_0$, which parameterizes the fatness of its throat. The fat throat
model also satisfies the axionic instanton solution, thus providing explicit another
example of an exact $N=(4,4)$ superconformal field theory block with $c=4$ to all
orders in $\alpha^{\prime}$. Furthermore, it can be related to the $\Delta_k$ model
by a $T$-duality transformation, thus completing the picture of interrelations
among all supersymmetric backgrounds with exact conformal field theory description
as Wess-Zumino-Witten models. In the thin throat limit, as $r_0 \rightarrow 0$,
one recovers the $W_k$ model, in which case the T-dual background becomes $C_k$,
as has been advocated above. Put it differently, the fat thoat model is the
most general axionic instanton background that all other exact $N=(4,4)$
Wess-Zumino-Witten models
relate to it by dualities or contractions in its moduli space.

More precisely, let us start from the bell-trumpet version of the model $\Delta_k$
and introduce the change of coordinates as in equation \eqn{coc} above.
The bell-black-hole version can also be treated along similar lines as described
below. Performing a T-duality transformation with respect to the Killing vector field
$\partial_{\tilde{\phi}}$, one obtains the following background that also
includes a non-trivial anti-symmetric tensor field:
\ba
ds^2 & = & d\rho^2 + d\theta^2 + {1 \over {\rm cosh}^2 \rho - {\rm cos}^2 \theta}
\left({\rm cos}^2 \theta {\rm cosh}^2 \rho d \tilde{\phi}^2 + {\rm sin}^2 \theta
{\rm sinh}^2 \rho d \tilde{\chi}^2 \right) , \nonumber\\
e^{-2\Phi} & = & e^{-2\Phi_0} \left({\rm cosh}^2 \rho - {\rm cos}^2 \theta \right) ,
\nonumber\\
B_{\tilde{\chi} \tilde{\phi}} & = & {{\rm sin}^2 \theta {\rm cosh}^2 \rho \over
{\rm cosh}^2 \rho - {\rm cos}^2 \theta} ~,
\ea
where the dilaton is determined up to a constant term written as ${\rm exp}(-2\Phi_0)$.
The background we have obtained in this way is the fat throat model and it can be
easily checked that it satisfies the axionic instanton condition \cite{bak1},
\cite{bak2}, \cite{sfet}.

In order to
describe the details of the geometry, as well as exhibit its complex structure,
it is useful to introduce the complex coordinates
\be
u = r_0 {\rm sin} \theta {\rm sinh}\rho e^{i\tilde{\chi}} ~, ~~~~
w = r_0 {\rm cos} \theta {\rm cosh}\rho e^{i\tilde{\phi}} ~,
\ee
where $r_0^2 = {\rm exp}(-2\Phi_0)$, but it can also be normalized to 1.
Then, following \cite{sfet}, the
background assumes the conformally flat form
\be
ds^2 = e^{2\Phi}(du d\bar{u} + dw d\bar{w}) ~, ~~~~~ e^{-2\Phi} =
\sqrt{\left(|u|^2 + |w|^2 + r_0^2 \right)^2 - 4r_0^2 |w|^2} ~,
\ee
whereas the anti-symmetric tensor can also be written in complex notation using
the axionic instanton condition. As for all axionic instantons, we have that
${\rm exp}(2\Phi)$ satisfies the Laplace equation in flat $u$, $w$ coordinates.
The model we are describing here reduces to the usual (semi)-wormhole space $W_k$
by taking the limit $r_0 \rightarrow 0$, provided that a rescaling of
the target space coordinates by $r_0$ is also taken into account; this rescaling
means that the geometry looks like an ordinary (semi)-wormhole from far away.
$W_k$ has an
$O(4)$ symmetry which is broken to $SO(2) \times SO(2)$ when the moduli $r_0$
is turned on, or else by looking close enough to the details of the new
background. It can be readily seen for $r_0 \neq 0$ that the dilaton blows up
when $u=0$ and $|w|^2 = r_0^2$, i.e., the singularities of the metric are not
concentrated on a single point but they rather spread out on a ring of radius $r_0$.
As a result, the background geometry, which still represents a (semi)-wormhole,
has a fat throat that can only become infinitely thin when $r_0 \rightarrow 0$.
Also, reversing the T-duality transformation that gave rise to the fat throat
model, we find that the model $C_k$ arises instead of $\Delta_k$ in the
contraction limit $r_0 \rightarrow 0$. Finally, we add for completeness that
in the other extreme limit, $r_0 \rightarrow \infty$, the axionic instanton
becomes simplified by suitable rescaling of the coordinates; thus by
zooming at its ring structure we arrive at a simpler supersymmetric background,
which is actually dual to the trivial flat space solution $F_4$.

Summarizing the present exposition, we note that {\it all exact $N=(4,4)$
superconformal field theories at hand with $c=4$, are either axionic instantons or
T-dual faces of them.} Of course, more general axionic instantons could also be
used in the transverse space by appealing even to other solutions with conformally
hyper-Kahler spaces and torsion , and their possible T-dual versions, but in
those cases there is no exact conformal field theory description as
Wess-Zumino-Witten models. An exact description is actually necessary in order to
demonstrate expliciltly the superconformal properties of such backgrounds, and
hence obtain a realization of the extended world-sheet supersymmetry that
is important for the light-cone formulation of string theory.
Is is useful to recall at this point that all axionic instanton backgrounds
exhibit manifest space-time supersymmetry as the dilatino variation vanishes
and Killing spinors can be constructed explicitly by solving the vanishing
condition of the gravitino variation. In these cases, the world-sheet
supersymmetry is locally realized and the usual theorems that relate it with
space-time supersymmetry are valid. On the other hand, T-duality with respect
to non-triholomorphic isometries, like $\partial_{\phi}$ that was used in the
examples above, does not commute with space-time supersymmetry, in the way that this is
usually realized in the lowest order effective theory \cite{bak3}. As a result, the
dual geometries, which only have a dilaton but no torsion fields, cannot be space-time
supersymmetric because the axionic instanton condition is violated after
the T-duality transformation, \cite{bs1}, \cite{bak1}.

This is not surprising as T-dual backgrounds generically exhibit less symmetries
than the original more symmetric geometries; in fact, only those isometries
that commute with a given Killing vector field are manifest in the T-dual
face of any model, whereas the remaining appear to be lost in the lowest order
effective geometry. However, since T-duality always yields an equivalent string
theory with the same amount of world-sheet supersymmetry, a paradox seems to
appear with regard to the supersymmetric properties of the corresponding
models. Its resolution is provided by the fact that the world-sheet
superconformal algebra simply changes realization from local to non-local,
thus evading the usual theorems that relate world-sheet with target space
symmetries \cite{bs1}. In the exact models we have presented above, it is possible
to use the (non-local) parafermion fields of the $\Delta_k$ and $C_k$ spaces
to provide the required representation of the $N=(4,4)$ world-sheet supersymmetry
inspite of the lack of manifest space-time supersymmetry \cite{koun}.
Since we will be mainly
interested in world-sheet actions in the light-cone gauge of string theory
for all backgrounds of pp-wave type, these remarks are certainly
important for understanding
the supersymmetric properties of the resulting interacting string models. Thus,
in general, it will make sense to search for supersymmetric pp-wave backgrounds
by solving the Killing spinor equations only in those case that the transverse
space is replaced by an axionic instanton background, but not by other spaces.

\section{R-R forms from $N=(4,4)$ superconformal blocks}
\setcounter{equation}{0}

In this section we consider the case where four of the transverse coordinates are
replaced by an exact $N=(4,4)$ superconformal field theory block, which can be
either $\Delta_k$, $C_k$ or $W_k$ or its fat throat generalization,
together with the dilaton and anti-symmetric NS-NS
tensor fields asscociated with them. The remaining four transverse dimensions
will be provided trivially
by the flat Euclidean space $F_4$, which will be parametrized using the complex
coordinates $v$ and $z$. Of course, one may consider more complicated possibilities where
the remaining four flat coordinates are also replaced by any exact superconformal
block with $N=(4,4)$ world-sheet supersymmetry, in all possible combinations,
but such hybrid models
will not be worked out in detail here, with the exception of a short discussion
at the end of this paper.

In any case, the results we describe in this section will be derived
by solving the second order equations \eqn{main} for appropriate (but educated)
choices of the front factor $F$, which give rise to some interesting integrable systems
in the two-dimensional light-cone action of string theory.
Hence, in this section,
we will not rely on supersymmetric conditions, i.e., the existence of Killing spinors in
target space, which typically
provide additional relations between the front factor $F$ and the R-R 5-form
${\cal F} = dx^+ \wedge \varphi$, and which are associated to first order
equations. The main reason is that the models $\Delta_k$ and $C_k$ do not exhibit
space-time supersymmetry, since their $N=(4,4)$ world-sheet supersymmetry is non-locally
realized.
On the other hand, the
(semi)-wormhole models have space-time supersymmetry, as
all axionic instanton backgrounds do. Here, we will only consider them briefly without
being concerned about space-time supersymmetry of the resulting pp-wave configurations.

In this section, we will derive some explicit pp-wave solutions and discuss in detail
their interpretation as integrable systems in the light-cone formulation of string theory.
In one instance we will demonstrate explicitly the emergence of the complex sine-Gordon
model, which can be viewed as integrable perturbation of the corresponding
superconformal building blocks of the transverse space. In other cases we will arrive at
the supersymmetric Liouville theory or at the complex sine-Liouville model. Thus,
some old results about integrable perturbations of two-dimensional conformal field
theories can be revisited and reinterpreted in the target space of string theory
due to the presence of non-trivial R-R fields. We will also examine some relations
among the resulting pp-wave backgrounds of type IIB theory by relying on the web of
known dualities that interconnect all exact superconformal field theory blocks
with $N=(4,4)$ superconformal invariance and $c=4$.
The construction of supersymmetric solutions for the (semi)-wormhole
models will be examined separately in section 6, where we will focus
exclusively on
the Killing spinor equations for the fields $g$, $\Phi$, $B$ and ${\cal F}$
in general axionic instanton backgrounds.
Only then, we will be able to obtain a systematic generalization of the Maldacena-Maoz
space-time
supersymmetric framework in the presence of non-trivial dilaton and NS-NS anti-symmetric
tensor fields.

\subsection{Solutions for the $\Delta_k$ model}

We start by considering solutions of the main second order equation \eqn{main} that
relates $F$ with $\varphi$ for the coset model $\Delta_k$.
Straightforward calculation shows that in this background we obtain the following
result
\ba
\nabla_i\left(e^{-2\Phi} \nabla^i F\right) & = & h^{u\bar{u}} h^{w\bar{w}}
\left( h^{u\bar{u}} \partial_u \partial_{\bar{u}} F + h^{w\bar{w}}
\partial_w \partial_{\bar{w}}F
+ \partial_v \partial_{\bar{v}} F +
\partial_z \partial_{\bar{z}} F\right) \nonumber\\
& + & {1 \over 2}
h^{u\bar{u}}h^{w\bar{w}} \left(w\partial_w F + \bar{w} \partial_{\bar{w}} F
- u\partial_u F - \bar{u} \partial_{\bar{u}} F \right)
\ea
for any generic choice of the front factor $F$. However, not any $F$ will produce a
quantity that can be written as the norm-squared of an anti-self-dual closed
4-form $\varphi$. We will describe
an interesting class of solutions that arise when $\Delta_k$ replaces
four of the transverse coordinates, using the following choice of front factors:
\be
F = -A^2 |u|^2 + B^2 |w|^2 + C^2 |v|^2 + D^2 |z|^2 ~.
\ee
Substituting into the general expression, we find, in particular, that the
corresponding $\varphi$ is determined by equation \eqn{main} to be
\be
{1 \over 2} |\varphi|^2 = h^{u\bar{u}} h^{w\bar{w}}
(2A^2 |u|^2 + 2B^2 |w|^2 + C^2 + D^2 -A^2 \pm B^2) ~,
\ee
where $h^{u\bar{u}} = 1 - |u|^2$ and $h^{w\bar{w}} = |w|^2 \pm 1$ are the components
of the inverse metric of the $SU(2)/U(1)$ and $SL(2)/U(1)$ cosets respectively;
here, the $\pm$
signs refer to the axial (respectively vector) gauging of the $SL(2)/U(1)$ part that
describes the geometry of a two-dimensional black-hole (respectively trumpet).

Choosing the real constants $A$, $B$, $C$ and $D$ so that
$\Delta^2 \equiv C^2 + D^2 - A^2 \pm B^2 \geq 0$,
we obtain a positive definite result for $|\varphi|^2$,
and therefore a solution to equation \eqn{main}
can be easily obtained by choosing the 4-form $\varphi$ as the following sum of
$(1,3)$-forms, plus their complex conjugates:
\ba
\varphi & = & 2Au du \wedge d\bar{w} \wedge d\bar{v} \wedge d\bar{z} +
2Bw dw \wedge d\bar{u} \wedge d\bar{v} \wedge d\bar{z} \nonumber\\
& + & \Delta dv \wedge d\bar{u} \wedge d\bar{w} \wedge d\bar{z} +
\Delta dz \wedge d\bar{u} \wedge d\bar{w} \wedge d\bar{v} + ~ cc ~.
\ea
This is an anti-self-dual closed 4-form that provides a solution to the equations
of type IIB supergravity in ten dimensions in the presence of a
self-dual (but non-constant) 5-form ${\cal F}$, as required. Note that the
component $\varphi_{w\bar{u}\bar{v}\bar{z}} \sim w$ vanishes for $w=0$, which is
possible only for the two-dimensional black-hole geometry; in this
case the geometry is that of an infinitely long cigar with scalar curvature
$R = 1/(|w|^2 + 1)$ that becomes maximal at its tip, where $w=0$. For the trumpet
geometry we have the constraint $|w|^2 \geq 1$, and hence $w$ can never vanish.
On the other hand, the component $\varphi_{u\bar{w}\bar{v}\bar{z}} \sim u$
always vanishes
at the symmetric top point $u=0$ of the bell
that describes the geometry of the $SU(2)/U(1)$ coset,
in which case the scalar curvature of the corresponding
two-dimensional space, $R = -1/(1-|u|^2)$, also assumes its maximum value.
Hence, for the bell-cigar geometry, the R-R 5-form ${\cal F}$ vanishes at
the point $(u, w) = (0,0)$ provided that the constants are chosen so that
$\Delta = 0$.

In the present case, the bosonic world-sheet action that describes
the ten-dimensional string model in the light-cone gauge, $x^+ = \tau$,
assumes the form
\be
S _{\rm lc} = S_{SU(2)/U(1)} + S_{SL(2)/U(1)}^{(\pm)} + S_v + S_z ~,
\ee
where the individual terms on the world-sheet $\Sigma$ are
\ba
S_{SU(2)/U(1)} & = & \int_{\Sigma} \left({\partial u \bar{\partial} \bar{u} +
\partial \bar{u} \bar{\partial} u \over 1 - |u|^2} + A^2 |u|^2 \right) ,
\nonumber\\
S_{SL(2)/U(1)}^{(\pm)} & = & \int_{\Sigma} \left({\partial w \bar{\partial} \bar{w} +
\partial \bar{w} \bar{\partial} w \over |w|^2 \pm 1} - B^2 |w|^2 \right) ,
\nonumber\\
S_v + S_z & = & \int_{\Sigma} \left(\partial v \bar{\partial} \bar{v} +
\partial \bar{v} \bar{\partial} v - C^2 |v|^2 \right) +
\int_{\Sigma} \left(\partial z \bar{\partial} \bar{z} +
\partial \bar{z} \bar{\partial} z - D^2 |z|^2 \right)
\ea
and $\pm$ refer to the two possible inequivalent geometries of the non-compact coset.
Clearly, $S_v$ and $S_z$ describe the action of the complex bosons $v$ and $z$ with
positive mass-squared terms $C^2$ and $D^2$ respectively. The first two terms on the
other hand, describe the semi-classical action of the $SU(2)/U(1)$ and $SL(2)/U(1)$
gauged Wess-Zumino-Witten models, which are also perturbed by potential terms of the
form $|u|^2$ and $|w|^2$ respectively.

It is instructive to write down the contribution from the two coset models in real
coordinates, choosing for definiteness the parameterization
$u = {\rm cos}\theta {\rm exp}(i \phi)$
and $w={\rm cosh}\rho {\rm exp}(i \chi)$ in the case of the bell-trumpet geometry.
We have, in particular,
\ba
S_{SU(2)/U(1)} & = & \int_{\Sigma} \left(\partial \theta \bar{\partial} \theta +
{\rm cot}^2 \theta \partial \phi \bar{\partial} \phi + A^2 {\rm cos}^2 \theta
\right) , \nonumber\\
S_{SL(2)/U(1)}^{(-)} & = & \int_{\Sigma} \left(\partial \rho \bar{\partial}\rho +
{\rm coth}^2 \rho \partial \chi \bar{\partial} \chi - B^2 {\rm cosh}^2 \rho
\right) ,
\ea
whereas the case of the bell-cigar geometry can be discussed in a similar fashion.
It is quite interesting to note that both these two-dimensional theories are
integrable and coincide with the so-called complex sine-Gordon and complex sinh-Gordon
models respectively \cite{regge}, \cite{pohl}; see also \cite{bak4} for a detailed
description of their conservation laws, among other things,
\cite{hollo} for some quantum mechanical calculations of their spectrum and scattering
properties,
and \cite{shin} for the construction of some periodic solutions.

First of all, in order to get a feeling about the nature of the resulting
two-dimensional models, we may set $\phi = 0 = \chi$, which is consistent with
the two-dimensional field equations\footnote{This choice, however, is not consistent
with the string background equations in target space as they reduce the dimensionality
of the supergravity background by two, thus turning the critical string theory at
hand into a non-critical theory.}. Then, we
observe that $S_{SU(2)/U(1)} + S_{SL(2)/U(1)}^{(-)}$ simply
becomes the action of the ordinary sine-Gordon model plus that of the sinh-Gordon model,
\be
S^{\prime} = \int_{\Sigma} \left(\partial \theta \bar{\partial}\theta + A^2 {\rm cos}^2
\theta \right) + \int_{\Sigma} \left(\partial \rho \bar{\partial}\rho - B^2
{\rm cosh}^2 \rho \right) ,
\ee
which, by the way, for the special choice of parameters $A=B$ coincides with the
bosonic action of the
$N=2$ sine-Gordon model for the field $\theta + i \rho$ and holomorphic superpotential
${\cal W} = {\rm cos}(\theta + i \rho)$. Then,
it becomes obvious that in the presence of the $\phi$ and $\chi$
components of the metric, the complete action $S_{SU(2)/U(1)}$ provides a generalization
of the sine-Gordon model in terms of a complex field $u$ with $|u|^2 \leq 1$;
as a result, the
configurations of the complex sine-Gordon model carry an additional $U(1)$
charge, which is
associated with the Noether current with respect to the symmetry $\phi \rightarrow
\phi + a$ of the action. Likewise, the complete action $S_{SL(2)/U(1)}^{(-)}$ provides a
generalization of the sinh-Gordon model in terms of a complex field $w$ with
$|w|^2 \geq 1$; its configurations also carry a $U(1)$ charge due to the invariance
of the action under $\chi \rightarrow \chi + a$.

The complete action $S_{\rm lc} = S_{SU(2)/U(1)} + S_{SL(2)/U(1)} + S_v + S_z$
can be made supersymmetric for $A=B$
by straightforward generalization of the supersymmetric treatment that
has already been discussed in the literature for the
complex sine-Gordon model \cite{gates}; in this case, the potential of the
two-dimensional theory is
\be
V  \equiv F = -|u|^2 + |w|^2 + |v|^2 + |z|^2 ~,
\ee
where we have also set $A=B=C=D=1$ (and hence $\Delta = 0$),
which in turn can be derived from a holomorphic superpotential,
\be
{\cal W} = u + w + {1 \over 2}v^2 + {1 \over 2}z^2 ~,
\ee
as follows:
\be
V= h^{u \bar{u}} \mid {\partial {\cal W} \over \partial u} \mid^2 +
h^{w \bar{w}} \mid {\partial {\cal W} \over \partial w} \mid^2 +
\mid {\partial {\cal W} \over \partial v} \mid^2 +
\mid {\partial {\cal W} \over \partial z} \mid^2 ~.
\ee
Furthermore, the kinetic terms can be derived from the Kahler potential
\be
{\cal K} = K_1(u, \bar{u}) + K_2(w, \bar{w}) + |v|^2 + |z|^2 ~,
\ee
where the Kahler potential terms of each coset model are defined
separately via the equations
\be
\partial_t K_1(t) = {1 \over t} {\rm log}{1 \over 1-t} ~~ {\rm for} ~~ t=|u|^2 ~,
~~~~~
\partial_s K_2(s) = {1 \over s} {\rm log}{1 \over s-1} ~~ {\rm for} ~~ s=|w|^2 ~.
\ee

The complex sine-Gordon model was originally introduced by Lund and Regge in their
classical treatment of vortices \cite{regge},
but it also has two other interesting interpretations
that clarify its integrability properties. The first one views this model as a
reduced $O(4)$
non-linear sigma-model by exploiting the classical conformal invariance of the
latter \cite{pohl}; as such,
it generalizes the derivation of the ordinary sine-Gordon model
from the $O(3)$ (i.e., $CP^1$) non-linear
sigma-model by a similar type of reduction. In this case, the coupling
constant of the potential term turns out to be positive,
exactly as it appears in the action term
$S_{SU(2)/U(1)}$ that has been obtained above
with parameter $A^2 > 0$\footnote{Note at this end that there
are two physically inequivalent regions of the coupling constant in the complex
sine-Gordon model depending on the sign of $A^2$; we cannot simply change the
sign of the coupling constant in the field equations by redefinition of
$\theta$ and $\phi$, as in the case of the ordinary sine-Gordon model where
$\theta \rightarrow \pi/2 - \theta$ does this job.}. Likewise, the sinh-Gordon model,
which appears in the action term $S_{SL(2)/U(1)}$, can arise as a reduced $O(2,2)$
non-linear sigma model. Therefore, due to this description, it is natural to expect
that massive mirror symmetry, which relates the $N=2$ sine-Gordon model to the
deformed $CP^1$ (or sausage) model \cite{fateev}, \cite{vafa},
may have a suitable generalization in the present
type IIB string background using a deformed $O(4)$ sausage model. We
expect to return to this point elsewhere by studying the more general relation
between multi-component sine-Gordon models on one hand, which arise classically as reduced
non-linear sigma-models (or as perturbed Wess-Zumino-Witten models \cite{bak5}),
and their deformed (or sausage model) cousins on the other hand, which
arise in quantum mechanical considerations by massive mirror symmetry.

A second interpretation of the complex sine-Gordon model (and its sinh-Gordon
companion) can be found within the context of perturbed coset models of two-dimensional
conformal field theory \cite{bak4}. Recall that the primary fields of the $SU(2)_k$
Wess-Zumino-Witten model are of the form $\Phi_{m, \bar{m}}^{(j)}$, where $m$ and
$\bar{m}$ run from $-j$ to $j$, each one taking $(2j+1)$ values, whereas $j$ takes
the values $0, 1/2, 1, \cdots , k/2$, \cite{wzw1}, \cite{wzw2}.
Their conformal dimensions are $j(j+1)/(k+2)$
and the quantum numbers $m$ and $\bar{m}$ describe the $U(1)$ charges of the primary
fields in the two chiral sectors of the model. The simplest non-trivial such
field is obtained for $j=1$, in which case its nine components can be naturally
represented by the matrix elements
\be
\Phi_{ab}^{(1)} = {\rm Tr}(g^{-1} T_a g T_b) ~,
\ee
where $g$, $T$ are $SU(2)$ group elements and Lie algebra generators respectively.
Clearly, their conformal dimension becomes zero in the semi-classical limit
$k \rightarrow \infty$. The $U(1)$-neutral fields $\Phi_{m=0, \bar{m}=0}^{(1)}$
are also primary fields of the $SU(2)_k/U(1)$ coset model with the simplest
example being the following $j=1$ neutral field from the complete list of nine
fields $\Phi_{ab}^{(1)}$,
\be
\epsilon_1 = {\rm Tr} (g^{-1} \sigma_3 g \sigma_3) ~.
\ee
This field can be identified with the so-called
first thermal operator
of the $SU(2)_k/U(1)$ coset model \cite{wzw2}, which is usually
denoted by $\epsilon_1$,  and its form can be worked out in terms of the
target space coordinates in the semi-classical limit. It turns out to be \cite{bak4}
\be
\epsilon_1(u, \bar{u}) = 2(2 |u|^2 - 1) ~,
\ee
and hence, modulo a shift in the vacuum energy, the perturbed conformal field theory
of the $SU(2)/U(1)$ coset model,
\be
S_{\rm pert} = S_{\rm cft} + {A^2 \over 4} \int_{\Sigma} \epsilon_1 ~,
\ee
can be readily seen to coincide with the the action of the complex sine-Gordon model,
as it appears in our present construction of the light-cone string theory.

This is a perturbation that drives the conformal field theory away from criticality,
but in a controlled way, as the resulting field theory is integrable both classically
and quantum mechanically. It is quite interesting that such a perturbation (together
with its sinh-Gordon companion) arise naturally in the context of type IIB
supergravity with non-trivial R-R 5-form. The particular choice of the front factor,
which has been boldly made in the present construction,
indeed describes an integrable
perturbation of the superconformal building block $\Delta_k$ in the transverse
space, as advertized.
Note, however, that this is only the simplest example one may consider in
the present framework, as there are other perturbations that can also appear in the
light-cone formulation of the world-sheet action, which are described by the higher
thermal operators $\epsilon_j$ of the coset model for
appropriate choices of $F$,
and hence ${\cal F}$. The details can be worked out by finding a description of the
higher operators $\epsilon_j$, which exist for all integer values $j$,
in terms of the target space coordinates that conform the two-dimensional field
theory solution into a type IIB supergravity background.

Such generalizations will not
be considered here, but we will only make a remark about their duality properties.
There is the Krammers-Wannier duality that acts on the world-sheet by exchanging
the spin variables $\sigma$ with the dual spin variables $\mu$ of the coset model
(order--disorder duality) \cite{wzw2}. Under this action, the thermal operators behave as
$\epsilon_j \rightarrow (-1)^j \epsilon_j$, and hence for odd values of $j$ they
flip sign. It is clear from the structure of our solution, which corresponds to $j=1$,
that the Krammers-Wannier duality cannot be interpreted as a duality symmetry within
type IIB supergravity because it amounts to changing $A^2$ to $-A^2$ in the
expression for the front factor $F$; such a change is quite severe, as it can be
easily seen that prevents the existence of solutions for the 4-form $\varphi$ in
our class of models.
Hence, the positivity of the coupling constant $A^2$ is a very
rigid requirement here that does not permit the use of the
Krammers-Wannier duality in the
context of the target space theory. One possibility to have a true manifestation
of this duality within the context of type IIB supergravity is to consider solutions
with front factors $F$ that correspond to perturbations by $\epsilon_j$ with even
values of $j$.
Thus, it will be interesting to explore further the existence of such solutions
and obtain the corresponding R-R 5-forms, if they can be defined consistently.

\subsection{Solutions for the $C_k$ model}

Next, we consider some interesting solutions that arise when four of the transverse
coordinates are replaced by the geometry of the $C_k$ conformal field theory in the
presence of the appropriate dilaton field. Straightforward calculation shows
that for generic choices of the front factor $F$, we have the result
\ba
\nabla_i\left(e^{-2\Phi} \nabla^i F \right) & = & e^{w+\bar{w}} h^{u\bar{u}}
\left(h^{u\bar{u}} \partial_u \partial_{\bar{u}} F + \partial_w \partial_{\bar{w}} F
+ \partial_v \partial_{\bar{v}} F + \partial_z \partial_{\bar{z}} F \right)
\nonumber\\
& + & {1 \over 2}
e^{w + \bar{w}} h^{u\bar{u}} \left(\partial_w F + \partial_{\bar{w}} F
- u \partial_u F - \bar{u} \partial_{\bar{u}} F \right) .
\ea
Thus, making the special choice of front factor
\be
F= A^2 e^{w + \bar{w}} - B^2 e^{w + \bar{w}}|u|^2 - C^2 |u|^2 + D^2|v|^2 + E^2|z|^2 ~,
\ee
we find, in particular, that the 4-form $\varphi$ has to satisfy the equation
\be
{1 \over 2} |\varphi|^2 = h^{u \bar{u}} e^{w + \bar{w}}
\left((2A^2 - B^2)e^{w+\bar{w}} + 2C^2|u|^2 +
D^2 + E^2 - C^2 \right) .
\ee
Setting $C^2=2D^2=2E^2$ in the sequel, in order to simplify the presentation,
we may solve equation \eqn{main}
by expressing the 4-form $\varphi$ as a suitable sum of $(1,3)$-forms,
plus their complex conjugates, provided that $2A^2 -B^2 \geq 0$.
Another choice of
parameters that will be useful later, while comparing with solutions of
the $W_k$ space, is $D^2 = -E^2$, which also requires $C=0$ for
positive definiteness of $|\varphi|^2$;
actually, this choice can also be incorporated in case (iii) below, provided that
the appropriate change of parameters takes place. We also note for
completeness that adding a term proportional to $|w|^2$ in $F$ appears to
be inconsistent with the existence of a closed anti-self-dual 4-form
$\varphi$ in the present background.

In the following, we investigate some special cases separately
that provide light-cone systems with increasing degree of complexity.

(i) \underline{$A=B=0$}: The solution for the 4-form is
\be
\varphi = 2C u e^{\bar{w}} du \wedge d\bar{w} \wedge d\bar{v} \wedge
d\bar{z} ~ + ~~ cc ~,
\ee
which vanishes at the symmetric point $u=0$ of the bell geometry $SU(2)/U(1)$.
Then, it is straightforward to see that the world-sheet
action in the light-cone formulation of the ten dimensional model becomes
\ba
S_{\rm lc} & = & \int_{\Sigma} \left({\partial u \bar{\partial} u
+ \partial \bar{u} \bar{\partial} u \over 1 - |u|^2} + C^2 |u|^2
\right) + \int_{\Sigma} \left(\partial w \bar{\partial} \bar{w} +
\partial \bar{w} \bar{\partial} w \right) \nonumber\\
& + & \int_{\Sigma} \left(\partial v \bar{\partial} \bar{v} + \partial \bar{v}
\bar{\partial} v + \partial z \bar{\partial} \bar{z} + \partial \bar{z}
\bar{\partial} z - {1 \over 2} C^2(|v|^2 + |z|^2) \right) .
\ea
It is thus clear that apart from the three complex bosons $w$, $v$ and $z$ with
their respective mass terms, we also have the contribution of a complex
sine-Gordon model associated to the field $u$ that parametrizes the
$SU(2)/U(1)$ part of the building block $C_k$.
We may refer to the previous discussion
for the properties of this integrable system.

(ii) \underline{$B=0$}: In this case the 4-form can be chosen as
\be
\varphi =  2A
e^{2w} dw \wedge d\bar{u} \wedge d\bar{v} \wedge d\bar{z} +
2C u e^{\bar{w}} du \wedge d\bar{w} \wedge d\bar{v} \wedge
d\bar{z}
~ + ~~ cc
\ee
and the light-cone action becomes,
\ba
S_{\rm lc} & = & \int_{\Sigma} \left({\partial u \bar{\partial} u
+ \partial \bar{u} \bar{\partial} u \over 1 - |u|^2} +C^2|u|^2 \right) +
\int_{\Sigma} \left(\partial w \bar{\partial} \bar{w} +
\partial \bar{w} \bar{\partial} w - A^2 \mid e^w \mid^2 \right) \nonumber\\
& + & \int_{\Sigma} \left(\partial v \bar{\partial} \bar{v} + \partial \bar{v}
\bar{\partial} v + \partial z \bar{\partial} \bar{z} + \partial \bar{z}
\bar{\partial} z -{1 \over 2} C^2(|v|^2 + |z|^2) \right) .
\ea
The first term corresponds again to the complex sine-Gordon model,
whereas the second term describes the bosonic part of the $N=2$ Liouville theory
with holomorphic superpotential ${\cal W} = {\rm exp}w$ \cite{liou};
as for the last two
terms, they describe again free complex bosons with their
respective mass terms. Clearly, for
$A=0$, the Liouville potential vanishes and one recovers case (i) above. Also,
$C$ can take any value, including zero, in which case the total potential
simplifies to the
Liouville term alone. In any case, for $B=0$, the resulting solution involves
the complex sine-Gordon and the Liouville models, which are decoupled from
each other. We will present next another choice of parameters that leads
to a coupled complex sine-Liouville model on the world-sheet.

(iii) \underline{$C=0$}: In this case we must also impose the restriction
$2A^2 \geq B^2$ in order to obtain a real solution for the 4-form $\varphi$. Then, we
have,
\be
\varphi = \sqrt{2(2A^2 -B^2)}
e^{2w} dw \wedge d\bar{u} \wedge d\bar{v} \wedge d\bar{z}
~ + ~~ cc
\ee
and the corresponding light-cone sigma-model action takes the form
\ba
S_{\rm lc} & = & \int_{\Sigma} \left({\partial u \bar{\partial} u
+ \partial \bar{u} \bar{\partial} u \over 1 - |u|^2} +
\partial w \bar{\partial} \bar{w} +
\partial \bar{w} \bar{\partial} w - \mid e^w \mid^2 (A^2 - B^2|u|^2)
\right) \nonumber\\
& + & \int_{\Sigma} \left(\partial v \bar{\partial} \bar{v} + \partial \bar{v}
\bar{\partial} v + \partial z \bar{\partial} \bar{z} + \partial \bar{z}
\bar{\partial} z \right) .
\ea
For $B=0$, we recover a special case of the solution (ii) above, where there
is only a Liouville potential for the field $w$. For $2A^2 \geq B^2 \neq 0$,
however, we
obtain a coupled system of the complex sine-Gordon and Liouville models,
which can be thought as a complex field generalization of the usual
sine-Liouville model. To see this, it is instructive to use the parametrization
of the $C_k$ model in terms of real variables and set $\phi = \chi =0$\footnote{As
before, this choice is only made here for illustrative purposes, as it is not
consistent with the string background equations in the critical dimension $D=10$.}
in the
notation that we introduced in the previous section. Then, the $(u, w)$-dependent
part of the two-dimensional action assumes the form
\be
S^{\prime} = \int_{\Sigma} \left( \partial \theta \bar{\partial}\theta +
\partial \rho \bar{\partial} \rho - A^2 e^{2\rho} {\rm sin}^2 \theta \right) ,
\ee
which is a suitable form of the sine-Liouville model that has been written here
for the special choice of parameters
$A=B$ (see, for instance, \cite{moore} and references therein);
note that, in general, for more arbitrary values of the
parameters, we will also have the appearance of a two dimensional
cosmological constant term in the action.

It seems that the
complex sine-Liouville model that we encounter here cannot be made supersymmetric
as it stands for any choice of $A$ and $B$,
as the potential term cannot be obtained from a holomorphic superpotential in $u$
and $w$. We note for completeness that a supersymmetric version of such models,
which has been known
for some time, is provided
by the complex sinh-Gordon-Liouville theory,
\be
\tilde{S}^{\prime} =  \int_{\Sigma} \left({\partial u \bar{\partial} u
+ \partial \bar{u} \bar{\partial} u \over |u|^2 - 1} +
\partial w \bar{\partial} \bar{w} +
\partial \bar{w} \bar{\partial} w - A^2 \mid e^w \mid^2 (2|u|^2 -1)
\right) ,
\ee
where $|u| \geq 1$, i.e., for a non-compact version of $C_k$, where the $SU(2)/U(1)$
coset is replaced by the Euclidean trumpet coset $SL(2)/U(1)$. The latter version
is known to coincide with the called non-abelian $B_2$ Toda system,
\cite{gervais}, and its potential
can be easily derived from the holomorphic superpotential ${\cal W} = u {\rm exp}w$.

\subsection{Solutions for the $W_k$ model}

Next in our list of examples, we consider solutions where
the superconformal field theory block $W_k$ replaces four of the transverse
space coordinates in ten-dimensional type IIB supergravity. The (semi)-wormhole
background, being an axionic instanton, possesses space-time supersymmetry and
therefore it can admit a large class of supersymmetric solutions with non-trivial
R-R 5-form and front factor $F$.
For the moment being,
we will construct a particular set of simple
solutions for the (semi)-wormhole geometry by
examining the second order string background equations \eqn{main}, and
also try to connect them with
the solutions of the $C_k$ model via T-duality. Later, we will revisit this model
from a more general viewpoint that provides solutions to its Killing spinor
equations.

It is straightforward to verify in the present case that
\be
\nabla_i \left(e^{-2 \Phi} \nabla^i F \right) = h^{u\bar{u}} h^{w\bar{w}}
\left( \partial_u \partial_{\bar{u}} F + \partial_{w} \partial_{\bar{w}} F\right)
+e^{-2\Phi}\left( \partial_v \partial_{\bar{v}} F +
\partial_z \partial_{\bar{z}} F \right) \label{wormy}
\ee
for all possible choices of the front factor $F$. Here, we only
consider front factors which are quadratic in the target space coordinates,
namely,
\be
F = A^2 |u|^2 + B^2 |w|^2 + C^2 |v|^2 - D^2 |z|^2 ~.
\label{examp1}
\ee
Explicit calculation shows in this case that
\be
{1 \over 2} |\varphi|^2 = (A^2 + B^2) \left(|u|^2 + |w|^2 \right)^2
+ (C^2 -D^2) \left(|u|^2 + |w|^2 \right)
\ee
and therefore, since $h^{u\bar{u}} = h^{w\bar{w}} = |u|^2 + |w|^2 ={\rm exp}(-2\Phi)$,
we conclude that
a solution to second order equation \eqn{main} can be easily obtained provided
that $C=D$.
With this choice of parameters, we find that $\varphi$ can be written as the
following sum of constant $(1, 3)$-forms, plus their complex conjugates:
\be
\varphi = \sqrt{2}A du \wedge d\bar{w} \wedge d\bar{v} \wedge d\bar{z} +
\sqrt{2}Bdw \wedge d\bar{u} \wedge d\bar{v} \wedge d\bar{z} ~ + ~~ cc ~ .
\label{examp2}
\ee

Then, the world-sheet action in the light-cone gauge of the corresponding string
background assumes a form, which is conveniently written here using the real
coordinates of $W_k$,
\ba
S_{\rm lc} & = & \int_{\Sigma} \left(\partial \rho \bar{\partial} \rho
+ \partial \theta \bar{\partial} \theta +
{\rm sin}^2 \theta \partial \psi \bar{\partial} \psi  + {\rm cos}^2 \theta
\left( \partial \phi \bar{\partial} \phi +\partial \phi \bar{\partial} \psi -
\bar{\partial} \phi \partial \psi \right) \right)\nonumber\\
&-& \int_{\Sigma} e^{2\rho}(A^2 {\rm cos}^2 \theta
+ B^2 {\rm sin}^2 \theta) \nonumber \\
& + & \int_{\Sigma} \left(\partial v \bar{\partial} \bar{v} + \partial \bar{v}
\bar{\partial} v + \partial z \bar{\partial} \bar{z} + \partial \bar{z}
\bar{\partial} z - C^2 |v|^2 + C^2 |z|^2 \right) .
\ea
Thus, we see the emergence of a Liouville term for the field $\rho$ that originates
from the non-compact boson $U(1)_Q$ of the model $W_k = SU(2)_k \times U(1)_Q$
with background charge, which is also coupled to $\theta$ for $A \neq B$, as in
the sine-Liouville model.
The first term in $S_{\rm lc}$ is the usual
semi-classical action of the $SU(2) \times U(1)$
Wess-Zumino-Witten model, where we have also
included the contribution from the anti-symmetric tensor field $B_{\phi \psi} =
{\rm cos}^2 \theta$, whereas the last term describes two complex bosons $v$, $z$
with mass-squared terms, one of which is tachyonic; these mass terms can also be
set equal to zero without drastic change of our solution.

Note that in heterotic and type II string theory,
the five-brane solution is constructed by
considering the ten-dimensional background $M_6 \times W_k$, where $M_6$ is the
six-dimensional Minkowski space. Here, we are making another
practical use of the (semi)-wormhole
geometry by embeding it into a ten-dimensional background of type IIB supergravity
of the form $\tilde{P}_6 \times W_k$, where
\be
ds^2 (\tilde{P}_6) = -4dx^+ dx^- - \left(A^2|u|^2 + B^2|w|^2 + C^2(|v|^2 - |z|^2)
\right) \left(dx^+ \right)^2 + dv d\bar{v} + dz d\bar{z}
\ee
is the six-dimensional
part of a pp-wave background that gives rise to a non-trivial constant
RR 5-form ${\cal F}$. It is worth mentioning that for $A= 0 =B$, the space
$\tilde{P}_6$ becomes the standard pp-wave backgroung in six dimensional space-time,
$P_6$, having the usual
quadratic front factor in $v$ and $z$ with zero trace;
if we also have $C=0$, which is certainly allowed
without threatening our final result with trivialization, $P_6$ will be the
Minkowski space $M_6$ written in light-cone coordinates.
Then, for $A, B \neq 0$, we are perturbing the front factor of the the six-dimensional
metric by the term $A|u|^2 + B^2|w|^2$, which depends on the
remaining (semi)-wormhole coordinates,
thus producing the relevant deformation of the pp-wave
background to $\tilde{P}_6$.

One may inquire at this point about the effect of the T-duality transformation
with respect to the isometry generated by the Killing vector field $\partial_\phi$,
which relates the superconformal field theory $W_k$ to $C_k$. Certainly, this has
no effect on the combination $A^2|u|^2 + B^2|w|^2 + C^2(|v|^2 - |z|^2) \equiv
e^{2 \rho}(A^2 {\rm cos}^2 \theta + B^2 {\rm sin}^2 \theta) + C^2 (|v|^2 - |z|^2)$ that
provides the front factor for the (semi)-wormhole background in transverse space;
therefore, it can be used without change in order to
describe solutions of type IIB supergravity
in ten dimensions, but with the superconformal field theory block $C_k$ living
now in their
transverse space. Writing the $\rho$-dependence of $F$ in terms of the new complex
coordinates $u$ and $w$ that parameterize
$C_k$, which should be distinguished from the corresponding complex coordinates
that parameterize $W_k$, it is easy to note that this front factor takes the following
form in the new (dual) background:
\be
F = B^2 e^{w +\bar{w}} -(B^2 - A^2) e^{w+\bar{w}} |u|^2 + C^2 (|v|^2 - |z|^2) ~.
\ee
This corresponds precisely to the general choice of front factor that was earlier
made for the model
$C_k$, provided that one sets there $C=0$ and rename $D^2 = -E^2$ as
the new $C^2$; we also have to rename $A^2$ and $B^2$ by comparing the two
different expressions,
in which case the condition $2A^2 \geq B^2$ for the $C_k$ models becomes
$A^2 + B^2 \geq 0$ for $W_k$. Since the dilaton also changes under the proposed T-duality,
one concludes that the operation of duality in transverse space yields the
R-R 5-form of the $C_k$ background as described by the special class of
solutions (iii) in the corresponding complex variables and for the appropriate
choice of parameters that turn the $z$-boson into a tachyon with mass-squared
equal to $-C^2$.

Thus, although T-duality relates in general type IIB to type IIA backgrounds in ten
dimensions, we see here that it can also be used as a trick in the transverse space to
provide a solution generating technique for type IIB supergravity with non-trivial
R-R 5-form fields. By the same token, we may attempt to
find other type IIB backgrounds for the
$W_k$ model, which are obtained by duality from the more general class of $C_k$ models
that have been studied above. Writing the front factor of these $C_k$ models in
terms of real coordinates and then reexpressing them in terms of the complex coordinates
of the $W_k$ model we find the following result,
\be
F= A^2 \left(|u|^2 + |w|^2 \right) - B^2 |u|^2 - C^2 {|u|^2 \over |u|^2 + |w|^2}
+ D^2 |v|^2 + E^2 |z|^2 ~.
\ee
In order to examine the existence of solutions of type
IIB supergravity for the $W_k$ model in transverse space, we apply the general
formula \eqn{wormy} for this particular choice of front factor, and find
\be
\nabla_i \left(e^{-2 \Phi} \nabla^i F \right) = (2A^2 - B^2)
\left( |u|^2 + |w|^2 \right)^2 + 2 C^2 |u|^2 ~.
\ee
Here, we have also set for convenience $C^2 = 2D^2 = 2E^2$, as in the corresponding
$C_k$ models with $2A^2 \geq B^2$.

We observe that a 4-form $\varphi$ exists for $C=0$, leading to a consistent type IIB
background for $W_k$ with quadratic front factor, as
before. For $C \neq 0$, however, there is no
consistent solution of type IIB theory with an R-R 4-form $\varphi$;
this is not surprising as T-duality is
not always meant to act consistently within type IIB theory,
but rather relate IIA to IIB.
In conclusion, only in certain cases the different classes of pp-wave solutions of the
$C_k$ and $W_k$ models can map to each other by T-duality.

\subsection{Solutions for the fat throat model}

In this case we find again for any front factor that
\be
\nabla_i \left(e^{-2 \Phi} \nabla^i F \right) = h^{u\bar{u}} h^{w\bar{w}}
\left( \partial_u \partial_{\bar{u}} F + \partial_{w} \partial_{\bar{w}} F\right)
+e^{-2\Phi}\left( \partial_v \partial_{\bar{v}} F +
\partial_z \partial_{\bar{z}} F \right)
\ee
taking into account the
corresponding expressions for the metric and dilaton fields, which also satisfy
$h^{u\bar{u}} = h^{w\bar{w}} = {\rm exp}(-2\Phi)$. We can repeat the previous analysis
and find solutions of the second order equation \eqn{main} for some simple, e.g., quadratic
choices of the front factor. However, since in the next section we will describe the
general class of supersymmetric pp-wave solutions for all axionic instanton backgrounds,
including the fat throat model, we will confine our analysis here
only to the construction of those solutions which can be obtained by T-duality from the
special solutions that are already available for the $\Delta_k$ model.
We will find once more that T-duality cannot
always act as a solution generating symmetry within type IIB theory alone.

Let us start by considering front factors of the $\Delta_k$ model written in real
coordinates as $F= -A^2 {\rm sin}^2 \theta + B^2 {\rm sinh}^2 \rho + C^2|v|^2 + D^2|z|^2$.
T-duality with respect to $\partial_{\phi}$ does not affect $F$ as it stands,
but in the fat throat
model that results by duality
the functional dependence of $F$ on the new complex coordinates
will be different. Using the appropriate change of coordinates, we find the following
expression for the fat throat model:
\ba
F & = & {1 \over 2r_0^2} (A^2 + B^2) \left(|u|^2 + |w|^2 - r_0^2 \right) + C^2|v|^2
+ D^2|z|^2 \nonumber\\
&-& {1 \over 2r_0^2} (A^2 - B^2) \sqrt{\left(|u|^2 + |w|^2 + r_0^2\right)^2 - 4r_0^2
|w|^2} ~.
\ea
Therefore, to determine $\varphi$, we first compute in this case
\be
\nabla_i\left(e^{-2\Phi} \nabla^i F\right) = h^{u\bar{u}} h^{w\bar{w}} {1 \over r_0^2}
(A^2 + B^2) + e^{-2\Phi} \left(C^2 + D^2 + {1 \over r_0^2} (B^2 - A^2) \left(|u|^2
+ |w|^2 \right) \right) .
\ee
We see that only if $A^2=B^2$ and $C^2 + D^2 = 0$, we will have a consistent solution for
the corresponding $\varphi$ in the fat throat model in terms of $(1,3)$ forms
and their complex conjugates. Otherwise, there will be no type
IIB background with a fat throat placed in its transverse space, which is T-dual to
$\Delta_k$ models for this particular class of front factors.

Conversely, as soon as we can find an entire class of supersymmetric solutions for
the fat throat model, as for all other axionic instantons, we may try to build consistent
type IIB backgrounds for all other spaces that can be obtained from it by T-duality
(and in some cases supplemented by a contraction with respect to $r_0$), provided that
its action is consistently implemented within the type IIB framework. This part
will not be analyzed here, but certainly it has the potential to produce some
interesting models, which will be very different in nature
from the complex sine-Gordon model.
We also leave this construction to future work.

\section{Solutions with space-time supersymmetry}
\setcounter{equation}{0}

In this section we focus only on axionic instanton solutions as building
blocks of the transverse space of gravitational backgrounds of pp-wave type.
First, we show that the second order equation \eqn{main}
for the front factor $F$ and the
anti-self-dual 4-form $\varphi$ can be easily mapped to the equation that
defines ordinary pp-wave solutions in flat transverse space by a conformal
rotation of the target space metric; in this case, we consider backgrounds
with front factor (and hence 4-form) that have support only on the four
transverse space coordinates provided by any axionic instanton solution.
This will also motivate the study of the Killing spinor equations that we
undertake in the sequel to characterize all possible supersymmetric solutions
by solving the Killing spinor equations. It will turn out that the supersymmetric
solutions can be obtained systematically from the class of solutions that
have already been described by Maldacena and Maoz in flat transverse space
with no dilaton nor anti-symmetric tensor fields, provided that one takes
into account the appropriate chiral projections imposed by the axionic instanton
condition.
The explicit construction of the supersymmetric solutions requires a number of
steps, which will be outlined in the following subsections in all detail.

\subsection{General structure of the solutions}

Consider gravitational
pp-wave backgrounds, where the front factor $F$ as well as the 4-form
$\varphi$ have support on the transverse space coordinates, and further
consider that $\varphi_{mn}$, $\varphi_{m\bar{n}}$ (and their complex conjugates) have a
block-diagonal form, using the short-hand notation,
\be
\varphi_{mn} = \left(\begin{array}{cccc}
\star & \star &  &  \\
\star & \star &  &  \\
  &  & \star & \star \\
  &  & \star & \star \\
\end{array} \right) , ~~~~~
\varphi_{m\bar{n}} = \left(\begin{array}{cccc}
\star & \star &  &  \\
\star & \star &  &  \\
  &  & \star & \star \\
  &  & \star & \star \\
\end{array} \right) ,
\ee
thus setting all components in the off-diagonal blocks equal to zero.
Actually, at it has already been pointed out in subsection 3.2, this is
in accordance with the $2+2$ decomposition of the four
space-time indices of
$\varphi$, in the sense that the two first indices of the allowed $(1,3)$ forms
$\varphi_{m \bar{n} \bar{p} \bar{q}}$, as well as of the $(2,2)$ forms
$\varphi_{m \bar{n} p \bar{p}}$, take values in the axionic instanton
block and the remaining two in the other. Notice that only in this case, {\it the
anti-self-duality condition for the R-R 4-form, $\varphi = -^{\star} \varphi$,
remains invariant with respect to conformal rescaling of each four-dimensional
building block of the transverse space separately},
and hence the same components of $\varphi$ can be used
in any axionic instanton block as in flat space. Furthermore, the closure of such
4-forms provides the same conditions as in flat space because $\varphi$ is totally
anti-symmetric and covariant derivatives translate into ordinary derivatives.
Thus, it is conceivable that flat space solutions can be lifted to the axionic
instanton block by conformal rotation, which, however, may yield pp-waves with
different front factor $F$, as the second order equation \eqn{main}
that relates
$F$ to the norm of of the R-R form, i.e., $|\varphi|^2$, is metric dependent.

We can be more explicit by looking at the general structure of the main equation
\eqn{main} in any four-dimensional axionic instanton block.
Using the
complex coordinates $u$ and $w$ (and their
conjugates) to parameterize a general axionic instanton background with
conformally flat metric, and $v$, $z$ (and their conjugates) to parameterize the
remaining  flat directions,
we find
\be
\nabla_i\left(e^{-2 \Phi} \nabla^i F \right) = e^{-4\Phi} (\partial_u \partial_{\bar{u}}
F + \partial_w \partial_{\bar{w}} F)
+ e^{-2\Phi} (\partial_v \partial_{\bar{v}}
F + \partial_z \partial_{\bar{z}} F) ~,\label{alter}
\label{flate}
\ee
where
\be
e^{-2 \Phi} = h^{u\bar{u}} = h^{w\bar{w}} ~, ~~~~ {\rm with} ~~
(\partial_u \partial_{\bar{u}} + \partial_w \partial_{\bar{w}} )e^{2\Phi} = 0 ~.
\ee
Then, it is clear that
pp-wave solutions can be easily constructed from all known pp-wave
solutions in flat transverse space by making the appropriate choices that
respect the anti-self-duality of $\varphi$, as well as its closure and co-closure
in the new gravitational backgrounds.
Note for this purpose that the
norm of the 4-form, $|\varphi|^2$, which
involves contractions of the transverse space indices, using the components of the
inverse metric $h^{u \bar{u}}$ and $h^{w \bar{w}}$, will then naturally be
related to the flat space norm by the conformal factor, since
\be
|\varphi|^2 = e^{-4\Phi} \sum_{u, w} \left(|\varphi_{mn}|^2 + |\varphi_{m \bar{n}}|^2
\right) +
\sum_{v, z} \left(|\varphi_{mn}|^2 + |\varphi_{m \bar{n}}|^2
\right) ,
\ee
which has to be compared with equation \eqn{flate} above. Thus, solutions in an
axionic instanton background can be easily mapped into flat space solutions by
conformal transformations, provided that one considers front factors without
$v$ or $z$ dependence, i.e., $F(u,w, \bar{u}, \bar{w})$, and likewise for
$\varphi$, with $\varphi_{mn} (u, w, \bar{u}, \bar{w})$ and
$\varphi_{m \bar{n}} (u, w, \bar{u}, \bar{w})$ satisfying the additional requirement
that the indices $m$, $n$ take values only in the first block, thus being zero
otherwise. Clearly, the latter requirement
eliminates the second bunch of terms in $|\varphi|^2$
that involve summation over $v$ and $z$, as they do not depend on the conformal factor
in the right way. Finally, we note for completeness that adding any real harmonic
function of $v$, $z$ (and their complex conjugates) to
the front factor $F$ can always yield more general
pp-wave solutions; these generalizations will not be of interest to us,
however,  as
they do not affect the R-R fields, which are inert to all such
modifications.

This general observation, which applies to any solution of the second order equation
\eqn{main}, clearly also includes the complete list of supersymmetric solutions
that have been constructed by Maldacena and Maoz in flat transverse space with the
additional restriction that all data is appropriately confined in the axionic
instanton block. Thus,
one may expect that the solutions of the Killing spinor equations for an
axionic instanton background in the presence of a R-R 5-form will also reduce to
the corresponding supersymmetric solutions in flat transverse space, leading to
new classes of space-time supersymmetric backgrounds. As we will see in the next
subsections, which only concern
the analysis of the supersymmetric configurations, this proceeds technically by making
use of the axionic instanton condition to simplify the Killing
spinor equations on pp-wave background and reduce them to
a much simpler set of flat space conditions.
However, due to the chiral properties of the Killing spinors imposed by the axionic
instanton condition, only flat space solutions with $(1,1)$ space-time
supersymmetry will lead to supersymmetric
pp-wave solutions for any axionic
instanton block living in the transverse space. Thus, solutions with more space-time
supersymmetry in flat space will only be solutions of the second order equations in the
present case.

\subsection{Supersymmetry variations}

We start by writing the
supersymmetry transformations of the type IIB dilatino and gravitino fields,
which are usually given in
the Einstein frame by the equations, \cite{gsw}, \cite{schw},
\ba
\delta \lambda^{\prime} &=& i \Gamma^{\prime \mu} {\cal P}_{\mu} \epsilon^{\prime \star}
- {i \over 24} \Gamma^{\prime \kappa \lambda \nu} {\cal G}_{\kappa \lambda \nu}
\epsilon^{\prime} ~, \nonumber\\
\delta \Psi_{\mu}^{\prime} & = & D_{\mu} \epsilon^{\prime} + {1 \over 96}
\left({\Gamma_{\mu}^{\prime}}^{\kappa \lambda \nu} {\cal G}_{\kappa \lambda \nu}
-9 \Gamma^{\prime \lambda \nu} {\cal G}_{\mu \lambda \nu} \right) \epsilon^{\prime \star}
\nonumber\\
& + & {i \over 16 \cdot 5!} \Gamma^{\prime \kappa \lambda \nu \rho \sigma}
\Gamma_{\mu}^{\prime} {\cal F}_{\kappa \lambda \nu \rho \sigma} \epsilon^{\prime}
\ea
omiting the fermionic terms, which are all set equal to zero for
bosonic solutions.
Here, the prime refers to the Einstein frame variables, the star denotes complex
conjugation, and the various quantities that
appear in the variations above are defined to be
\ba
D_{\mu} \epsilon^{\prime} &=& \left(\partial_{\mu} + {1 \over 4}
{\omega_{\mu}^{\prime}}^{\nu \rho} \Gamma_{\nu \rho} - {i \over 2} {\cal Q}_{\mu}
\right) \epsilon^{\prime} ~, \nonumber\\
{\cal Q}_{\mu} &=& = -i \epsilon_{\alpha \beta} V_{-}^{\alpha} \partial_{\mu}
V_{+}^{\beta} ~, \nonumber\\
{\cal P}_{\mu} &=& -\epsilon_{\alpha \beta} V_{+}^{\alpha} \partial_{\mu} V_{+}^{\beta}
~, \nonumber\\
{\cal G}_{\kappa \lambda \nu} &=& -\epsilon_{\alpha \beta} V_{+}^{\alpha}
F_{\kappa \lambda \nu}^{\beta} ~,
\ea
where $\alpha$ and $\beta$ take the values 1 and 2, with $\epsilon_{12} = 1$, and
$F_{\kappa \lambda \nu}^1 = (F_{\kappa \lambda \nu}^2)^{\star}$.
These equations are written down using the $SU(1,1)$ formulation of the
theory, which relates $V_{\pm}^{\alpha}$ to the dilaton $\Phi$ and the R-R scalar
$C^{(0)}$ by the equation
\be
\left(\begin{array}{ccc}
1 &  & 1 \\
  &  &   \\
-i&  & i \\
\end{array} \right)
\left(\begin{array}{ccc}
V_-^1 &  & V_+^1 \\
      &  &       \\
V_-^2 &  & V_+^2 \\

\end{array} \right)
= e^{\Phi/2}
\left(\begin{array}{ccc}
-\bar{\tau} &  & -\tau \\
            &  &       \\
1 &  & 1 \\
\end{array} \right)
\ee
where $\tau = C^{(0)} + i {\rm exp}\Phi$. Furthermore, in the most general situation,
the field strengths of the NS-NS and R-R 2-forms $B$ and $C^{(2)}$ respectively are
related to $F_{\kappa \lambda \nu}^{\alpha}$ as follows,
\be
\left(\begin{array}{c}
-dC^{(2)} \\
     \\
dB \\
\end{array} \right) =
\left(\begin{array}{c}
{\rm Re} F^1 \\
             \\
{\rm Im} F^1 \\
\end{array} \right) = {1 \over 2}
\left(\begin{array}{ccc}
1 &  & 1 \\
  &  &   \\
-i&  & i \\
\end{array} \right)
\left(\begin{array}{c}
F^1 \\
    \\
F^2 \\
\end{array} \right) ~.
\ee

In this paper we focus on configurations with only a R-R 5-form turned on, setting the
potentials of the remaining R-R fields equal to zero, i.e., $C^{(0)} = 0 = C^{(2)}$. In this case,
we have the identifications
\ba
V_+^1 & = & i {\rm sinh} {1 \over 2} \Phi ~, ~~~~ V_+^2 = -i {\rm cosh} {1 \over 2} \Phi ~,
\nonumber\\
V_-^1 & = & i {\rm cosh} {1 \over 2} \Phi ~, ~~~~ V_-^2 = -i {\rm sinh} {1 \over 2} \Phi ~.
\ea
As a result, we obtain immediately that
\ba
{\cal Q}_\mu & = & 0 ~, ~~~~ {\cal P}_\mu = {1 \over 2} \partial_\mu \Phi ~, \nonumber\\
{\cal G}_{\kappa \lambda \nu} &=& e^{-\Phi/2} H_{\kappa \lambda \nu}
\ea
and the corresponding supersymmetry variations simplify considerably.

We will present the first order equations that arise from the vanishing condition of the
dilatino and gravitino equations in the {\em sigma} {\em model}
{\em frame}, which is more appropriate for describing our solutions. Since the two frames
are related to each other by $g_{\mu \nu} = {\rm exp} (\Phi/2) g_{\mu \nu}^{\prime}$,
and hence
$g^{\mu \nu} = {\rm exp}(-\Phi/2) g^{\prime \mu \nu}$,
we have the relation among the gamma matrices
$\Gamma^\mu = {\rm exp} (-\Phi/4) \Gamma^{\prime \mu}$, so that
$\{\Gamma^{\mu} , \Gamma^{\nu} \} = 2 g^{\mu \nu}$ in the sigma model frame. As for the
spinors appearing in the supersymmetry transformations, we introduce the
following redefinitions
(see, for instance, \cite{hassan} for a recent discussion of this and related issues):
\ba
\epsilon & = & e^{\Phi/8} \epsilon^{\prime} ~, ~~~~ \lambda = e^{-\Phi/8} \lambda^{\prime} ~,
\nonumber\\
\Psi_\mu &=& e^{\Phi/8} \left(\Psi_{\mu}^{\prime} + {i \over 4} \Gamma_{\mu}^{\prime}
\lambda^{\prime \star} \right) .
\ea
Using the previous field identifications, it is straightforward to find the following equation
arising from the dilatino variation in the sigma model frame,
\be
\delta \lambda = 0 = {i \over 2} \Gamma^{\mu} (\partial_\mu \Phi) \epsilon^{\star} -
{i \over 24} \Gamma^{\kappa \lambda \nu} H_{\kappa \lambda \nu} \epsilon ~.
\ee
As for the gravitino variation, it yields the following equation in the
sigma model frame,
\be
\delta \Psi_\mu = 0 = \left(\pa_\mu +{1\over 4}{\omega_\mu}^{\nu\rho}
\G_{\nu\rho}
+{i\over 16} e^{\Phi} \scF ~ \G_\mu\right)\e - {1 \over 8} \Gamma^{\lambda \nu}
H_{\mu \lambda \nu} \epsilon^{\star} ~,
\ee
where we set $\scF = \Gamma^{\kappa \lambda \nu \rho \sigma}
{\cal F}_{\kappa \lambda \nu \rho \sigma}/5!$.
This result is simply derived using the relation between the
components of the spin connection in the two different frames,
${\omega_{\mu}^{\prime}}^{ab} \Gamma_{ab} = {\omega_{\mu}}^{ab}
\Gamma_{ab} - (\partial^{\nu} \Phi) (\Gamma_\mu \Gamma_\nu - g_{\mu \nu})/2$, as
well as the identity ${\Gamma_{\mu}}^{\kappa \lambda \nu} H_{\kappa \lambda \nu}
= \Gamma_\mu \Gamma^{\kappa \lambda \nu} H_{\kappa \lambda \nu} -
3 \Gamma^{\lambda \nu} H_{\mu \lambda \nu}$.

The first order equations above describe the supersymmetric solutions
of type IIB supergravity in
the presence of a non-trivial R-R
5-form ${\cal F}$; as such, they yield solutions of the second order field equations
in the sigma model frame by squaring their action.
Note that for axionic instanton backgrounds, the dilatino
condition $\delta \lambda = 0$ is not modified by R-R fields, and hence
it remains the same in our general class of pp-wave string backgrounds as in the
usual description of ordinary type IIB five-brane backgrounds \cite{cal}.
Hence, the only non-trivial task here is
to solve the Killing spinor equations provided by the vanishing condition of the
gravitino field equations, $\delta \Psi_{\mu} = 0$, in the presence of the R-R form
${\cal F}$, by constructing explicit supersymmetric solutions that generalize the
usual five-brane geometries.

It is important to note the
simultaneous appearance of the Killing spinor $\epsilon$ and its complex
conjugate $\epsilon^{\star}$ in the presence of a non-trivial NS-NS antisymmetric tensor
field, which in turn
make the Killing spinor equations of type IIB supergravity look different
than the corresponding equations of type IIA or heterotic theory. Of course, this is not
surprising given the $(2, 0)$ chiral nature of type IIB theory in ten dimensions, but
it has the important consequence that in five-brane solutions (and their generalizations
we are seeking here) one arives at a non-chiral world-brane theory; we will say more
about this later. At this point, we only rewrite the Killing spinor equations using the
decomposition of the complex Weyl spinor in terms of two real Majorana-Weyl spinors
$\epsilon^{(\alpha)}$, as $\epsilon = \epsilon^{(1)} + i \epsilon^{(2)}$, which will be
useful in the sequel. In particular, we obtain from the dilatino condition
\be
\left(\Gamma^{\mu} \partial_{\mu} \Phi -
{1\over 12} H_{\mu \nu \rho} \Gamma^{\mu \nu \rho} \right)\epsilon^{(1)} - i
\left(\Gamma^{\mu} \partial_{\mu} \Phi +
{1\over 12} H_{\mu \nu \rho} \Gamma^{\mu \nu \rho} \right)\epsilon^{(2)} = 0 ~,
\label{dilaeq}
\ee
whereas from the gravitino condition we have
\ba
& & \left(\pa_\mu +{1\over 4}({\omega_\mu}^{\nu\rho} -
{1\over 2} {H_\mu}^{\nu\rho})\G_{\nu\rho} \right) \epsilon^{(1)}
-{1\over 16} e^{\Phi} \scF ~ \G_\mu \epsilon^{(2)} +  \nonumber\\
& & i\left(\pa_\mu +{1\over 4}({\omega_\mu}^{\nu\rho}+
{1\over 2} {H_\mu}^{\nu\rho})\G_{\nu\rho} \right) \epsilon^{(2)}
+{i\over 16} e^{\Phi} \scF ~ \G_\mu \epsilon^{(1)} = 0
\label{gravieq}
\ea
in terms of Majorana-Weyl spinors. This decomposition is not necessary in the original
work of Maldacena and Maoz, as they only consider solutions with vanishing torsion, in
which case the Killing spinor equation involves only $\epsilon$ and not its complex
conjugate $\epsilon^{\star}$.

\subsection{Chiralities and projections}

Before we proceed further, let us briefly discuss the chiral properties of the
Killing spinor $\epsilon$, and its real components, with respect to the chirality
operator of the eight-dimensional transverse space, as well as its individual
four-dimensional building blocks, by recalling some basic properties of the spinorial
representations that are involved.
The supersymmetries of type IIB supergravity are generated by a
complex chiral spinor $\epsilon$
with sixteen components, which is taken here to have positive chirality with respect to
$\Gamma_{11}$, defined as,
\be
\Gamma_{11} = -\Gamma^{0}\Gamma^{1 \cdots 8} \Gamma^{9} \equiv
[\Gamma^+ , \Gamma^-] \Gamma^{1 \cdots 8} ~,
\ee
and which may also be written down using the light-cone $\Gamma$-matrices,
\be
\Gamma^{\pm} = {1 \over 2} \left(\Gamma^0 \pm \Gamma^9 \right) ; ~~~~
\{\Gamma^+ , \Gamma^- \} = 2g^{+-} \equiv -1 ~.
\ee
It is convenient to decompose the complex chiral spinor $\epsilon$, with
$\Gamma_{11} \epsilon = + \epsilon$, into two components
$\epsilon = \epsilon_+ + \epsilon_-$, which are defined by
\be
\epsilon_+ = - \Gamma^- \Gamma^+ \epsilon ~, ~~~~~
\epsilon_- = - \Gamma^+ \Gamma^- \epsilon ~.
\ee
Then, since $[\Gamma^+ , \Gamma^-] \epsilon_{\pm} = \pm \epsilon_{\pm}$, it follows that
$\epsilon_+$ has positive $SO(1,1)$ chirality and hence it also has positive $SO(8)$
chirality with respect to the transverse space operator $\Gamma^{1 \cdots 8}$;
likewise, $\epsilon_-$ has negative $SO(1,1)$ and $SO(8)$ chiralities. Put it
differently, using the embeding $SO(9,1) \supset SO(1,1) \times SO(8)$, we are
decomposing the sixteen component complex chiral spinor $\epsilon$ as
\be
16 \rightarrow (1_+ , 8_+) \oplus (1_- , 8_-) ~.
\ee
As the transverse space is built from two block-diagonal four dimensional superconformal
theories,
it is also useful to decompose the resulting
complex spinors further by considering the embedding
$SO(8) \supset SO(4) \times SO(4)$. Then, using the chirality operators of each separate block,
say $\Gamma^{1 \cdots 4}$ and $\Gamma^{5 \cdots 8}$, which are parameterized by the
corresponding transverse space coordinates, we arrive at a more refined decomposition
of the complex chiral spinors given by
\be
8_+ \rightarrow (2_+ , 2_+) \oplus (2_- , 2_-) ~, ~~~~
8_- \rightarrow (2_+ , 2_-) \oplus (2_- , 2_+) ~.
\ee

On the other hand, we may decompose the complex Weyl spinors into Majorana-Weyl spinors by
introducing real and imaginary parts for $\epsilon_{\pm}$, and hence $\epsilon$, as
follows:
\be
\epsilon_{\pm} = \epsilon_{\pm}^{(1)} + i \epsilon_{\pm}^{(2)} ~.
\ee
>From now on, we will use the notation $\epsilon_+^{(\alpha) ++}$ to
denote the Majorana-Weyl spinors
with positive chiralities with respect to both $SO(4)$ factors, and $\epsilon_+^{(\alpha)--}$
when both $SO(4)$ chiralities are negative, for each $\alpha = 1, 2$. Likewise,
$\epsilon_-^{(\alpha)+-}$ and $\epsilon_-^{(\alpha)-+}$ will denote the real components of the
$SO(8)$ Weyl spinor with opposite $SO(4) \times SO(4)$ chiralities, as indicated by the
signs in superscript.
In the presence of non-trivial dilaton and anti-symmetric tensor fields, one has to
impose the
condition $\delta \lambda = 0$ on all supersymmetric solutions,
which, as we will see next, restricts the allowed choices of $\epsilon_{\pm}$
with respect to the
$SO(4) \times SO(4)$ decomposition. In fact, the axionic instanton condition will naturally
select opposite $SO(4) \times SO(4)$ chiralities for the two real components
$\epsilon_+^{(1)}$ and $\epsilon_+^{(2)}$ of $\epsilon_+$, and likewise for the two real
components of $\epsilon_-$. This chirality flip has the same origin as in the standard
construction of type IIB five-brane solutions via the axionic instanton embeding, which in
turn leads to a non-chiral world-brane theory. This is an important fact that has to be taken
into account while solving the Killing spinor equations $\delta \Psi_\mu = 0$ for our class
of generalized pp-wave solutions. Actually, the chiral decomposition of the Killing spinor
equations with respect to $SO(4) \times SO(4)$, as well as their Majorana decomposition into
real and imaginary components will be instrumental for the construction of explicit solutions
by making appropriate use of the Fock space representation that was initially introduced by
Maldacena and Maoz for the systematic description of all
supersymmetric pp-wave solutions with flat
transverse space, \cite{malda2}.
The precise details of the construction of supersymmetric pp-wave solutions with an
axionic instanton block sitting in the transverse space involves a number of steps, however,
which we will present in detail; only at the very end we will combine them all together
in order to present the desirable result in closed form.

Recall at this point the precise way that the dilatino
equation is solved for a four dimensional axionic instanton background, \cite{cal}. Since
\be
{1 \over 2} H_{ijk} = \pm {\epsilon_{ijk}}^l \partial_l \Phi ~,
\ee
where, here, for definiteness, the sign $\pm$ refers to the axionic instanton or
anti-instaton solution respectively, and all indices take the values $1, \cdots, 4$
in the first superconformal block, we have
\be
\Gamma^i \partial_i \Phi - {1 \over 12} H_{ijk} \Gamma^{ijk} =
\Gamma^i (\partial_i \Phi) (1 \pm \Gamma^{1234}) ~.
\ee
This follows easily from the algebra of Dirac $\Gamma$-matrices and it implies that
\be
\left(\Gamma^i \partial_i \Phi - {1 \over 12} H_{ijk} \Gamma^{ijk} \right) \epsilon =
2 \Gamma^i (\partial_i \Phi) P_{\pm}^{(I)} \epsilon ~,
\ee
and likewise
\be
\left(\Gamma^i \partial_i \Phi + {1 \over 12} H_{ijk} \Gamma^{ijk} \right) \epsilon =
2 \Gamma^i (\partial_i \Phi) P_{\mp}^{(I)} \epsilon ~,
\ee
for all spinors $\epsilon$,
where $P_{\pm}^{(I)}$ is the $SO(4)$ chirality projection operator of the (first)
four dimensional building block provided by any axionic instanton or anti-instanton
background respectively,
\be
P_{\pm}^{(I)} = {1 \over 2} (1 \pm \Gamma^{1234}) ~.
\ee
Consequently, the dilatino equation of type IIB supergravity will be satisfied if the
Majorana component $\epsilon^{(1)}$ of the spinor $\epsilon$
has negative (respectively positive) $SO(4)$ chirality, whereas the remaining
$SO(4)$ chiral component
of the spinor is zero; likewise, the other Majorana component of the spinor, $\epsilon^{(2)}$,
must be restricted to have positive (respectively negative) $SO(4)$ chirality, depending on
the axionic instanton or anti-instanton condition.

Applying this observation in the eight dimensional context, where
the second superconformal block is taken to be the four dimensional flat space with no dilaton
and anti-symmetric tensor fields, we conclude immediately that the two $SO(8)$
chiral spinors $\epsilon_{\pm}$ should be restricted to the following $SO(4) \times SO(4)$
chiral forms in terms of their Majorana components:
\ba
\epsilon_+ & = & \epsilon_+^{(1)--} + i \epsilon_+^{(2)++} ~, ~~~~ \epsilon_- =
\epsilon_-^{(1)-+} + i \epsilon_-^{(2)+-} ~~~~
{\rm axionic} ~ {\rm instantons} , \nonumber\\
\epsilon_+ & = & \epsilon_+^{(1)++} + i \epsilon_+^{(2)--} ~, ~~~~ \epsilon_- =
\epsilon_-^{(1)+-} + i \epsilon_-^{(2)-+}
~~~~
{\rm axionic} ~ {\rm anti-instantons} ,
\ea
which, thus, also affect the $SO(4)$ chirality of the spinors in the second superconformal block.

\subsection{The Killing spinor equations}

With these restrictions in mind, we may proceed to study the Killing spinor equations
arising from the gravitino equation. First, it is necessary to compute the components
of the generalized spin connection that also includes the contribution from the
torsion term $H$. Introducing vierbeins, as usual, we have for the two superconformal
building block of the transverse space, which are conformally flat and flat
respectively, that
\be
{e^{\hat{i}}}_j = e^{\Phi} \delta_j^{\hat{i}} ~, ~~~~~
{e^{\hat{i}^{\prime}}}_{j^{\prime}} = \delta_{j^{\prime}}^{\hat{i}^{\prime}} ~,
\ee
where $i$, $j$ run from $1$ to $4$ and $i^{\prime}$, $j^{\prime}$ run from $5$ to $8$,
whereas hats denote the tangent space indices of the corresponding vierbeins.
Then, it is straightforward to compute the components of the spin connection in the
transverse space, which turn out to be
\be
{\omega_k}^{\hat{i} \hat{j}} = \left(\delta_k^{\hat{i}} \delta^{\hat{j} l}
- \delta_k^{\hat{j}} \delta^{\hat{i} l} \right) \partial_l \Phi ~, ~~~~
{\omega_{k^{\prime}}}^{\hat{i}^{\prime} \hat{j}^{\prime}} = 0
\ee
and obviously all cross terms among the two blocks vanish. It is then easy to see
for the first superconformal block that
\be
\left({\omega_i}^{\hat{j} \hat{k}} - {1 \over 2} {H_i}^{\hat{j} \hat{k}} \right)
\Gamma_{\hat{j} \hat{k}} = \left(2{\Gamma_i}^l \mp {\epsilon_i}^{\hat{j} \hat{k} l}
\Gamma_{\hat{j} \hat{k}} \right) \partial_l \Phi ~,
\ee
by computing the contribution of the non-trivial torsion term in the axionic instanton or
anti-instanton background respectively. Using the algebra of
Dirac $\Gamma$-matrices, we also have the identity
\be
{\epsilon_i}^{\hat{j} \hat{k} l} \Gamma_{\hat{j} \hat{k}} = -2 {\Gamma_i}^l \Gamma^{1234} ~,
\ee
and as a result we find
\be
{1 \over 4} \left({\omega_i}^{jk} - {1 \over2} {H_i}^{jk} \right) \Gamma_{jk} =
{\Gamma_i}^l (\partial_l \Phi) P_{\pm}^{(I)} ~,
\ee
in terms of the $SO(4)$ chirality projection operators of the first building block.
Likewise, we find
\be
{1 \over 4} \left({\omega_i}^{jk} + {1 \over2} {H_i}^{jk} \right) \Gamma_{jk} =
{\Gamma_i}^l (\partial_l \Phi) P_{\mp}^{(I)} ~.
\ee
Obviously, the components of the (generalized) spin connection
are all zero for the second block.

We also include for
completeness the light-cone zwiebeins,
\be
{e^{\hat{+}}}_+ = 2 ~; ~~~~ {e^{\hat{-}}}_- = 2 ~, ~~~~
{e^{\hat{-}}}_+ = {1 \over 2} F ~,
\ee
for which the only non-vanishing components of the spin connection are
\be
{{\omega_+}^{\hat{-}}}_{\hat{i}} = -{1 \over 2} (\partial_j F) {E_{\hat{i}}}^j ~, ~~~~
{{\omega_+}^{\hat{-}}}_{\hat{i}^{\prime}} = -{1 \over 2} (\partial_{j^{\prime}} F)
{E_{\hat{i}^{\prime}}}^{j^{\prime}}
\ee
given in terms of the inverse vierbeins ${E_{\hat{i}}}^j$ and
${E_{\hat{i}^{\prime}}}^{j^{\prime}}$.
Thus, in our normalization of the ten dimensional metric, we have
\be
ds^2 = -e^{\hat{+}} e^{\hat{-}} + \sum_{i=1}^4 e^{\hat{i}} e^{\hat{i}}
+ \sum_{i^{\prime} = 5}^8 e^{\hat{i}^{\prime}} e^{\hat{i}^{\prime}} ~.
\ee

The Killing spinor equations take a very simple form for $\mu = -$, due to
$(\Gamma^+)^2 = g^{++} =0$, which in turn implies that
$\scF ~\Gamma_- = \Gamma^+ \sv ~\Gamma_- = 0$.
Then, using the $SO(8)$ chiral components, we obtain the equations
\be
\partial_- \epsilon_+ = 0 = \partial_- \epsilon_- ~.
\ee
These equations are written using the complex chiral spinors, but they can also be
written using the Majorana-Weyl components that are constrained to have definite
$SO(4) \times SO(4)$ chiralities imposed by the axionic instanton conditions.

For $\mu = +$, taking into account the non-vanishing components of the spin connection with
light-cone indices, we have
\be
\left(\partial_+ - {1 \over 8} \Gamma_{\hat{-}} \sp F +
{i \over 16} e^{\Phi} \Gamma^+ \sv ~ \Gamma_+ \right) \epsilon = 0 ~,
\ee
where both $\sp F$ and $\sv$ involve summation over all eight coordinates in transverse space.
Since the field strength of the NS-NS anti-symmetric tensor field has no components in the
$+$ direction, this Killing spinor equation involves only the complex spinor $\epsilon$ and
not its complex conjugate.
Simple algebra of the $\Gamma$-matrices implies that for anti-self-dual 4-forms $\varphi$
one has the
identity $2 \sv = \sv ~ (1- \Gamma^{1 \cdots 8})$, and therefore we obtain $\sv ~ \epsilon =
\sv ~\epsilon_-$ with zero contribution from the positive $SO(8)$ chiral component; then, we have
\be
\Gamma^+ \sv ~ \Gamma_+ \epsilon = -2 \Gamma^+ \Gamma^- \sv ~ \epsilon = -[\Gamma^+ , \Gamma^-]
\sv ~ \epsilon + \sv ~ \epsilon = 2 \sv ~ \epsilon_- ~,
\ee
as $[\Gamma^+ , \Gamma^-] \epsilon_- = - \epsilon_-$, due to its negative light-cone chirality.
On the other hand, we have $\Gamma_{\hat{-}} \epsilon_- = 0$, as
$\epsilon_- = -\Gamma^+ \Gamma^- \epsilon$ and $(\Gamma^+)^2 = 0$. Taking all these
facts into account,
we arrive at the equations
\be
\left(i\partial_+ - {1 \over 8}
e^{\Phi} \sv \right) \epsilon_- = {i \over 8} \Gamma_{\hat{-}} (\sp F)
\epsilon_+ ~, ~~~~
\partial_+ \epsilon_+ = 0
\ee
by comparing terms of the same $SO(8)$ net chirality\footnote{For this, note that the term
$\Gamma_{\hat{-}} (\sp F) \epsilon_+$ has negative $SO(1,1)$ chirality, due to the identity
$[\Gamma^+ , \Gamma^-] \Gamma_{\hat{-}} = - \Gamma_{\hat{-}}$; therefore, $\Gamma_{\hat{-}}$
acting on any state yields a state with negative $SO(1,1)$ chirality, and hence, in our case,
with negative $SO(8)$ chirality.}. These equations will be studied later in more detail by
comparing terms of the same $SO(4) \times SO(4)$ chirality, as well as by comparing their real
and imaginary parts when the Majorana components of $\epsilon_{\pm}$ are introduced; then,
as we will see, some non-trivial constraints will be introduced by incorporating the
axionic instanton condition into them.

For $\mu = i$, where $i$ runs over the indices of the first conformally flat block,
we have, in particular, taking into account the expression for the generalized spin connection
above, that
\be
\left(\partial_i + {\Gamma_i}^l (\partial_l \Phi) P_{\pm}^{(I)} + {i \over 16} e^{\Phi}
\Gamma^+ \sv ~ \Gamma_i \right) \epsilon^{(1)} + i
\left(\partial_i + {\Gamma_i}^l (\partial_l \Phi) P_{\mp}^{(I)} + {i \over 16} e^{\Phi}
\Gamma^+ \sv ~ \Gamma_i \right) \epsilon^{(2)} = 0 ~,
\ee
where the $\pm$ sign depends on the axionic instanton or anti-instanton condition respectively.
Since $\Gamma_- \epsilon_- = 0$, as before, we have $ 2\Gamma^+ \sv ~ \Gamma_i \epsilon =
- \Gamma_- \sv ~\Gamma_i \epsilon_+$ by also using the fact that $2\Gamma^+ = - \Gamma_-$ in
our normalization of the light-cone components of the metric; this term, as before, has
negative $SO(8)$ net chirality. Recall at this point that the Majorana components of
the $SO(8)$ spinors are both
restricted to satisfy the special conditions $P_{\pm}^{(I)} \epsilon^{(1)} = 0$ and
$P_{\mp}^{(I)} \epsilon^{(2)} = 0$ with respect to the $SO(4)$ chirality projection operator
of the axionic instanton (respectively the anti-instanton) block
imposed by the vanishing condition of the
dilatino variation. Thus, the action of the generalized spin connection on those particular
spinors vanishes identically, and as a result we arrive at the simpler equations
\be
\partial_i \epsilon_- = {i \over 32} e^{\Phi} \Gamma_- \sv ~\Gamma_i \epsilon_+ ~, ~~~~
\partial_i \epsilon_+ = 0 ~; ~~~~ i = 1, 2, 3, 4 ~,
\ee
by separating terms of the same $SO(8)$ net chirality.

As for the Killing spinor equations of the second building block, which follow by
setting $\mu = i^{\prime}$, we immediately obtain
\be
\partial_{i^{\prime}} \epsilon_- = {i \over 32} e^{\Phi} \Gamma_- \sv ~
\Gamma_{i^{\prime}} \epsilon_+ ~, ~~~~
\partial_{i^{\prime}} \epsilon_+ = 0 ~; ~~~~ i^{\prime} = 5, 6, 7, 8 ~,
\ee
as the corresponding components of the spin connection are zero from the very beginning
and there are no components of the NS-NS anti-symmetric tensor field in the second block.
These equations, as well as the previous ones arising for $\mu = i$, are conveniently written
here using complex spinors; introducing the Majorana components of $\epsilon_{\pm}$, and
taking into account their individual $SO(4) \times SO(4)$ chiralities,
we will see later that further simplifications occur.

Summarizing the results we have obtained so far, {\it we observe that all components of the
Killing spinor equations are identical to those arising for supersymmetric pp-wave solutions
with flat transverse space without dilaton and anti-symmetric tensor fields, as these
fields
have been conformally rotated away by imposing the axionic instanton condition.}
Actually, the only difference is the appearance
of the dilaton factor ${\rm exp} \Phi$, which can be removed
from the Killing spinor equations by absorbing it completely into the
redefinition of the $\Gamma$-matrices, letting
\be
\tilde{\Gamma}^i = e^{\Phi} \Gamma^i ~, ~~~~~
\tilde{\Gamma}^{i^{\prime}} = \Gamma^{i^{\prime}} ~,
\label{redefga}
\ee
so that $\{\tilde{\Gamma}^i , \tilde{\Gamma}^j \} = 2 \delta^{ij}$ and
$\{\tilde{\Gamma}^{i^{\prime}} , \tilde{\Gamma}^{j^{\prime}} \} =
2 \delta^{{i^{\prime}}{j^{\prime}}}$, as
it should be in flat transverse space.

To prove the consistency of this rescaling, it is first important to note that $\sv$
involves
the product of two $\Gamma$-matrices from each four dimensional block, due to the
block-diagonal nature of its space-time indices; this clearly takes care of the
dilaton factor in the Killing spinor equations associated with the components
$\mu = -$ and $\mu = i$. The Killing spinor equation arising for
$\mu = i^{\prime}$
still appears to exhibit an explicit dependence on the dilaton factor, but
it will be shown in subsection 6.6 below that both sides of this Killing spinor equation
are equal to zero, i.e., $\partial_{i^{\prime}} \epsilon_- = 0 =
({\rm exp}\Phi) \Gamma_- \sv ~ \Gamma_{i^{\prime}} \epsilon_+$, due to some remarkable
identities that also originate from the block-diagonal structure of the 4-form $\varphi$;
therefore, the presence of the dilaton factor in this equation becomes irrelevant.
The proof of these identities requires some background material, however,
which is included in
subsection 6.5 below, before they can be applied to the final construction of
axionic pp-wave solutions with manifest space-time supersymmetry.
Finally, the Killing spinor equation for $\mu = +$ also turns out to be
independent of ${\rm exp} \Phi$ after performing the rescaling \eqn{redefga} into
flat space $\Gamma$-matrices, because $\epsilon_-$ can be taken independent
of the light-cone coordinate $x^+$ and, moreover, $\partial_{i^{\prime}} F$
will be shown to be zero by imposing the appropriate axionic instanton chiral
projection on this equation;
the proof will also be discussed in detail in subsection
6.6 below and it amounts to $\sp F = \Gamma^i \partial_i F$, which only
receives contribution from the first (axionic instanton) block.
Thus, for all practical purposes, the rotation of all Killing spinor equations into flat
pp-wave equations is correct, as advertized, by also taking into account the
appropriate chiral projections.

We may use the
redefinition of the gamma matrices \eqn{redefga}, but for
convenience the tilde will be dropped later
by always refering to the flat space space
equations.
Then, using the explicit solutions of the Killing spinor
equations, as given by Maldacena and Maoz in flat transverse
space, the solutions in an axionic instanton or anti-instanton
background can be subsequently obtained by imposing the
appropriate $SO(4) \times SO(4)$ chiral projections
of the flat $SO(8)$ spinors.
These constraints, which have to be solved simultaneously with the Majorana decomposition
of all Killing spinor equations,
will provide restrictions on the allowed form of the front factor $F$ and
the components of the 4-form $\varphi$, and yield the complete classification of all
supersymmetric solutions from the known flat space construction. This is precisely what we
will describe in the remaining part of this section in detail.

\subsection{Fock space representation}

It is convenient for our purposes to introduce complex coordinates in the transverse
space, say $u$, $w$ in the axionic instanton block and $v$, $z$ in the flat space block,
and define the complexification of the corresponding $\Gamma$-matrices as
\ba
& & \Gamma^{u} = {1 \over \sqrt{2}} (\Gamma^1 + i\Gamma^2) ~, ~~~~~
\Gamma^{\bar{u}} = {1 \over \sqrt{2}} (\Gamma^1 - i\Gamma^2) ~, \nonumber\\
& & \Gamma^{w} = {1 \over \sqrt{2}} (\Gamma^3 + i\Gamma^4) ~, ~~~~~
\Gamma^{\bar{w}} = {1 \over \sqrt{2}} (\Gamma^3 - i\Gamma^4) ~,
\ea
and likewise for $\Gamma^v$ and $\Gamma^z$ (and their complex conjugates) in terms of
the coordinates $x^5$, $x^6$ and $x^7$, $x^8$ respectively. Then, the chirality
operators of each four dimensional block assume the form
\be
\Gamma^{1 \cdots 4} = -{1 \over 4} [\Gamma^u , \Gamma^{\bar{u}}]
[\Gamma^w , \Gamma^{\bar{w}}] ~, ~~~~~
\Gamma^{5 \cdots 8} = -{1 \over 4} [\Gamma^v , \Gamma^{\bar{v}}]
[\Gamma^z , \Gamma^{\bar{z}}] ~.
\ee

Following \cite{malda2}, we may describe all flat
space spinors using the Fock space representation.
In particular, we introduce the vacuum states $|0>_{SO(4)}^{(I), (II)}$
associated with the $SO(4)$ spinors of each
superconformal block, with the properties
\be
\Gamma^{u} |0>_{SO(4)}^{(I)} = 0 = \Gamma^{w} |0>_{SO(4)}^{(I)} ~; ~~~~~
\Gamma^{v} |0>_{SO(4)}^{(II)} = 0 = \Gamma^{z} |0>_{SO(4)}^{(II)} ~,
\ee
and similarly we introduce the vacuum state of $SO(1,1)$ spinors, $|0>_{SO(1,1)}$,
so that
\be
\Gamma^- |0>_{SO(1,1)} = 0 ~.
\ee
Then, the vacuum state of the spinor space in ten dimensional
space-time is represented by the
tensor product
\be
|0> = |0>_{SO(1,1)} \otimes |0>_{SO(4)}^{(I)} \otimes |0>_{SO(4)}^{(II)} ~.
\ee
Other states in the fermionic Fock space are described as excited states by applying
creation operators on the vacuum.
In particular, for the $SO(1,1)$ part, we can have the additional state
\be
|+> = \Gamma^+ |0>_{SO(1,1)} ~,
\ee
which is fully filled as $(\Gamma^+)^2 = 0$. Likewise, for each
$SO(4)$ part, we obtain the additional states
\ba
& & |\bar{u}> = \Gamma^{\bar{u}} |0>_{SO(4)}^{(I)} ~ , ~~~~
|\bar{w}> = \Gamma^{\bar{w}} |0>_{SO(4)}^{(I)} ~, \nonumber\\
& & |\bar{v}> = \Gamma^{\bar{v}} |0>_{SO(4)}^{(II)} ~ , ~~~~
|\bar{z}> = \Gamma^{\bar{z}} |0>_{SO(4)}^{(II)} ~ ,
\ea
and furthermore we have the fully filled states
\be
|\bar{u} \bar{w}> = \Gamma^{\bar{u}} \Gamma^{\bar{w}} |0>_{SO(4)}^{(I)} ~, ~~~~
|\bar{v} \bar{z}> = \Gamma^{\bar{v}} \Gamma^{\bar{z}} |0>_{SO(4)}^{(II)} ~.
\ee

It is straightforward to assign $SO(1,1)$ and $SO(4)$ chiralities to all these states
by acting with the corresponding chirality operators. We find that $|0>_{SO(1,1)}$ has
positive $SO(1,1)$ chirality, whereas the fully filled state $|+>$ has negative
$SO(1,1)$ chirality.
Likewise, $|0>_{SO(4)}^{(I)}$ and $|\bar{u} \bar{w}>$ have
negative $SO(4)_{\rm I}$ chirality, whereas $|\bar{u}>$ and $|\bar{w}>$ have positive
$SO(4)_{\rm I}$ chirality.
Similarly, $|0>_{SO(4)}^{(II)}$ and $|\bar{v} \bar{z}>$ have
negative $SO(4)_{\rm II}$ chirality, whereas $|\bar{v}>$ and $|\bar{z}>$ have positive
$SO(4)_{\rm II}$ chirality. Then, the chiral properties of the $SO(8)$ (or
$SO(1,1) \times SO(8)$) spinor states, which are described by taking suitable tensor
products of all possible states above, will follow by multiplying the chiralities of
each separate factor; for instance, the ten dimensional spinor $|0>$ described above
has positive $SO(1, 1) \times SO(8)$ chirality.
The assignment of these block chiralities is very important for
constructing supersymmetric solutions in an axionic instanton background from the
known solutions of the Killing spinor equations in flat transverse space.

The Killing spinor equations for pp-waves with flat transverse space have been solved
by Maldacena and Maoz by choosing the following general expressions for the two $SO(8)$
chiral components of the complex Weyl spinor $\epsilon$, $\epsilon_{\pm}$,
\be
\epsilon_+ = \alpha |0> + \zeta |\Delta> ~, ~~~~
\epsilon_-  =  \Gamma^+ \left( \beta_{\bar{k}} \Gamma^{\bar{k}} |0> +
\delta_{k} \Gamma^k |\Delta> \right) ,
\ee
where
\be
|\Delta> =  {1 \over 4} \Gamma^{\bar{u} \bar{w} \bar{v} \bar{z}}|0> ~.
\ee
Here, summation is implicitly assumed over all holomorphic indices $k$ taking values
$u$, $w$, $v$ or $z$, and similarly for the anti-holomorphic indices that are
parameterized collectively by $\bar{k}$. The spinors
$\epsilon_{\pm}$ are written down as shown by including
all possible terms with the correct
$SO(8)$ net chirality in either case. Actually, $\epsilon_+$ can also contain
constant terms of the form $\gamma_{\bar{m} \bar{n}} \Gamma^{\bar{m} \bar{n}} |0>$ with
positive $SO(8)$ chirality, but they can always be set equal to zero
in flat transverse space by
performing a suitable $SO(8)$
rotation\footnote{Recall at this point that $\Gamma^{\mu \nu}$, which are anti-symmetric in
the indices $\mu$, $\nu$, are the generators of the orthogonal group $SO(D)$ in the
spinorial representation. Hence, one may use group elements of the general form
${\rm exp}(\alpha_{\mu \nu} \Gamma^{\mu \nu})$ for appropriately chosen canonical
parameters $\alpha_{\mu \nu}$ in order to implement $SO(D)$ rotations of the corresponding
spinors. In our case, it is sufficient to choose $\alpha_{\bar{m} \bar{n}} =
- \gamma_{\bar{m} \bar{n}}$ to rotate way all constant terms of the form
$\gamma_{\bar{m} \bar{n}} \Gamma^{\bar{m} \bar{n}} |0>$ in $\epsilon_+$, as indicated above.}.
However, there are some subtle points regarding this rotation in the case that
the transverse space has a block-diagonal form, which in effect breaks $SO(8)$ to
the subgroup $SO(4) \times SO(4)$, that will be addressed later.
In any case, rotating these additional constant terms away proves helpful for finding
explicit solutions of the Killing spinor equations rather easily, but the resulting
expressions are parameterized by only two complex parameters $\alpha$ and $\zeta$, which
are constant, since $\epsilon_+$ is a constant spinor independent from all target space
coordinates.
In this case, one describes all solutions with $(2,2)$ supersymmetry, but not
be able to distinguish solutions with more supersymmetry, as their complex parameters
have been rotated away to zero. Likewise, if $|\alpha| = |\zeta|$, which is a
special case that will be shortly discussed, the solutions only depend
on a single complex parameter, and hence exhibit $(1, 1)$ supersymmetry.
As for the other spinor, $\epsilon_-$,
it depends on all target space coordinates but $-$, and
therefore, the coefficients $\beta_{\bar{k}}$ and $\delta_{k}$ depend on the remaining
transverse coordinates of space-time.

Substituting these general expressions into the Killing spinor
equations, and assuming further that $\epsilon_-$ is also independent from the
other light-cone coordinate $x^+$, one arrives at a system of first order equations
for the coefficients of $\epsilon_-$, which are solved as follows, \cite{malda2}:

\noindent

(i) \underline{$|\alpha| \neq |\zeta|$}: In this case the coefficients assume the form
\be
\beta_{\bar{k}} = {i \over 4} \left(\alpha \varphi_{l \bar{k}} z^l - \zeta \partial_{\bar{k}}
\bar{\cal W} \right) ~, ~~~~
\delta_{k} = {i \over 4} \left(\zeta \varphi_{k \bar{l}} z^{\bar{l}} - \alpha \partial_k
{\cal W} \right) ~,
\ee
and they are determined in terms of a holomorphic superpotential ${\cal W}$ that, in principle,
can depend on all transverse space coordinates. Then, the solutions have $(2,2)$ supersymmetry
(or more) and the particular expressions for the front factor $F$ and all components
$\varphi_{mn}$ and $\varphi_{m \bar{n}}$ (using the short-hand notation) are detailed in
section 2.2.

\noindent
(ii) \underline{$|\alpha| = |\zeta|$}: In this case, choosing without loss of generality
$\alpha = - \zeta^{\star}$, we obtain the solution\footnote{In the original work of
Malacena and Maoz, \cite{malda2}, there has been a different choice of complex
parameters with $|\alpha| = |\zeta|$, namely $\alpha = - \zeta$, which results into
some slightly different expressions for the coefficients $\beta_{\bar{k}}$ and
$\delta_k$. Here, we choose $\alpha = - \zeta^{\star}$ instead, as it arises more naturally
in connection with the reality conditions that will be implemented later in the
form of the Killing spinors in axionic instanton backgrounds. In any case, any
particular choice amounts to redefining the complex coordinates by a constant phase.}
\be
\beta_{\bar{k}} = -{i \over 4} \zeta^{\star} \partial_{\bar{k}} U ~, ~~~~~
\delta_k = {i \over 4} \zeta
\partial_k U ~,
\ee
which is parameterized by an arbitrary real harmonic function $U$ that, in principle, can
depend on all transverse space coordinates. Then, the corresponding solutions of type IIB
supergravity have only $(1,1)$ supersymmetry and the particular expressions for $F$,
$\varphi_{mn}$ and $\varphi_{m \bar{n}}$ are also detailed in section 2.2.

\subsection{Axionic pp-wave solutions}

These results can be adapted to pp-wave backgrounds with an axionic instanton placed
in one of the two four dimensional building blocks of the transverse space,
by implementing the appropriate $SO(4)$ chiral projections on the flat space
solutions of the Killing spinor equations. We choose for definiteness the axionic
instanton condition that selects certain chiral components of the Majorana-Weyl
spinors associated to the complex Weyl spinors $\epsilon_+$ and $\epsilon_-$, namely

\be
P_+^{(I)} \epsilon_{\pm}^{(1)} = 0 = P_-^{(I)} \epsilon_{\pm}^{(2)} ~,
\ee
as opposed to the axionic anti-instanton condition that selects the opposite chiralities.
Of course, the
physical interpretation of the end result will not really depend on
the particular choice that one makes, as the
condition of self-duality or anti-self-duality is just a matter of convention.
Then, according to this, we have
\be
\epsilon_+ = \epsilon_+^{(1) --} + i \epsilon_+^{(2) ++} ~, ~~~~
\epsilon_- = \epsilon_-^{(1) -+} + i \epsilon_-^{(2) +-} ~.
\ee
Next, we may decompose each Killing spinor equation into pairs
of equations with definite $SO(4) \times SO(4)$ chirality by acting with $P_+^{(I)}$
or $P_-^{(I)}$ from their left on each side. The calculation can be easily performed
bearing in mind the following obvious identities:
$[\sv ~, P_{\pm}^{(I)}] = 0$, which follows
from the block-diagonal form of $\varphi$,
$P_{\pm}^{(I)} \Gamma^i = \Gamma^i  P_{\mp}^{(I)}$ when $i$ runs in the first (axionic
instanton) block, and $P_{\pm}^{(I)} \Gamma^{i^{\prime}} = \Gamma^{i^{\prime}} P_{\pm}^{(I)}$
when $i^{\prime}$ runs in the second (flat space) block.

Working out the $SO(4) \times SO(4)$ chiral components of the $\mu = +$ Killing spinor equation
and assuming that $\epsilon_-$ is also independent of $x^+$, we arrive at the following pair
of equations written in terms of the non-vanishing Majorana-Weyl components:
\ba
e^{\Phi} \sv ~ \epsilon_-^{(2) +-} & = &
- \Gamma_{\bar{-}} \left((\Gamma^i \partial_i F)
\epsilon_+^{(1) --} + i (\Gamma^{i^{\prime}} \partial_{i^{\prime}} F) \epsilon_+^{(2) ++}
\right) , \nonumber\\
e^{\Phi} \sv ~ \epsilon_-^{(1) -+} & = &
\Gamma_{\bar{-}} \left((\Gamma^i \partial_i F)
\epsilon_+^{(2) ++} - i (\Gamma^{i^{\prime}} \partial_{i^{\prime}} F) \epsilon_+^{(1) --}
\right) ,
\ea
where the dilaton factor ${\rm exp} \Phi$ still appears explicitly.
Comparing the imaginary parts we immediately obtain
\be
\partial_{i^{\prime}} F = 0 ~; ~~~~~ i^{\prime} = 5, 6, 7, 8
\ee
and so the front factor $F$ can only depend on the coordinates of the axionic instanton block.
As for the real parts, they can be collectively written in complex form,
\be
\sv ~ \left(\epsilon_-^{(1) -+} +
i \epsilon_-^{(2) +-} \right) = -i \Gamma_{\bar{-}}
(\Gamma^i \partial_i F) \left(\epsilon_+^{(1) --} + i \epsilon_+^{(2) ++} \right),
\ee
where, here, the rescaling $\tilde{\Gamma}^i = ({\rm exp \Phi}) \Gamma^i$
has also been taken into account in order to
absorb the dilaton factor consistently, and the tilde has been further dropped from the
resulting flat $\Gamma$-matrices for convenience. The result
states that the $SO(4)$ projection of this Killing spinor equation is identical to the
same equation for
the $SO(4)$ projected Weyl spinors, as anticipated, plus the additional restriction that
$\partial_{i^{\prime}} F = 0$.

In a similar way, we may work out the two $SO(4)$ chiral components
of the Killing spinor equations for
$\mu = i$ taking values in the first block. We arrive at the following pair of
real equations, using the Majorana components of the Weyl spinors,
\be
\partial_i \epsilon_-^{(2) +-} = {1 \over 32} \Gamma_- \sv ~
\Gamma_i \epsilon_+^{(1) --} ~, ~~~~
\partial_i \epsilon_-^{(1) -+} = -{1 \over 32} \Gamma_- \sv ~
\Gamma_i \epsilon_+^{(2) ++}
\ee
or equivalently at the following equation for the corresponding complex spinors,
\be
\partial_i \left(\epsilon_-^{(1) -+} + i \epsilon_-^{(2) +-}\right) =
{i \over 32} \Gamma_- \sv ~
\Gamma_i \left(\epsilon_+^{(1) --} + i \epsilon_+^{(2) ++} \right)
\ee
as it is anticipated, without the need to impose any additional restrictions for the consistency
of this particular projection; here, ${\rm exp} \Phi$ has also been absorbed into the
definition of the flat space $\Gamma$-matrices \eqn{redefga}
from the very begining and the tilde has been further
dropped for convenience.

The two $SO(4)$ chiral components of the Killing spinor equations that correspond to
$\mu = i^{\prime}$ running
in the second block, require special attention, as they introduce certain additional constraints
that are a bit trickier to analyze. In particular, we arrive at the following pair of equations
by imposing the two chiral projections, using $P_+^{(I)}$ and $P_-^{(I)}$,
\be
\partial_{i^{\prime}} \epsilon_-^{(2) +-} = {i \over 32}
e^{\Phi} \Gamma_- \sv ~ \Gamma_{i^{\prime}}
\epsilon_+^{(2) ++} ~, ~~~~
\partial_{i^{\prime}} \epsilon_-^{(1) -+} = {i \over 32}
e^{\Phi} \Gamma_- \sv ~ \Gamma_{i^{\prime}}
\epsilon_+^{(1) --} \label{killi}
\ee
and therefore, by comparing the real and imaginary parts we arrive immediately at the
special conditions
\be
0 = \partial_{i^{\prime}} \left( \epsilon_-^{(1) -+} + i \epsilon_-^{(2) +-} \right) ,
~~~~
0 = e^{\Phi} \sv ~ \Gamma_{i^{\prime}} \left(\epsilon_+^{(1) --} +
i \epsilon_+^{(2) ++}\right) ,
\ee
which are written here collectively using the corresponding complex Weyl spinors.
The first one implies that $\epsilon_-$ is independent from the coordinates of the second
four dimensional block, in accordance to the independence of the front factor $F$ from
them, whereas the second equation imposes some new constraints on the block diagonal
structure of $\varphi$.

We can examine its consequences by recalling the following
identities
\ba
\sv ~ \Gamma^{\bar{m}} |0> & = & 4\left({\varphi^{\bar{m}}}_n \Gamma^n |\Delta> -
{\varphi^{\bar{m}}}_{\bar{n}} \Gamma^{\bar{n}} |0>\right) ~,  \nonumber\\
\sv ~ \Gamma^{m} |\Delta> & = & 4\left({\varphi^m}_{\bar{n}} \Gamma^{\bar{n}}|0> -
{\varphi^m}_n \Gamma^n |\Delta>\right) ~,
\ea
which are fairly straightforward to prove. Here, $m$, $n$ can take values on all holomorphic
indices in either block and $\bar{m}$, $\bar{n}$ are the anti-holomorphic indices.
These identities in turn imply that
\be
\sv ~ \Gamma^v \left(\alpha |0> + \zeta |\Delta> \right) =4\zeta \left({\varphi^v}_{\bar{n}}
\Gamma^{\bar{n}} |0> - {\varphi^v}_{n} \Gamma^n |\Delta> \right),
\ee
since $\Gamma^v |0> = 0$, and therefore the left-hand side vanishes, as required,
provided that $\varphi_{\bar{v} \bar{n}} = 0 =
\varphi_{\bar{v} n}$ for all possible values of the indices $n$ and $\bar{n}$.
Likewise, we have
\be
\sv ~ \Gamma^z \left(\alpha |0> + \zeta |\Delta> \right) =4\zeta \left({\varphi^z}_{\bar{n}}
\Gamma^{\bar{n}} |0> - {\varphi^z}_{n} \Gamma^n |\Delta> \right) ,
\ee
which vanishes if $\varphi_{\bar{z} \bar{n}} = 0 = \varphi_{\bar{z} n}$ for all $n$ and
$\bar{n}$.
In a similar fashion, since $\Gamma^{\bar{v}} |\Delta> = 0$, we easily find that
\be
\sv ~ \Gamma^{\bar{v}} \left(\alpha |0> + \zeta |\Delta> \right) =4\alpha \left(
{\varphi^{\bar{v}}}_n \Gamma^n |\Delta> - {\varphi^{\bar{v}}}_{\bar{n}} \Gamma^{\bar{n}} |0>
\right) ,
\ee
implying that it vanishes when $\varphi_{v n} = 0 = \varphi_{v \bar{n}}$ for all $n$ and $\bar{n}$,
and finally we have
\be
\sv ~ \Gamma^{\bar{z}} \left(\alpha |0> + \zeta |\Delta> \right) = 4\alpha \left(
{\varphi^{\bar{z}}}_n \Gamma^n |\Delta> - {\varphi^{\bar{z}}}_{\bar{n}} \Gamma^{\bar{n}} |0>
\right) ,
\ee
which vanishes when $\varphi_{z n} = 0 = \varphi_{z \bar{n}}$.
Putting all these together, we conclude that the conditions $\sv ~ \Gamma_{i^{\prime}} \epsilon_+ = 0$
above are true for all $i^{\prime}$ running in the second four dimensional block, provided that
$\varphi_{mn}$, $\varphi_{m \bar{n}}$ and their complex conjugates are zero when the indices take values
in the second block; of course, there can be no components with indices from the off-diagonal
blocks. Thus, the 4-form $\varphi$, which can only depend on the coordinates of the axionic instanton
block (as can be seen by inspection of the projected Killing spinor equations), has non-vanishing
components, in the short-hand notation, only in the first block, as required by
the consistency of the Killing
spinor equations. In this case, the appearance of the dilaton factor ${\rm exp} \Phi$ becomes
irrelevant.

Let us now turn to the explicit solution of the resulting Killing spinor equations for
$\epsilon_+ =\epsilon_+^{(1) --} + i\epsilon_+^{(2) ++}$ and
$\epsilon_- =\epsilon_+^{(1) -+} + i\epsilon_-^{(2) +-}$, which can only depend on the
coordinates of the axionic instanton block.
For this, it is important to note that the
general parameterization of $\epsilon_+$, which was chosen by Maldacena and Maoz by
rotating away all other possible terms with positive $SO(8)$ chirality,
describes a complex Weyl spinor with definite
$SO(4) \times SO(4)$ chirality that is $(-, -)$. Therefore, we have to set
the Majorana component $\epsilon_+^{(2)++} = 0$ in order to respect the
block-diagonal chiral properties of this
spinor. In turn, this implies that $\epsilon_+ = \epsilon_+^{(1) --}$ is real,
as we are only restricted to the other Majorana component. Majorana spinors can be easily
constructed in the Fock space representation by noting that
\be
|0>^{\star} = - |\Delta> ~, ~~~~~ |\Delta>^{\star} = - |0> ~,
\ee
which follow by taking the complex conjugate of the defining relations for the vacuum state, i.e.,
$\Gamma^n |0> = 0$ for all holomorphic indices. Indeed, working with the Majorana representation
of the gamma matrices, i.e., $(\Gamma^{\mu})^{\star} = - \Gamma^\mu$ for all real space-time
indices, we obtain
$0 = (\Gamma^n |0>)^{\star} = - \Gamma^{\bar{n}} |0>^{\star}$, and therefore $|0>^{\star}
= \pm |\Delta>$; here, we choose the minus sign without loss of generality. Then, according to
this particular reality condition, we have that $(\alpha |0> + \zeta |\Delta>)^{\star} =
-\zeta^{\star} |0> - \alpha^{\star} |\Delta>$, and therefore
\be
\epsilon_+^{\star} = \epsilon_+  ~~~~~ {\rm for} ~~ \alpha = -\zeta^{\star} ~.
\ee

On the other hand, since $\epsilon_+^{(2)++}$ is taken to be zero, consistency of the Killing
spinor equations \eqn{killi}
implies that $\epsilon_-^{(1) -+} = 0$; actually, it can be any constant spinor, but all
values other than zero can be removed by an appropriate shift of the coordinates in the final
form of the solution. Thus, $\epsilon_- = i \epsilon_-^{(2) +-}$ is purely imaginary with
$SO(4) \times SO(4)$ chirality $(+, -)$. A close inspection of the terms appearing in the
general parameterization of $\epsilon_-$, as given by Maldacena and Maoz, reveals that
the terms proportional to $\beta_{\bar{u}}$,
$\beta_{\bar{w}}$, $\delta_u$ and $\delta_w$ have $(+ , -)$ block-diagonal chilarities,
whereas the remaining terms that depend on the coordinates $v$, $z$ and their complex
conjugates have $(-,+)$ chiralities. Setting the latter equal to zero by the chiral
projection is also in agreement with the independence of the Killing spinor equations
from the coordinates of the second superconformal building block.
The purely imaginary character of $\epsilon_-$ can be easily implemented in this case
by noting that $(\Gamma^{\bar{n}} |0>)^{\star} = \Gamma^n |\Delta>$ for all $n$ in the
Fock space representation. Thus, since $(\Gamma^+)^{\star} = - \Gamma^+$ in the Majorana
representation, we have,
\be
\epsilon_-^{\star} = - \epsilon_- ~~~~~ {\rm for} ~~ \delta_u^{\star} =  \beta_{\bar{u}} ~, ~~
\delta_w^{\star} =  \beta_{\bar{w}}
\ee
with the additional restriction that $\beta_{\bar{v}} = 0 = \beta_{\bar{z}}$ and
$\delta_v = 0 = \delta_z$. This is precisely the reality condition satisfied by the
coefficients $\beta_{\bar{k}}$ and $\delta_k$ in the flat space solution of the
Killing spinor equations with $|\alpha| = |\zeta|$, provided that one chooses
$\alpha = - \zeta^{\star}$ as it is presently required.

Combining these results together, we conclude that only {\it the flat space pp-wave
solutions with $(1,1)$
supersymmetry provide supersymmetric pp-wave gravitational
backgrounds in the presence of an axionic
instanton block in their transverse space}. Namely, for any real harmonic function
$U(u, w, \bar{u}, \bar{w})$, we have
\be
\varphi_{mn} = \partial_m \partial_n U ~,
~~~~ \varphi_{\bar{m} \bar{n}} = \partial_{\bar{m}} \partial_{\bar{n}} U ~,
~~~~ \varphi_{m \bar{n}} = \partial_m
\partial_{\bar{n}} U ~,
\ee
where all derivatives are taken with respect to the flat space coordinates in the
axionic instanton block. As for the front factor of the corresponding pp-wave solutions,
it is given by
\be
F = {1 \over 2} \mid \partial U \mid^2 ~,
\ee
where the contraction is also taken here using the flat space norm; therefore, the resulting
solutions fit precisely into the general class of
solutions that were already anticipated in section 6.1 by studying the
general structure of the
second order equations. A particularly simple example is obtained by choosing
\be
U = {1 \over \sqrt{2}} \left(A(u^2 + {\bar{u}}^2) + B(w^2 + {\bar{w}}^2) \right) ,
\ee
which reproduces the solution obtained in section 5.3 with quadratic front factor
\eqn{examp1} and constant 4-form $\varphi$ \eqn{examp2}, provided that one also sets
there $C= 0 =D$.

>From a geometrical point of view, the meaning of our result
is that the type IIB five-brane
configuration admits a supersymmetric generalization in the presence of a non-trivial
R-R self-dual 5-form ${\cal F} = dx^+ \wedge \varphi$ by simply
turning its world-volume from
$M_6$ into a six dimensional pp-wave. However, since the front factor $F$ depends only
on the coordinates of the remaining four directions that parameterize the axionic
instanton block in the transverse space of the full ten dimensional pp-wave solution,
the ``world-volume theory" is rather special and does not correspond to a genuine
six dimensional pp-wave. It certainly requires a more detailed
investigation in order to find applications of such deformed solitonic configurations
in modern day string theory.

We conclude our discussion of the supersymmetric pp-wave configurations
in the presence of axionic
instantons in their transverse space by examining the possibility to have more general
choices for the constant complex spinor $\epsilon_+$, which could
be provided by additional
terms of the form $\gamma_{\bar{m} \bar{n}} \Gamma^{\bar{m} \bar{n}}|0>$. All such terms,
which can be rotated away in flat transverse space,
have positive $SO(8)$ chirality and they can be grouped in two classes depending on their
$SO(4) \times SO(4)$ block-diagonal chiralities: mixed terms of the form
$\Gamma^{\bar{u} \bar{v}}$, $\Gamma^{\bar{u} \bar{z}}$, $\Gamma^{\bar{w} \bar{v}}$ and
$\Gamma^{\bar{w} \bar{z}}$ acting on the vacuum $|0>$ have (+, +) chiralities, thus
forming complex spinors of the
type $\epsilon_+^{++}=\epsilon_+^{(1)++} + i \epsilon_+^{(2)++}$, whereas
terms of the form $\Gamma^{\bar{u} \bar{w}}$
and $\Gamma^{\bar{v} \bar{z}}$ acting on the
vacuum have $(-, -)$ chiralities, thus forming complex spinors
of the type $\epsilon_+^{--} = \epsilon_+^{(1)--} + i \epsilon_+^{(2)--}$.
The Majorana component $\epsilon_+^{(1)++}$ has been projected away by the axionic
instanton condition, while $\epsilon_+^{(2)++}$ also turns out to be zero by the Killing
spinor equation that relates it to $\partial_i \epsilon_-^{(1)-+}$. It is sufficient
to note for this purpose, following the previous assignment of $SO(4) \times SO(4)$
chiralities to all possible terms that appear in the general parameterization of
$\epsilon_-$, that $\epsilon_-^{-+}$ depends on the coordinates of the second block
through the coefficients $\beta_{\bar{v}}$, $\beta_{\bar{z}}$, $\delta_v$ and $\delta_z$;
but on the other hand, since we also have the Killing spinor equation
$\partial_{i^{\prime}} \epsilon_- = 0$, it turns out that this can only happen if
$\epsilon_-^{(1)-+} = 0$ (modulo constants that can be removed by a change of
coordinates) and hence $\epsilon_+^{(2)++} = 0$. As for the two extra terms
that can be added to the parameterization of $\epsilon_+^{--}$, they can be removed by
an $SO(4) \times SO(4)$ rotation, due to their block-diagonal structure
$\Gamma^{\bar{u} \bar{w}} |0>$ and $\Gamma^{\bar{v} \bar{z}} |0>$, which respects
the $SO(4) \times SO(4)$ structure of the two building blocks of the transverse space.
These remarks clarify the issue that was raised earlier regarding the use of
$SO(4) \times SO(4)$ versus $SO(8)$ rotations in our generalized class of
type IIB gravitational backgrounds, and, furthermore, they prove
that the solutions we have described in the presence of an axionic instanton block
are the most general supersymmetric ones.

It is not surprising that the new gravitational backgrounds we have constructed exhibit
less supersymmetry than the corresponding solutions in flat transverse space.

\section{Further considerations}
\setcounter{equation}{0}

In this section, we collect some further ideas and partial results that are related to
the general class of pp-wave gravitational backgrounds in string theory. First, we
briefly examine the possibility to have double axionic instanton backgrounds in the
transverse space, in the sense that both four dimensional building blocks are
replaced by axionic instanton solutions, and we assume that $F$ and $\varphi$ can have
support on all eight transverse space coordinates. We will present some
special solutions of the second order equations and comment on the (non)-existence
of space-time supersymmetric configurations of this particular type.
Another topic that we will also
briefly discuss concerns some aspects of the general duality between type IIA and type
IIB backgrounds for pp-wave backgrounds with non-trivial R-R fields, and their
possible IIA counterparts. Our study of this particular subject is very limited;
it is merely included here for completeness, in order to illustrate some simple
results and motivate some interesting
directions for future work.

\subsection{On double axionic instanton solutions}

Let us first examine the general structure of the
second order equation \eqn{main} in the presence of two different superconformal
blocks in the transverse space, which nevertheless are both taken to be axionic
instantons. Since the fields are now supposed to have support on all transverse
directions, we obtain the following general result
\be
\nabla_i\left(e^{-2 \Phi} \nabla^i F \right) =
e^{-4\Phi_1} e^{-2 \Phi_2}
(\partial_u \partial_{\bar{u}}
F + \partial_w \partial_{\bar{w}} F)
+ e^{-2\Phi_1} e^{-4 \Phi_2}
(\partial_v \partial_{\bar{v}}
F + \partial_z \partial_{\bar{z}} F) ~,
\ee
where $(1)$ and $(2)$ are indices labeling the two different four dimensional
blocks and $\Phi = \Phi_1(u, w; \bar{u}, \bar{w}) + \Phi_2(v,z; \bar{v}, \bar{z})$ is
the dilaton field of the entire background under consideration.

In this case, one may seek solutions of the main equation
\eqn{main} for front factors $F$ that depend on all eight transverse coordinates,
so that
\ba
(\partial_u \partial_{\bar{u}} + \partial_w \partial_{\bar{w}}) F & = &
e^{-2\Phi_2 (v,z; \bar{v}, \bar{z})} A ~, \nonumber\\
(\partial_v \partial_{\bar{v}} + \partial_z \partial_{\bar{z}}) F & = &
e^{-2\Phi_1 (u,w; \bar{u}, \bar{w})} A ~,
\ea
where $A$ is also a function of all eight transverse coordinates. In particular, if
$A$ can be written in the form
\be
A = (\partial_u \partial_{\bar{u}} + \partial_w \partial_{\bar{w}}
+ \partial_v \partial_{\bar{v}} + \partial_z \partial_{\bar{z}}) f ~,
\ee
where $f$ is another (yet unknown)
function of the transverse coordinates, then solutions can
be easily constructed from the flat space models with front factor $f$. For this, it
is sufficient to note that $h^{u\bar{u}} = h^{w\bar{w}} = {\rm exp}(-2\Phi_1)$ and
$h^{v\bar{v}} = h^{z\bar{z}} = {\rm exp}(-2 \Phi_2)$, where each one depends on the
coordinates of the corresponding superconformal blocks; then, the norm of the 4-form
$\varphi$ can become conformally related to the norm of a 4-form in flat space
associated to a front factor $f$.

A simple example of double axionic solutions is provided by the model $W_k \times W_k$,
where two independent (semi)-wormholes are placed in the transverse space so that
\be
h^{u\bar{u}} = h^{w\bar{w}} = |u|^2 + |w|^2 ~, ~~~~
h^{v\bar{v}} = h^{z\bar{z}} = |v|^2 + |z|^2 ~,
\ee
with their respective dilaton and anti-symmetric tensor fields. Then, it can be easily
verified, as a consequence of the previous analysis, that the following choice
of front factor and 4-form,
\ba
F & = & \left(|u|^2 + |w|^2 \right) \cdot \left(|v|^2 + |z|^2 \right) , \nonumber\\
\varphi & = & du \wedge d\bar{w} \wedge d\bar{v} \wedge d\bar{z} +
dw \wedge d\bar{u} \wedge d\bar{v} \wedge d\bar{z} \nonumber\\
& + & dv \wedge d\bar{u} \wedge d\bar{w} \wedge d\bar{z} +
dz \wedge d\bar{u} \wedge d\bar{w} \wedge d\bar{v} ~ + ~~ cc
\ea
provides a new non-trivial background of pp-wave form, which is also supersymmetric.
It can be readily seen that this
background is obtained by appropriate embedding of the maximal
supersymmetric pp-wave solution with flat transverse space.
More complicated solutions of this type may be constructed if necessary, but the details
will not be presented here.

The supersymmetric properties of gravitational pp-waves solutions with double axionic instantons
in their transverse space can also be investigated by treating the Killing spinor equations, as
in the case of single axionic instanton backgrounds.
Note at this point that the structure of the Killing spinor equations is such that the
$SO(4) \times SO(4)$ chiral projections restrict all fields to have support on any chosen
axionic instanton block and no dependence on the other block. Thus, if two axionic
instanton blocks are put together in the transverse space, in a diagonal form, the only
consistent supersymmetric solution is the null configuration with $\varphi = 0$ and
constant factor $F$. Thus, we conclude that there can be no space-time supersymmetric
solutions of this kind.

\subsection{Type IIA-IIB duality}

It is well known that in ten dimensions there exist two different theories of type II
supergravity. First, we have type IIA theory whose bosonic sector contains the
metric, a vector field, a scalar field (dilaton), as well as a 2-form and a 3-form,
\be
{\rm IIA}: ~~~~ \{g_{\mu \nu}, A_{\mu}, \Phi, B_{\mu \nu}, C_{\mu \nu \rho} \} ~.
\ee
This theory can be obtained by dimensional reduction of $N=1$ eleven dimensional
supergravity, where the first three fields of type IIA theory arise
from the components of the eleven dimensional metric, whereas the remaining two forms
arise from the eleven dimensional 3-form. On the other hand, type IIB theory
contains in general the metric, a complex anti-symmetric tensor field, a
complex scalar field, as well as a 4-form with self-dual field strength,
\be
{\rm IIB}: ~~~~ \{g_{\mu \nu}, {\cal B}_{\mu \nu}, \tilde{\Phi},
C_{\mu \nu \rho \sigma}^{(4)} \} ~.
\ee
The complex scalar parameterizes the coset space $SU(1,1)/U(1)$, but in our case we
have only considered a real field associated with the dilaton, setting its R-R part
$C^{(0)}$ equal to zero; likewise, we have
imposed the reality condition on the complex anti-symmetric tensor field by
considering only a non-vanishing NS-NS 2-form field $B$, setting its R-R part
$C^{(2)}$ also equal to zero.

Both type IIA and IIB supergravities give rise to the same $N=2$ theory in nine
dimensions upon dimensional reduction, and therefore, provided that one makes
correctly the field identifications in nine dimensions in either case, the
T-duality rules can be seen to relate type IIA with type IIB backgrounds in
ten dimensions \cite{dine}; we also refer the reader to \cite{berg}, where the details
have been worked out carefully in the most general terms. Setting the R-R fields
equal to zero in both type of theories, T-duality relates the remaining NS-NS fields
as usual, as in type I theories. In a type II context, the small-large radius
duality has to be replaced by the map $R_{\rm IIA} \rightarrow
{\alpha}^{\prime}/R_{\rm IIB}$, where $R_{\rm IIA}$ is the radius characterizing the
type IIA decompactification to ten dimensions, and likewise $R_{\rm IIB}$ for the
decompactification of the type IIB theory; more generally, type II T-duality maps
the symmetries of each individual ten dimensional theory into one another.
In the most general situation, when R-R fields are also present, the T-duality
rules extend naturally to all ten dimensional bosonic fields, and therefore can be
used to construct consistent type IIA backgrounds from all type IIB pp-wave solutions
with non-trivial R-R 4-form $\varphi$. In some special examples, however, the T-duality
transformation rules among the NS-NS fields can also be used exclusively within the
type IIB context to relate consistent gravitational backgrounds with R-R fields, as
we have already seen in section 5 for
certain pp-wave solutions, and for appropriate choices of their
front factors. The occurrence of dual pp-wave solutions is accidental,
however, as it does not rely on any
symmetry principles; they rather arise by the mere existence of some special solutions of
the second order field equations. In more general situations,
T-duality can not be consistently implemented within the type IIB theory alone,
and hence pp-wave solutions with non-trivial R-R fields will only give rise to
consistent solutions with a type IIA field content.

It is a very interesting exercise to perform the T-duality transformation on all
supersymmetric pp-wave solutions based on the presence of axionic instantons in their
transverse space, and work out the details of the corresponding type IIA
backgrounds. In simpler cases, without having any R-R fields turned on, it is known that
supersymmetric string waves of type IIB theory map into (generalized) fundamental
string solutions of type IIA, and vice versa (see, for instance, \cite{berg} and
references therein). For example, working in the appropriate frame and combining
T-duality with S-duality along the way, one finds that the supersymmetric string wave
\be
ds^2 = -4dX^+ dX^- - F(y) \left(dX^+\right)^2 + \delta_{ij}dy^i dy^j
\ee
with front factor satisfying the Laplace equation in the flat transverse space,
$\nabla^2 F = 0$, as required by the absence of R-R fields,
can be rotated by dualities to the following
solution of type IIA theory,
\ba
ds^2 & = & -4 e^{2 \Phi} dX^+ dX^- - \delta_{ij} dy^i dy^j ~, \nonumber\\
B_{+-} & = & 4 \left(e^{2 \Phi} -1\right) dX^+ \wedge dX^- ~, ~~~~
e^{-2 \Phi} = 1 + {1 \over 2}F ~,
\ea
which is a very special case of the more general class of fundamental string solutions.

Thus, it is natural to expect that in the presence of R-R fields, the duality rotation
of more general supersymmetric pp-wave solutions will admit an interesting interpretation
in type IIA theory. It will also be interesting to revisit in this context some recent
results that have been obtained for (twisted) toroidal compactifications of pp-waves
and their type IIA dual faces, \cite{jmick},
by extending their applicability to the more general
gravitational configurations that have been encountered in the present work.
We plan to return elsewhere to this problem
for a more systematic exposition, following the general rules of dualities for R-R
potentials,
together with
the physical interpretation of the resulting solutions that can be constructed in this
fashion.

\section{Conclusions and discussion}
\setcounter{equation}{0}

We have presented a systematic study of string theory models that propagate
on geometries of pp-wave type in the presence of non-trivial metric, dilaton
and NS-NS anti-symmetric tensor fields, which all reside in the transverse
space. These backgrounds, which also admit a non-trivial R-R 5-form
${\cal F} = dx^+ \wedge \varphi$ for suitably chosen anti-self-dual closed
4-forms in the transverse space, give rise to interacting string models in the
light-cone formulation of the theory. The light-cone gauge,
$x^+ = \tau$, is
an eligible choice for this general class of gravitational string backgrounds,
as the ten dimensional geometry of type IIB supergravity admits by construction
a null Killing vector field provided by $\partial_{x^-}$ in adapted coordinates.
In those cases that the front factor $F$ corresponds to non-linear, but integrable,
interactions among the bosonic fields on the string world-sheet, the quantization
program can be in principle carried out in the light-cone gauge by relying on
properties of the underlying integrable two dimensional quantum field theories.
Thus, one obtains new classes of tractable superstring models that also include
interactions.

Of course, there are some technicalities that could obscure the exact quantum
mechanical treatment of this problem at first sight, which originate from the fact
that the two dimensional light-cone action is defined on a cylinder rather than
a plane, as the spatial coordinate $\sigma$ of the world-sheet is compact.
Integrable systems on a cylinder are less studied than their planar counterparts,
and even in the classical theory the structure of their solutions can be quite
complicated, but yet tractable by methods of algebraic geometry. A key point in
such a description is provided by the fact that although integrable systems can
be highly non-trivial in their self-interactions, their integrability properties
may be used to device a superposition principle, as in all free field
models. Thus, classical solutions on a cylinder may be written as suitable
linear superpositions of infinitely long chains of planar solutions, such as
alternating sums of kinks and/or anti-kinks, which create an infinitely
long lattice
on the plane with spacing equal to the period of the $\sigma$ direction.
However, it can be easily understood that such superpositions, which can be
resumed by using theta functions, may destroy the stability properties
of the resulting configurations upon small fluctuations, if
strictly periodic configurations are only taken into account;
they will be oscillatory and hence
non-monotonic, which prevents their stability.
Other topological sectors, which can be constructed by
winding or orbifolding on the string world-sheet, are nevertheless stable
(see, for instance, \cite{bak6}).
Consequently, the quantum mechanical treatment of integrable systems on a
cylinder, and their solitonic configurations that one usually employs
in the computation of the spectrum, turn into a rather difficult problem
that has not been investigated systematically to this day. In any case,
the emergence of integrable systems on a cylinder in
the light-cone formulation of string theory
in pp-wave backgrounds with non-trivial R-R 5-form, makes it even more
pressing to complete this task in favor of constructing exactly solvable
interacting string models.

Motivated by the recent construction of all supersymmetric solutions in
purely metric geometries of pp-wave type with Ricci flat Kahler transverse space
\cite{malda2},
and some generalizations that were subsequently studied \cite{tsey3},
we have considered
solutions with non-trivial dilaton and NS-NS anti-symmetric tensor fields
that also carry a R-R 5-form. We proposed the use of exact superconformal
building blocks in the transverse space with $N=(4,4)$ world-sheet
supersymmetry, which all have central charge $c=4$ exact to all orders in
$\alpha^{\prime}$. These models, which are well studied, and in many cases
they admit an exact conformal field theory description as Wess-Zumino-Witten
models, have a large amount of supersymmetry that makes them exact solutions
of string theory beyond the lowest order effective theory of type IIB
supergravity. If the superconformal blocks put in the
transverse space are axionic instantons with conformally flat metric, the
pp-wave solutions can also exhibit space-time supersymmetry and generalize
the usual five-brane solutions into new backgrounds with non-trivial
R-R 5-form. We have shown that axionic instanton
backgrounds, due to the conformally flat nature of
their target space metric, allow for
solutions of the defining Killing spinor equations by rotating them to
flat space and imposing the appropriate chiral projections.

In other cases, where the superconformal properties of the blocks
do not manifest in target space, as the
dilatino variation
cannot be made zero, we were able to find some
sporadic, yet quite interesting, solutions by relying directly on the second
order field equations. Some of them can also be described by
performing T-duality on
simpler axionic instanton solutions, which are space-time supersymmetric.
In any case, for appropriate choices of the front factor in the pp-wave
geometry, we have encountered a variety of integrable interacting light-cone
models that correspond to the complex sine-Gordon model, the supersymmetric
Liouville theory or to the complex sine-Liouville model. Since some of these
models correspond to integrable perturbations of the superconformal
building blocks that drive them away from criticality, we think that it
will be interesting to construct more general solutions by making appropriate
choices of the front factor, which can yield
the Lagrangian description of other integrable perturbations
by world-sheet operators in the light-cone gauge.
For example, the effect of turning on magnetic fields
should be examined in this context.

There is a number of other questions that have been raised from our study,
but they still remain unanswered. We will briefly mention a few, since
their answers may
help to make sharper our current understanding of the subject and
even more complete. First,
there is a question regarding the use of T-dualities in the transverse space
to generate new solutions of ten dimensional supergravity. Since T-duality
is a symmetry that relates type IIA to type IIB backgrounds in general, it will be
interesting to develop a systematic interpretation of our solutions within
the type IIA side, in
particular for the supersymmetric ones that arise from axionic instanton
backgrounds, and construct the corresponding
background fields. Also, it will be interesting to determine the general
circumstances under which the T-dual backgrounds of the supersymmetric
solutions based on axionic instantons are also consistent backgrounds
of type IIB theory for appropriately chosen R-R 5-forms, thus allowing
T-duality transformations to act as solution generating symmetries within
type IIB in some cases.
Second, it will be interesting to study further the structure
and the properties of the double axionic instanton solutions in transverse
space and extend the classification of the corresponding
solutions in a systematic way. Third, one may more generally consider
superconformal theories with $N=4$ supersymmetry that also admit torsion, as
they naturally generalize the usual axionic instanton solutions to spaces that are
conformally equivalent to hyper-Kahler manifolds. In this more general setting,
it is natural to expect that the Killing spinor equations, which determine
all possible supersymmetric solutions, can also be solved by rotating them
to the Maldacena-Maoz equations for purely gravitational backgrounds
with curved hyper-Kahler (i.e., Ricci flat and Kahler) transverse space.
Finally, it will
be interesting to introduce D-branes in the light-cone
quantization of the resulting interacting light-cone models and study
their effect in string theory computations, as it was done for the
simpler case of the maximally supersymmetric pp-wave background \cite{green},
where only quadratic (i.e., mass) terms arise in the potential of the
world-sheet bosons, or even more recently for other supersymmetric pp-wave
backgrounds with $(2,2)$ supersymmetry \cite{hiki}.

We hope to return to all these problems elsewhere together with the interpretation
of (some of) our solutions as Penrose limits of more complicated supergravity
backgrounds, if this is at all possible.

\vskip1cm

\centerline{\bf Acknowledgments}
\no
This work was supported in part by the European Research and Training Network
``Superstring Theory" (HPRN-CT-2000-00122). One of us (I.B.) also acknowledges
partial support by the European Research and Training Network ``Quantum Structure
of Space-time" (HPRN-CT-2000-00131), the Greek State Foundation Award
``Quantum Fields and Strings" (IKYDA-2001-22), and the NATO Collaborative Linkage
Grant ``Algebraic and Geometric Aspects of Conformal Field Theories and
Superstrings" (PST.CLG.978785).
He is also thankful to CERN/TH for hospitality and financial support during the early
stages of the present work and to the {\em Nestor Institute of Physics} in Pylos
for providing a stimulating enviroment and excellent facilities during the 2002
European School of High Energy Physics. The other author (J.S.) is grateful to
CERN/TH for hospitality and financial support during his subbatical leave from
Israel. The work of J.S was supported in part by the Israel Science
fundation,
US-Israel Binational Science Foundation and the German-Israeli
Foundation For Scientific Research and Development. J.S also
 wishes to thank J. Maldacena, L. Maoz and S. Theisen  for useful discussions.

\end{document}